\documentclass[12pt]{article}
\usepackage{framed}
\usepackage{amsmath}
\numberwithin{equation}{section}
\usepackage{amssymb,amscd}
\usepackage{bm}
\usepackage{wrapfig}
\usepackage{color}
\usepackage{amsbsy} 
\usepackage{theorem}

\def\be{\begin{equation}}
\def\ee{\end{equation}}
\def\bea{\begin{eqnarray}}
\def\eea{\end{eqnarray}}

\newcommand{\rd}{\mathrm{d}}

\def\cM{{\cal M}}

\newcommand*{\R}{{\mathbb R}}

\newcommand{\bx}{\boldsymbol{X}}

\def\ba{\mbox{\boldmath $A$}}

\def\cba{\mbox{\boldmath $\check{A}$}}

\def\bbd{\mbox{$D$}}

\newcommand{\sbv}[2]{{({{#1},{#2}})_{BV}}}
\newcommand{\ssbv}[2]{{({{#1},{#2}})}}

\def\xzero{X{}}
\def\tilx{\widetilde{X}{}}
\def\tila{\widetilde{A}{}}

\newcommand\Sbvcov{{S_{BV}}{}}

\def\beqa{\begin{eqnarray}}
\def\eeqa{\end{eqnarray}}
\def\beq{\begin{equation}}
\def\eeq{\end{equation}}
\def\be{\begin{equation}}
\def\ee{\end{equation}}

\newcommand{\Gammaz}{\stackrel{\circ}{\Gamma}}

\newcommand{\uX}{{{X}}{}}
\newcommand{\uAplus}{{\underline{A}^+}{}}
\newcommand{\ucplus}{{\underline{c}^+}{}}
\newcommand{\uc}{{\underline{c}}{}}
\newcommand{\uA}{{\underline{A}}{}}
\newcommand{\uXplus}{{\underline{X}^+}{}}
\newcommand{\uphi}{{\underline{\phi}}{}}
\newcommand{\upsi}{{\underline{\psi}}{}}
\newcommand{\uxi}{{\underline{\xi}}{}}

\newcommand{\udx}{{\underline{\rd x}}{}}
\newcommand{\upartial}{{\underline{\partial}}{}}

\def\bbd{{\rd}}
\def\sd{\mbox{\boldmath $\mathrm{d}$}}

\begin{document}

%%%%%%%%%%%%%%%%%%%%%%%%%%%%%%%%%%%%%%%%%%%%%%%%%%%%%%%%%%%%%%%%%%
%%%%%%%%%%%%%%%%%%%%%%%% Title %%%%%%%%%%%%%%%%%%%%%%%%%%%%%%%%%%%
%%%%%%%%%%%%%%%%%%%%%%%%%%%%%%%%%%%%%%%%%%%%%%%%%%%%%%%%%%%%%%%%%%

\baselineskip 6mm %TO HAVE A BIGGER DISTANCE BETWEEN LINES

\begin{titlepage}
\begin{flushright}
\end{flushright}

%\vskip 1cm

\begin{center}
{\Large \bf
BV \& BFV for the H-twisted Poisson sigma model}
\vskip 1cm
Noriaki Ikeda${}^{a}$
\footnote{E-mail:\ 
nikedaATse.ritsumei.ac.jp}
and 
Thomas Strobl${}^{b}$
\footnote{E-mail:\ 
stroblATmath.univ-lyon1.fr
}

\vskip 0.4cm
{
\it
${}^a$
Department of Mathematical Sciences,
Ritsumeikan University \\
Kusatsu, Shiga 525-8577, Japan \\

\vskip 0.4cm
${}^b$
Institut Camille Jordan,
Universit\'e Claude Bernard Lyon 1 \\
43 boulevard du 11 novembre 1918, 69622 Villeurbanne cedex,
France

}
\vskip 0.4cm

{
November 30, 2020
}
\vskip 1.6cm
\vskip 1.6cm

\begin{abstract}
We present the BFV and the BV extension of the Poisson sigma model (PSM) twisted by a closed 3-form $H$. There exist superfield versions of these functionals such as  for the PSM and, more generally, for the AKSZ sigma models. However, in contrast to those theories, they depend on the Euler vector field of the source manifold and contain terms mixing data from the source and the target manifold.

Using an auxiliary connection $\nabla$ on the target manifold $M$, 
we obtain alternative, purely geometrical expressions without the use of superfields, which are new also for the ordinary PSM and promise adaptations to other Lie algebroid based gauge theories: The BV functional, in particular,  is the sum of the classical action, the Hamiltonian lift of the (only on-shell-nilpotent) BRST  differential, and a term quadratic in the antifields which is essentially  the basic curvature and measures the compatibility of $\nabla$ with the Lie algebroid structure   on $T^*M$. %Except for the classical action, all the $H$-dependence is now contained in the torsion of the connection and the structural functions of the Lie algebroid.
We finally construct a %find a %($\nabla$-dependent) 
$\mathrm{Diff}(M)$-equivariant isomorphism between the two BV formulations.

\end{abstract}
keywords: Gauge symmetry; sigma models; topological field theories;  BV formalism; BFV formalism; AKSZ theories; superspace; supergeometry; Lie algebroids.
\end{center}
\end{titlepage}

\newpage

\tableofcontents
%%%%%%%%%%%%%%%%%%%%%%%%%%%%%%%%%%%%%%%%%%%%%%%%%%%%%%%%%%%%%%%%%%%%%%%%%%%
%%%%%%%%%%%%%%%%%%%%%%%%%%%%%%%%%%%%%%%%%%%%%%%%%%%%%%%%%%%%%%%%%%%%%%%%%%%
%%%%%%%%%%%%%%%%%%%%%%%%%%%%%%%%%%%%%%%%%%%%%%%%%%%%%%%%%%%%%%%%%%%%%%%%%%%
\newpage 
\section{Introduction and brief description of the results}

The Poisson sigma model (PSM) \cite{Ikeda,Schaller-Strobl} can be viewed as an ideal toy model for introducing the BV formalism \cite{BV1,BV2}: Most notably, one has a relatively simple action functional giving rise to structure functions in the commutator of  gauge transformations. This is absent in standard Yang-Mills (YM) type gauge theories, where then the less sophisticated BRST formalism is already absolutely sufficient---although with the exception of BF-theories in at least four dimensions, where one encounters a good example for the necessity of ghosts for ghosts already in the abelian YM-setting (see, e.g., \cite{Henneaux-Teitelboim}). Structure functions appear otherwise typically 
within (super)gravity theories, the definition of which requires much more technical knowledge and the ensuing BV extension considerably more calculational efforts.
In addition,  for the PSM, the original fields, the ghosts, and the antifields combine beautifully into superfields such that the BV extension takes almost miraculously the form of the original classical action, just reinterpreted in terms of fields living on a super-extension of the worldsheet. The latter feature does not come as a surprise, however, if the PSM is viewed as the AKSZ model to which it reduces for the choice of a two-dimensional source manifold. 
A brief summary of the respective formulas can be found in Appendix \ref{sec:PSMBV}---for the original literature on this subject we refer to \cite{AKSZ,Cattaneo-Felder,CFAKSZ} and for related reviews to  \cite{RoytenbergAKSZ,Ikeda:2012pv}.

The twisting of the PSM is obtained by adding  a closed 3-form $H$ as a Wess-Zumino term to the action functional of the form of the PSM (or,  if this 3-form is exact, $H=\rd B$, just the pullback of $B$ by the map from the worldsheet $\Sigma$ to the target manifold $M$) \cite{Klimcik-Strobl}. To keep the theory topological, the Poisson condition for the bivector field $\pi$ on $M$ is then modified to \cite{Klimcik-Strobl,Park} 
\begin{equation} \label{PiPi}
 \tfrac{1}{2}[\pi,\pi] = \langle \pi \otimes \pi \otimes \pi , H \rangle \, ,
\end{equation}
where the contraction with $H$  on the right-hand side is over the first, third, and fifth factor of the 6-tensor $\pi^{\otimes 3}$. While Poisson manifolds correspond to Dirac structures in the standard Courant algebroid $TM \oplus T^*M$ which are projectable to the second factor, it was shown in \cite{Severa-Weinstein} that this applies equally well to $(\pi,H)$ satisfying \eqref{PiPi} if the standard Courant algebroid is deformed by the 3-form $H$ to a split exact Courant algebroid.

It is natural to expect that twisting the PSM by a closed 3-form will not change its BV structure very much. Nonetheless, the problem to construct the BV extension of the $H$-twisted Poisson sigma model (HPSM) has resisted all previous attempts for quite some while already. Also, the result we find in this paper turns out to not  meet the above expectation of a simple deformation, see, in particular, Eq.\ \eqref{superfieldBVaction} below. 

In this context, it is important to clarify how this relates to \cite{Park}:  In that paper, the naive super-extension of the HPSM appears as Eq.\ (3.51)---Eq.\ \eqref{naiveBVPSM} in the notation of our paper---and is advocated as a deformation of the PSM. Moreover, Eq.\ \eqref{PiPi} is obtained as a consistency condition there as well---albeit within the AKSZ-BV formalism of a three-dimensional sigma model which is only implicitly related to the HPSM super-action functional \eqref{naiveBVPSM}, namely after integrating out some fields and notably after putting to zero some BV field equations (see also Example 5.1 in \cite{Ikeda:2013wh}). It is  already emphasized in \cite{Park} that \eqref{naiveBVPSM}  is only an on-shell-valid expression of the theory considered there. In particular, it does \emph{not} provide the BV extension of the HPSM; we show this also explicitly in Appendix \ref{sec:naive}:  \eqref{naiveBVPSM}  satisfies  the master equation only upon usage of precisely those BV field equations mentioned above (or if $H$ is put to zero).

There is, in fact, a simple direct argument, without a calculation, which shows that the super-extension \eqref{naiveBVPSM} of the action functional of the HPSM cannot satisfy the classical BV master equation: The defining twisted Poisson condition, Eq.\ \eqref{PiPi}, is up to cubic in $\pi$, the master equation $(S_{BV},S_{BV})=0$  is at most quadratic in the bivector if  the functional $S_{BV}$ depends on $\pi$ at most linearly and if the BV bracket is independent of $\pi$. Since use of the condition \eqref{PiPi} will be necessary for a BV master equation to hold true---after all it was shown in \cite{Klimcik-Strobl} to be precisely this condition that ensures the maximal gauge symmetry of the model---either the BV functional $S_{BV}$ or the BV bracket $(\cdot ,\cdot)$ need to be at least quadratic in $\pi$. As the result \eqref{superfieldBVaction} shows, the functional is even a cubic polynomial in $\pi$, at least when keeping the standard BV symplectic form in Darboux coordinates. 

The BFV form of a gauge theory  \cite{Batalin:1977pb}\cite{Batalin:1983pz}  is usually easier to construct than its BV extension. This is in part due to the fact that the dimension is lowered by one, but also due to particularities of the Hamiltonian formalism. In addition, we will be able to take recourse to \cite{Ikeda-Strobl} for its construction in the present context.
It turns out that the resulting BFV-theory of the HPSM can be formulated in terms of super-fields on the super-circle $T[1]S^1$. This is possible only by means of the use of the Euler vector field $\tilde{\varepsilon}$ of the source manifold, however, 
see \eqref{superBFVBRSTcharge} below.

This feature is absent for the ordinary PSM, resulting from putting $H$ to zero, and is shared by the BV  and the BFV form of the HPSM  (as we will show). In fact, in the AKSZ framework, the BV functional and the BFV charge are the sum of a differential on the source manifold and a differential on the target manifold, both lifted in a Hamiltonian way to the mapping space. The differential on the source manifold is the de Rham differential, the differential on the target manifold results from a (possibly higher) Lie algebroid structure which is compatible with a symplectic form. The two differentials commute (in a graded sense) since essentially they operate on different spaces and thus they can be added to yield the total differential, which then is identified with $S_{BV}$ or $S_{BFV}$ for an AKSZ sigma model. The twisting of a Poisson structure by a 3-form $H$ still defines a Lie algebroid structure on $T^*M$,\footnote{As recalled above, it was shown by \v Severa-Weinstein that twisted Poisson structures correspond to particular Dirac structures. Dirac structures are Lie algebroids. Since these Dirac structures inside $TM \oplus T^*M$ are projectable to $T ^*M$,  the Lie algebroid structure can be transferred to $T ^*M$.} but the corresponding nilpotent vector field $Q$ is no more compatible with the canonical symplectic structure on $T^*[1]M$; thus, the AKSZ procedure cannot be applied here, at least not for a two-dimensional source manifold. This reflects itself also in the resulting BV and BFV functionals: they necessarily contain terms mixing both, the data of the target manifold ($H$ and $\pi$ here) and the data of the source manifold (the de Rham differential $\sd$ and the Euler vector field $\bm{\varepsilon}$), both lifted to field space in a canonical way. The latter fact and the possibility to use superfields to express the BV and BFV functionals show yet in another way that the HPSM is as close to an AKSZ theory as one can come without being an AKSZ theory.

%\vspace{3mm}

When dealing with a sigma model, %such as the PSM or its generalization, the HPSM, 
one is naturally confronted with the question of how the theory---and in particular also its extensions  containing information about its gauge symmetries---behave with respect to a change of coordinates on the target manifold $M$. Here an interesting feature of the (H)PSM appears: while on-shell, i.e.\ on solutions of the Euler-Lagrange (EL) equations, the gauge symmetries are globally well-defined without any additional structure on $M$ except for the (twisted) Poisson data, their off-shell description is described best by means of an additional auxiliary connection $\nabla$. 

Let us provide some more details about this: the fields of the PSM are the vector bundle morphisms from the tangent bundle $T\Sigma$ of a two-dimensional, orientable manifold $\Sigma$ to the cotangent bundle $T^*M$ of the Poisson manifold $M$. This is essentially the same for the HPSM, just that $\Sigma$ must not have a boundary in this case and there are the usual subtleties of Wess-Zumino terms, which effect the definition of the functional, but are of no relevance for the EL field equations and into which we therefore do not want to enter here. $T\Sigma$ as well as $T^*M$ are both Lie algebroids, for the PSM and the HPSM, and the EL equations are satisfied iff the vector bundle morphisms are Lie algebroid morphisms (see \cite{Bojowald-Kotov-Strobl,Kotov-Schaller-Strobl}, also for the following statement). Moreover, two such solutions are gauge equivalent, iff they are related by a Lie algebroid homotopy. This is a globally well-defined notion evidently, which can be expressed in terms of the data defining the sigma model. 

Certainly, the group $\mathrm{Symm}$ of all symmetries of a functional is completely defined by means of this functional as well. However, for every gauge theory, there is the following exact sequence of  symmetry groups \cite{Henneaux-Teitelboim}
\beq 0 \to \mathrm{Symm}_{triv} \to  \mathrm{Symm} \to  \mathrm{Symm}_{onshell} \to 0 \, . \label{seq}
\eeq 
Here $\mathrm{Symm}_{triv}$ denotes the group of symmetries of the functional which does not move any solution of the EL equations. More importantly, every functional has such trivial gauge symmetries and one talks of a gauge theory only if the on-shell-part $\mathrm{Symm}_{onshell}$, the quotient of all symmetries by these trivial ones, is non-trivial. For the BV formalism, in particular, one needs generators of the non-trivial gauge symmetries acting on the space of fields. This requires a splitting of the sequence \eqref{seq} (as sets, not necessarily as groups). It turns out that, for the PSM,  such a splitting is singled out  by the choice of a torsion-free connection $\nabla$ on $M$. More generally, for the HPSM, every choice of a connection $\nabla$ provides a splitting of  \eqref{seq} provided only that the torsion $\Theta$ of the connection is given by the formula\footnote{Certainly, in the formula for the gauge transformations which contain the corresponding connection coefficients, one can replace them by connection coefficients of a torsionless connection plus a contorsion contribution and view the latter term as an independent one to the gauge transformations. However, this is also not only unnatural, it renders the calculations more complicated.}
\beq \Theta = \langle \pi, H \rangle , \label{torsionpiH}
\eeq 
where,  on the right-hand side, the second factor of the bivector field $\pi$ is contracted with the first entry of the 3-form $H$. 

In the traditional literature \cite{Ikeda,Schaller-Strobl,Cattaneo-Felder} about the PSM, the generators of gauge symmetries were provided in particular target space coordinates  without the use of a connection. These formulas do not transform properly with respect to changes of such coordinates  \cite{Bojowald-Kotov-Strobl}. In the language introduced above, they correspond to the local choice of a flat connection and adapted coordinates $x^i$  such that the connection coefficients $\Gamma^i_{jk}$ vanish.
%They may be viewed as resulting from the choice of a flat connection $\nabla$ in the description provided above, being valid only in a local patch $U \subset M$  
%(i.e.\ for maps with an image lying inside $U$) and for a choice of coordinates $x^i$ such that the locally defined connection coefficients $\Gamma^i_{jk}$ vanish. 
In this way  the description of the gauge symmetries does not only lose covariance with respect to target space diffeomorphisms but, at least for a manifold $M$ not admitting a flat connection, also its global validity. 

Subsequently several works used auxiliary connections in different ways so as to obtain a globally valid description of the PSM and its applications to deformation quantization, see, e.g., \cite{Losev,CFglobal,PSMglobal1,PSMglobal2} (but see also \cite{Bojowald-Kotov-Strobl}  for a global alternative of defining the off-shell symmetry generators without the use of a connection). It is still important to note in this context that---from this perspective somewhat miraculously---the BV-functional of the PSM \cite{Cattaneo-Felder}  is globally well-defined even without the use of a connection. This follows immediately when it is obtained from the AKSZ procedure \cite{CFAKSZ}, where no reference to coordinates or a connection is needed for its definition. (As pointed out in \cite{Losev}, the situation may change again when turning to the subsequent gauge fixing.)

As remarked above, the AKSZ formalism is not applicable to describe the HPSM. Thus, for the construction of the BV extension of the HPSM, we will start directly with the gauge symmetries containing the connection coefficients. 
Since these formulas depend only on the Lie algebroid structure (for $E:=T^*M$) and the connection $\nabla$, this has also the advantage that one may expect similar results for the BV extension of other algebroid based gauge theories, including non-topological ones (see, e.g., \cite{StroblAYM,Zucchini}).

In this geometrical approach, the resulting BV functional looks relatively simple: the terms extending the classical action to first order in the antifields (the momenta in the BV formalism) are nothing but the Hamiltonian lift of the gauge transformations---together with the Lie algebroid structure on the ghosts---to the BV phase space. They correspond to a  BRST  operator squaring to zero on-shell only; thus the extension requires further terms of higher order in the antifields. There is only precisely one more term, quadratic in the antifields and proportional to the basic curvature $S$ of $(E,\nabla)$.
 %, see \eqref{SBV2} below. 
 The tensor $S$ measures the deviation of $\nabla$ to be compatible with the Lie algebroid structure on $E$ \cite{Blaom,Mayer-Strobl,Abad-Crainic,Kotov-Strobl2}. The resulting  BV functional $S_{BV}^\nabla$ is given by \eqref{Mainresult2} or \eqref{Mainresult2prime} below. We show, moreover, that this BV formulation of the HPSM is globally well-defined and that different choices of the auxiliary connection $\nabla$ satisfying \eqref{torsionpiH} provide isomorphic BV descriptions. 
 
To obtain the superfield formulation $S_{BV}$ of the HPSM, we will return to the a local patch representative of $S_{BV}^\nabla$ for a choice of local coordinates on the target manifold and choose the corresponding connection coefficients such that their torsion-free part vanishes. At this point then nothing guarantees that there exists a superfield reformulation of this locally defined functional or even that it has any global meaning. In fact, since one dismissed connection coefficients, one would rather expect that this would not be the case. But it is. The covariance with respect to changes of coordinates on $M$ is achieved by a different behavior of the fields in this picture: For example, while within $S_{BV}^\nabla$ the classical fields transform as they do within the classical action---corresponding to vector bundle morphisms in particular---within $S_{BV}$ the 1-form part of these fields receive ghost contributions upon a change of target space coordinates. 

In fact, we will even establish a global equivalence of the two BV descriptions: 
Denoting by $(\cM_{BV},\omega_{BV},S_{BV}^\nabla)$ the data obtained in the context of a connection and by $(\cM,\omega,S_{BV})$ those from the superfield formalism, 
there exists a  $\nabla$-dependent isomorphism
\beq (\cM,\omega,S_{BV}) \xrightarrow{\sim} (\cM_{BV},\omega_{BV},S_{BV}^\nabla) \, ,
\label{iso}\eeq
which intertwines the BV-symplectic group action of  $\mathrm{Diff}(M)$ on the two spaces. %Here the left-hand side of \eqref{iso} denotes the BV data that one obtains in the superfield formalism and its right-hand side those obtained in the context of the choice of a connection.

We think that the HPSM is a very nice extension of the PSM and AKSZ theories as a toy model for a better understanding and development of the BV and the BFV formalism and we expect that still much more can be learned by its study. For example, as we intend to show elsewhere, while applying the recipe \cite{BBD1,BBD2,GrigorievD,Grigoriev,Ikeda-Strobl}  for a transition from BFV to BV leads to the known BV formulation of the PSM, for non-vanishing $H$ the resulting BV functional breaks worldsheet covariance, i.e.\ the covariance with respect to $\Sigma$.

The structure of the paper is as follows: In Sec.\ \ref{sec:HPSM} we recall the action functional of the HPSM, its gauge symmetries, and its Hamiltonian formulation. In Sec.\ \ref{sec:BFV} we construct the BFV formulation and present the result in different forms,  with and without an auxiliary connection in particular.
In Sec.\ \ref{sec:BV1} we construct  the BV action   $S_{BV}^\nabla$ ab initio using a connection and in Sec.\ \ref{sec:BV2} we provide and study the alternative superfield functional $S_{BV}$. 

Appendix \ref{sec:PSMBVBFV} summarizes  known formulas for the PSM. In Appendix \ref{sec:naive} we show explicitly that a naive superfield extension of the classical theory does not yield the correct BV extension for the HPSM. Appendix \ref{appC} contains the full extension in which the BV symplectic form carries an $H$-contribution. In Appendix \ref{sec:BianchiofS}, finally, we establish an identity of the basic curvature $S$, which is needed in proving the master equation for $S_{BV}^\nabla$.

 \newpage 
\section{Twisted Poisson sigma models}\label{sec:HPSM}
\subsection{Lagrangian formulation in a local patch on the target}
In this section we briefly summarize the classical formulation of the  3-form twisted Poisson sigma model. The target space of this topological model is a twisted Poisson manifold 
$(M,\pi,H)$, i.e.\ an n-dimensional manifold $M$ equipped with a closed 3-form $H$ and a bivector field $\pi$ such that 
 the couple $(\pi,H)$ satisfies the defining identity of twisted Poisson geometry \eqref{PiPi}. The source space of the theory is an oriented 3-manifold $N$, whose boundary $\Sigma=\partial N$ is thus a closed 2-manifold with the induced orientation. The fields of the model are smooth maps $X  \colon N \to M$ which, upon restriction to the boundary $\Sigma$, are enhanced to vector bundle morphisms $a \colon T\Sigma  \to T^*M$.  So, if we choose local coordinates $(x^i)_{i=1}^{n}$ on the target manifold $M$, the fields are $n$ functions $X^i=X^*(x^i)$ on $N$ (and thus also on its boundary $\Sigma$) and an equal amount of 1-forms $A_i= A_{\mu i} \rd\sigma^{\mu} $ living only on $\Sigma$, on which we will use the coordinates $(\sigma^{\mu})\equiv (\sigma^0,\sigma^1)$.
 
 Then the classical action functional of the theory takes the following form  
\begin{eqnarray}
\! \!\!S = \int_{\Sigma=\partial N} 
%\left(
A_i \wedge \rd \xzero^i + \tfrac{1}{2} X^* \pi^{ij} A_i \wedge A_j 
%\right)
+ \int_{N} X^*H \, ,
\label{classicalactionofHPSM}
\end{eqnarray}
which is a non-local functional for the fields living on the boundary $\Sigma$ due to the Wess-Zumino term induced by $H$. This non-locality is very mild, however: The variation of $S$ leads to following local Euler-Lagrange equations, only restricting the fields on $\Sigma$:
\begin{eqnarray}
F^i := \rd X^i + \pi^{ij} A_j &=& 0  \, ,
 \label{eom1}\\
G_i := \rd A_i + \tfrac{1}{2} \pi^{jk},_{i}
A_j \wedge A_k + \tfrac{1}{2} H_{ijk} \rd X^j\wedge \rd X^k &=&0  \, ,
\label{eom2}
\end{eqnarray}
where the dependence of $\pi^{ij}$, its partial derivative $\pi^{ij},_{k}\equiv \partial \pi^{ij}/\partial x^k$, and $H_{ijk}$ on $X(\sigma)$ is understood (we could have equally added a symbol expressing the pullback by $X$,  as in \eqref{classicalactionofHPSM}, restricted to the boundary $\Sigma$, but we prefer to not overload the notation; sometimes we will, however, write, for example, $\pi^{ij}(X)$ so as to emphasize the $X$-dependence).

The equations \eqref{eom1} and \eqref{eom2} have dependences among one another, $n$ local ones in a standard counting: one verifies that\footnote{We are grateful to A.\ Hancharuk for making us aware of a mistake in a first version of this formula.} 
\beq \rd F^i + (\pi^{ij},_k + \pi^{il} \pi^{jm} H_{klm}) A_j \wedge F^k - \pi^{ij} G_j - \tfrac{1}{2} \pi^{ij} H_{jkl} F^k \wedge F^k  \equiv 0 ; \label{Bianchi}
\eeq the combination on the left-hand side vanishes identically due to \eqref{PiPi}, thus not restricting the fields in any way. This implies that the theory has gauge symmetries (the inverse statement, namely that gauge symmetries imply dependencies of the field equations, is known under the name of  Noether's second theorem). Here, inside a local patch on the target, these symmetries can be parametrized by means of a set of functions $\epsilon_i(\sigma)$ 
\cite{Kotov:2014dqa}:
\begin{eqnarray}
\delta \xzero^i &=& - \pi^{ij} \epsilon_j,
\label{trafoX}
\\
\delta A_{i}  
&=& 
\rd \epsilon_i
+  \pi^{jk},_{i} A_{j} \epsilon_k
+ \tfrac{1}{2} \pi^{jk} H_{ijl} ( \rd \xzero^l 
- \pi^{lm} A_{m}) \epsilon_k.
\label{gaugetransformation02a}
\end{eqnarray}
In the particular case where $H=0$, the third term in the second transformation equation disappears and one obtains the gauge transformations of the Poisson sigma model as given in their original form \cite{Ikeda,Schaller-Strobl}.
A direct calculation shows that the classical action $S$ is strictly invariant under the transformations induced by \eqref{trafoX} and \eqref{gaugetransformation02a}. 
This is due to the fact that $\Sigma$ cannot have any boundary itself, being the boundary of a 3-manifold. If, on the other hand, one considers $\int_\Sigma X^* B$ instead of the third term in \eqref{classicalactionofHPSM}, and one works on a general oriented $\Sigma$, the equations of motion \eqref{eom1}, \eqref{eom2} remain still valid when taking $H= \rd B$; now the transformations \eqref{trafoX} and \eqref{gaugetransformation02a}, with this choice of $H$ and \eqref{PiPi} holding true, are still gauge transformations, since, as one shows, $S$ changes by the following  boundary term only:
\begin{eqnarray}
\delta S 
&=& 
\int_{\Sigma} \rd \left[ \epsilon_i\left(\rd \xzero^i 
+\pi^{ik} B_{kj}  \rd \xzero^j\right) \right].
\end{eqnarray} 

The transformation equation \eqref{gaugetransformation02a} becomes  more transparent, if rewritten as follows:
\begin{equation}\label{trafoAf}
\delta A_{i}  = \rd \epsilon_i
+ f_i^{jk} A_{j} \epsilon_k
+ \tfrac{1}{2} \pi^{jk} H_{ijl} \, F^l \, \epsilon_k.
\end{equation}
Here, as before, $F^i \equiv \tfrac{\delta S}{\delta A_i}$ and we introduced the abbreviation
\begin{equation}\label{f}
f_k^{ij} \equiv \pi^{ij},_{k}
+ \pi^{il} \pi^{jm} H_{klm} 
 \, .
%\right)
\end{equation}
%and, as already before, the pullback by $X$ is understood in \eqref{symmf}.
At this point it is useful to recall the geometrical reinterpretation  \cite{Severa-Weinstein} of twisted Poisson structures  in terms of Dirac geometry: The $H$-twisted Courant-Dorfman bracket between sections of $TM \oplus T^*M$ is of the form
\begin{equation}\label{CDbracket}
[u+\alpha,v+\beta]_{H} = [u,v] + {\cal L}_u\beta - \iota_v \rd \alpha - \iota_u\iota_v H \, .
\end{equation}
Consider sections of the subbundle $D$ given by the graph of a bivector field $\pi$, i.e.\ where the vector field part $v$ is determined uniquely by the 1-form part $\alpha$ in terms of $v=\iota_\alpha \pi$. The bracket \eqref{CDbracket} of two such sections gives again a section of this subbundle, iff \eqref{PiPi} holds true. The vector bundle $D$ is in fact a particular Lie algebroid over $M$. To determine its structure functions $C^{ij}_k$ in a holonomic basis induced by coordinates $x^i$ on $M$, one needs to calculate 
\begin{equation}\label{ijH}
[\pi^{im} \partial_m + \rd x^i,\pi^{jn} \partial_n + \rd x^j]_H =C^{ij}_k \, (\pi^{ks} \partial_s + \rd x^k) \, .
\end{equation}
A direct computation, using \eqref{PiPi}, yields $C^{ij}_k = f_k^{ij}$. $D$ being projectable to $T^*M$ implies that these two bundles can be identified. We thus deal with a Lie algebroid on $E=T^*M$. Its anchor $\rho$ is induced by the bivector $\pi$, $\rho(\alpha) := 
\iota_\alpha \pi$, and the Lie bracket in the holonomic basis is an $H$-twisted version of the Koszul bracket \cite{Koszul} of the standard Poisson case, 
\begin{equation}\label{ijE}
[\rd x^i, \rd x^j] =f_k^{ij} \, \rd x^k \, .
\end{equation}
%with $f_k^{ij}$ given by formula \eqref{f} above.

In view of the above discussion, it is useful to introduce the quantity 
\beq F_i := \rd A_i + f^{jk}_i A_j \wedge A_k \label{newF}
\eeq 
and to observe that the field equations \eqref{eom1} and \eqref{eom2} can be equivalently expressed as $F^i=0$, $F_i=0$. Equation \eqref{Bianchi} now simplifies to $  \rd F^i + \pi^{ij},_k  A_j \wedge F^k - \pi^{ij} F_j = 0 $ ,
which can be viewed as one of two Bianchi identities for the ''field strengths'' (or generalized curvatures), see Eqs.\ (2.10) and (2.11) in \cite{Gruetzmann-Strobl}.

%We finally want to draw the reader's attention to the fact that the third term in \eqref{gaugetransformation02a} is not proportional to the field equations, due to an important difference in relative signs. \eqref{trafoAf} is obtained by changing this sign to the one appearing in \eqref{eom1}, which then automatically transforms the partial derivatives of the bivector field into the correct combination \eqref{f} for the structural functions of the Lie algebroid governing the sigma model.

\subsection{Target space covariance by means of an auxiliary connection}\label{sec:cov}

Thus we now see that the gauge transformations of the $H$-twisted Poisson sigma model \eqref{classicalactionofHPSM} are closely related to the Lie algebroid $E \cong D$ recalled above: it is its anchor map that acts on the scalar fields $X^i$ and the transformations of the 1-form gauge fields are governed by the Lie algebroid structure functions \eqref{ijE}, up to a term vanishing on-shell, i.e.\ vanishing for fields satisfying the field equations, here concretely equation \eqref{eom1}.

The gauge symmetries \eqref{trafoX} and \eqref{trafoAf} are now essentially of the form of the general discussion of gauge symmetries of Lie algebroid theories as started in \cite{Bojowald-Kotov-Strobl,StroblAYM} and further developed in \cite{Mayer-Strobl,Gruetzmann-Strobl,Kotov-Strobl1,Carow-Watamura:2016nwj,Carow-Watamura:2016lob,Zucchini,Wright,Kotov-Strobl2}. The setting is as follows. The gauge fields correspond to vector bundle morphisms $a \colon T \Sigma \to E$ where $\Sigma$ is $d$-dimensional spacetime (in our case $d=2$) and $E \to M$ a Lie algebroid. Then $a$ corresponds to a map $X \colon \Sigma \to M$ and 1-form gauge fields $A \in \Omega^1(\Sigma, X^*E)$. After choosing a local frame $e_a$ of sections of $E$, these give rise to the 1-form fields $A^a$. One consistent way of parametrizing gauge symmetries is then parametrized by a connection $\nabla$ on $E$ and is of the form $\delta^\nabla X^i = \rho^i_a(X) \epsilon^a$ together with\footnote{In \cite{Bojowald-Kotov-Strobl,StroblAYM}, the focus was put  more on a Lie derivative version of the gauge symmetries, mentioning the connection version only as a side remark; but the considerations are very similar and one of the two versions is needed for consistency, as recalled also in the paragraph below. The Dirac sigma model \cite{Kotov-Schaller-Strobl} provides an example of a Lie algebroid theory, where the gauge symmetries of this form are not sufficient, requiring a second connection on $E$ and an additional term proportional to $\ast F^i$ in the parametrization of the infinitesimal symmetries. See also the more recent development about Dirac sigma models as a universal gauge theory in two dimensions and geometrical interpretations of the two connections: \cite{youngpeople1,youngpeople2,youngpeople3,transverse,Ikeda:2019pef,Wright}.}
\begin{equation} \label{deltaAa}
\delta^\nabla A^a = \rd \epsilon^a + C^a_{bc}(X) A^b \epsilon^c + \omega^a_{bi}(X) \epsilon^b F^i \, .
\end{equation}
Here $\rho \colon E \to TM$ is the anchor of the Lie algebroid, $\rho(e_a)=\rho^i_a (x) \partial_i$, and 
$C^a_{bc}$ are the Lie algebroid structure functions in the chosen frame, $[e_a,e_b]=C^c_{ab}(x) e_c$, 
$F^i \equiv \rd X^i - \rho^i_a(X) A^a$, and $\omega^a_{bi}(x)$ are the coefficients of the connection $\nabla$ on $E$, $\nabla e_a = \omega^b_{ai} \rd x^i \otimes e_b$.

The connection is needed to specify the gauge symmetries in such a way that they glue correctly from one patch underlying one choice of frame $e_a$ in $E$ to another one. Note that this still does \emph{not} mean that $\delta A^a$ is tensorial in the index $a$ due to the following Leibniz rule: $\delta^\nabla \!(M^a_b(X) A^b) = \delta^\nabla \! (M^a_b(X)) A^b + M^a_b(X) \delta^\nabla \!(A^b)$, where $\delta^\nabla \! (M^a_b(X)) \equiv M^a_{b,i}(X) \rho^i_a(X) \epsilon^a$. Although \eqref{deltaAa} already contains  connection coefficients, it still does not contain the change of frame induced by a change of the base point for $A = A^a \otimes \underline{e}_a \in \Omega^1(\Sigma, X^*E)$. Here $\underline{e}_a$ denotes the basis in $X^*E$ corresponding to a basis $e_a$ in $E$. If one sets 
\beq  \label{deltatensor}\delta^\nabla \!A := \bar{\delta}^\nabla \! A^a \otimes \underline{e}_a,
\eeq 
 then  
\begin{equation} \label{bardeltaAa}\bar{\delta}^\nabla \! A^a = \delta^\nabla \!A^a +\omega^a_{bi}(X) A^b \delta^\nabla \! X^i \, ,
\end{equation}
which is now a tensorial object in the free index. Note that $\bar{\delta}^\nabla$ still satisfies the Leibniz rule when applied to a product like $M^a_b(X) A^b$, but now, in this index-notation, one needs to treat $M^a_b$ as the components of an endomorphism $M \in \Gamma(X^*\mathrm{End}(E))$; then $\bar{\delta}^\nabla \! (M^a_b(X)) \equiv M^a_{b;i}(X) \rho^i_a(X) \epsilon^a$, where the semicolon indicates the covariant derivative. Combining  \eqref{deltaAa}, \eqref{deltatensor}, and \eqref{bardeltaAa}, one finds the inherently tensorial 
expression\footnote{T.S.\ is grateful to S.\ Fischer for clarifying discussions in this context. --- See also \cite{Kotov-Schaller-Strobl}.}
\begin{equation} \label{deltaA}\delta^\nabla \!A = \mathrm{D} \epsilon - ({}^ET\circ X)(A,\epsilon) \, .
\end{equation}
Here $\epsilon$ is a section in $\Gamma(X^*E)$ and $ \mathrm{D}$ denotes the exterior covariant derivative induced in this bundle by $\nabla$.  ${}^ET \in \Gamma(E \otimes \Lambda^2 E^*)$, finally, is the \emph{negative} of the $E$-torsion of the Lie algebroid covariant derivative ${}^E\widetilde{\nabla}$ which is defined for all $s,s' \in \Gamma(E)$ simply by means of ${}^E\widetilde{\nabla}_s s' := \nabla_{\rho(s)} s'$: \beq {}^ET(s,s') =  - \nabla_{\rho(s)} s' + \nabla_{\rho(s')} s + [s,s'].\label{defEtorsion}\eeq

Let us now apply these general ideas to the present context for the Lie algebroid $E=T^*M$ induced by the Dirac structure $D=\mathrm{graph}(\pi)$. We have not introduced a connection for describing the gauge symmetries of the $H$-twisted PSM; this implies that the expressions found above,  equations \eqref{trafoX} and \eqref{gaugetransformation02a} or \eqref{trafoAf}, can hold true only in one coordinate patch in which, in addition, the connection coefficients have a particular form. Let us introduce a connection $\nabla$ on $TM$ with its connection symbols  $\Gamma^k_{ij}$ in a holonomic frame, $\nabla \partial_i = \Gamma^k_{ij} \rd x^j \otimes \partial_k$; as usual, a connection on $TM$ induces one on $T^*M$ and\footnote{Our conventions are such that a covariant derivative  of a vector field $v$ along  $\partial_i$ has components $\nabla_i v^j \equiv v^j_{;i} = v^j_{,i} + \Gamma^j_{ki} v^k$.} 
\beq \nabla \rd x^i = - \Gamma^i_{kj} \rd x^j \otimes \rd x^k. \label{conn}\eeq 
Specifying the equation \eqref{deltaAa} 
to our case $E=T^*M$ and taking into account \eqref{ijE}, we obtain
\begin{equation} \label{deltaAnabla}
\delta^\nabla \! A_{i}  = \rd \epsilon_i
+ f_i^{jk} A_{j} \epsilon_k
- \Gamma^k_{ij} \, F^j \, \epsilon_k.
\end{equation}
The comparison now shows that in a frame underlying Equation  \eqref{trafoAf}, one has 
$\Gamma^k_{ij}=- \tfrac{1}{2} \pi^{km} H_{mij}$. Note that this expression is antisymmetric in the lower two indices, which means that it corresponds to a non-zero torsion of $\nabla$ (for a non-vanishing contraction of $H$ with $\pi$). More generally, calculating the change of the action 
\eqref{classicalactionofHPSM} under the infinitesimal transformation \eqref{trafoX} and \eqref{deltaAnabla}, one finds that it is invariant under these transformations iff 
\begin{equation} \label{Gamma}
\Gamma^k_{ij}=\stackrel{\circ}{\Gamma^k_{ij}}- \tfrac{1}{2} \pi^{km} H_{mij} \, 
\end{equation}
where 
 $\Gammaz$ are the connection coefficients of an \emph{arbitrary torsion-free} connection. We thus find also a geometrical interpretation of the last term in the local form of the gauge transformations  \eqref{trafoAf}: it is the (con)torsion of the connection $\nabla$. In contrast to the case $H=0$, i.e.\ to the ordinary PSM, there now does not exist a patch where the connection can be made to disappear, only its torsion-free part $\stackrel{\circ}{\nabla}$ can be made to vanish while the torsion of $\nabla$ is completely fixed by the required gauge invariance.  
 
We now return to the discussion in the introduction around \eqref{seq}. An important part of the trivial gauge transformations of the HPSM is generated by means of 
 \beq  \delta^{triv} X^i = \Upsilon^{ij} \ast G_j \, , \qquad \delta^{triv} A_i = \Lambda_{ij} F^j + \Xi_{ij} * F^j  \eeq  
 where $*$ denotes the Hodge duality operator for an auxiliary metric on $\Sigma$ and the matrices $\Upsilon$, $\Lambda$, and $\Xi$ are antisymmetric, symmetric, and antisymmetric, respectively. This is the case since, for example, $F^i = \delta S/\delta A_i$ and $\Lambda_{ij} F^i \wedge F^j \equiv 0$. If we require $\Lambda$ to be proportional to the infinitesimal parameters $\epsilon_i$ from before, one arrives at the identification
 \beq  \Lambda_{ij} = \stackrel{\circ}{\Gamma^k_{ij}} \epsilon_k . \label{Lambda} \eeq 
Thus, for the (H)PSM, a splitting of \eqref{seq}  (as sets) is provided by means of the choice of a torsion-free connection on $M$ (together with canonical choices such as $\Upsilon=0$ and $\Xi=0$).
 
We finally specify the covariant formula \eqref{deltaA} to our setting. The gauge parameter $\epsilon$ is a section of  $X^*T^*M$,  $A$ and $\delta^\nabla \! A$ are 1-forms on $\Sigma$ with values in that bundle, $\delta^\nabla \! A \in \Omega^1(\Sigma,X^*T^*M)$, and one has\footnote{Also \eqref{newF} has covariance problems, but $F^\nabla_i := F_i + \Gamma_{ij}^kA_k \wedge F^j$ results into 
$F^\nabla = \mathrm{D} A - \tfrac{1}{2}T(A,A).$}
\beq  \label{deltaAhere}\delta^\nabla \!A = \mathrm{D} \epsilon - T(A,\epsilon) \, .
\eeq 
Here we denoted the $E$-torsion for $T^*M$ simply by $T$ and consider the composition with $X$ to be understood in such formulas. For the components of $T \equiv \tfrac{1}{2} T^{ij}_k \partial_i \wedge \partial_j \otimes\rd x^k $ one  finds 
\beq  T^{ij}_k = - f^{ij}_k - \pi^{il} \Gamma^j_{kl} + \pi^{jl} \Gamma^i_{kl}  \, . \label{Tcomp}
\eeq 
The Lie bracket of a  Lie algebroid $E$---here $E=T^*M$---is not tensorial due to the Leibniz rule it satisfies; the tensor $T$ provides a tensorial form of the structure functions at the expense of introducing a connection.  Implementing our expressions for the structure functions \eqref{f} and 
 the connection coefficients \eqref{Gamma}, one obtains the simple formula
 \beq T = - \stackrel{\circ}{\nabla} \!\pi \, . \label{nablapi}
 \eeq
 So, the $E$-torsion turns out to be independent of the 3-form $H$  and we 
 see that  $H$ enters the gauge transformations \eqref{deltaAhere} only by means of the torsion of the (ordinary) exterior covariant derivative $\mathrm{D}$.

 \subsection{Hamiltonian formulation}
 \label{Hamform}
 
We conclude this section with the Hamiltonian formulation of \eqref{classicalactionofHPSM}. It is obtained by the standard Dirac procedure \cite{Dirac} or, more easily, by the Faddeev-Jackiw \cite{Faddeev-Jackiw} approach. It is well-defined for the closed 1-manifold $S^1$, i.e.\ describing the (topological)  propagation of closed strings. This is because the Wess-Zumino term only twists the canonical symplectic form on the cotangent space of loop space, which we identify with bundle maps from $TS^1$ to $T^*M$; denoting the ``spatial''  component $A_{1i}$ of $A_i$ by $p_i$, one has:
\begin{eqnarray}
\omega &=& \oint_{S^1}  \rd \sigma 
\left(\delta \xzero^i \wedge \delta p_{i}
+ \tfrac{1}{2} H_{ijk} \, \partial \xzero^i 
\delta \xzero^j \wedge \delta \xzero^k
\right)\, ,
\label{classicalsymplecticform}
\end{eqnarray}
where  $\sigma \equiv \sigma^1$  denotes a 2$\pi$-periodic coordinate around the circle $S^1$ and $\partial \equiv \partial/\partial \sigma$. 
This corresponds to the following fundamental classical Poisson brackets
\begin{eqnarray}
\{\xzero^i(\sigma), \xzero^{j}(\sigma^{\prime})\}  &=&0 \nonumber \\
\{\xzero^i(\sigma), p_{j}(\sigma^{\prime})\}  &=& \delta^i{}_j \delta(\sigma - \sigma^{\prime}),
%\\
%\{c_i, A_{1}^{+j} \}  &=& \delta_i{}^j \delta(\sigma - \sigma^{\prime}),
\label{classicalPB}
\\
\{p_{i}(\sigma), p_{j}(\sigma^{\prime})\}  &=& 
%- %\tfrac{1}{2} 
- H_{ijk}(\xzero) \partial \xzero^k \delta(\sigma - \sigma^{\prime}). \nonumber
%\label{classicalPB2}
\end{eqnarray}
The model is a constrained system, with the fields  $A_{0 i}$ serving as Lagrange multipliers for the following constraints:
\begin{eqnarray} \label{G}
J^i \equiv \partial_1 \xzero^i + \pi^{ij}(\xzero) p_{j} \approx 0.
\end{eqnarray}
These are of the first class, i.e.\ forming a coisotropic submanifold in the (infinite-dimensional) phase space, precisely according to \eqref{PiPi} \cite{Klimcik-Strobl} (cf. also \cite{Alekseev-Strobl} for a more general and detailed discussion of such constrained systems).  An explicit calculation yields
\begin{eqnarray} \label{constraintalgebra}
\{J^i(\sigma), J^j(\sigma^{\prime})\}  = - 
f^{ij}_k(X(\sigma)) \, J^k(\sigma) \delta(\sigma - \sigma^{\prime})\, ,
\end{eqnarray}
where $f^{ij}_k$ are again the structure functions of our Lie algebroid, see equations \eqref{f} and \eqref{ijE} above, pulled back here by $X \colon S^1 \to M$. 
According to the fact that the theory is parametrization invariant and ``topological'', the Hamiltonian is a combination of the constraints and vanishes on-shell, i.e.\ on the constraint surface:
\begin{eqnarray}
{\cal H} =  \oint_{S^1} \rd \sigma A_{0i} J^i  \approx 0.
\label{TPSMhamiltonian}
\end{eqnarray}
%-----Begin MAYBE TO BE ADAPTED-------
%In Appendix \ref{sec:gaugetransformation} we present a possible derivation of the gauge symmetries \eqref{trafoX} and \eqref{trafoAf} when starting from the Hamiltonian formulation of the theory.
%-----End MAYBE TO BE ADAPTED-------

\newpage 
\section{BFV formulation of twisted Poisson sigma models}\label{sec:BFV}
\subsection{Traditional formulation}
In this section, we construct the BFV form of the HPSM, which, as always, has the Hamiltonian theory---see Section \ref{sec:HPSM}---as a starting point. The constraint algebra \eqref{constraintalgebra} is governed by the structure functions of the Lie algebroid 
$E=T^*M$. In addition, we have
\beq \label{anchorfield} \{ J^i(\sigma), f(X(\sigma'))\} = -\delta(\sigma-\sigma') \, (\pi^{ij}f_{,j})(X(\sigma)) \, ,
\eeq
recognizing the vector field $\rho(\rd x^i)$ applied to the function $f$ on the right-hand side, where $\rho$ is the anchor map of $E$. Thus we are in a situation as  described  in  \cite{Ikeda-Strobl}, up to some changes of signs which originate from different conventions and which we will take into account below. As shown in \cite{Ikeda-Strobl}---at least in the case that the constraints are irreducible---the BFV functional of such a theory takes the minimal form
\beq \label{SBFVfield} S_{BFV} = \oint_{S^1} \rd \sigma \left(
c_i\,J^i +\tfrac{1}{2} f^{ij}_k \,  c_i c_j b^k \right). \eeq
Here
we introduced two Grassmann odd fields $c_i(\sigma)$ and $b^i(\sigma)$,
of ghost number $1$ and $-1$, respectively, such that the fundamental Poisson brackets \eqref{classicalPB} are extended by
\begin{eqnarray}
\{c_i(\sigma), 
%A_{1}^{+j}
b^j(\sigma^{\prime}) \}&=& 
\delta_i{}^j \delta(\sigma - \sigma^{\prime}).
\end{eqnarray}
The resulting graded Poisson bracket is called the  BFV bracket 
and corresponds to the  BFV symplectic form
\begin{eqnarray}
\omega_{BFV} = \oint_{S^1} \rd \sigma
\left(\delta \xzero^i \wedge \delta p_{i}
+ \delta c_i \wedge \delta b^i
%A_{1}^{+i}
+ \tfrac{1}{2} H_{ijk}(\xzero) \partial \xzero^i
\delta \xzero^j \wedge \delta \xzero^k \right).
\label{BFVsymplecticform}
\end{eqnarray}
The functional \eqref{SBFVfield} satisfies the classical BFV master equation  %\cite{Ikeda-Strobl}
\beq \{ S_{BFV},S_{BFV}\} = 0 \, ,
\eeq
which is one of the central equations in this context. 

There is one slightly delicate point, however, that we still need to address. It concerns eventual dependencies of the constraints, impeding their irreducibility. This happens here when the bivector field $\pi$ does not have full rank (see also \cite{Schaller-Strobl} for the likewise observation in the context of the PSM). To display the resulting dependencies, we  restrict, for simplicity, to a patch in $M$: 
Assume that we have chosen the first coordinate $x^1$ of the coordinate system such that $\pi^{1j}=0$ for all $j$. Then  $J^1 = \partial X^1$ and 
\beq	\oint_{S^1} J^1(\sigma) \rd \sigma \equiv 0  \, . \eeq
According to the standard rules of quantization, one needs to add further canonically conjugate ghost pairs to the system, called ghosts for ghosts, which here are not fields but \emph{global} modes. %, and for eventual dependencies between dependencies add further such ghost pairs. 
Since $\pi$ does not need to have constant rank, this can be rather intricate. Usually such issues are ignored, but one needs to keep in mind that then the degree zero BFV cohomology %at degree zero 
will not reproduce correctly the functions on the reduced phase space.

In the following, we will follow the standard conventions and disregard this issue and call the functional \eqref{SBFVfield} \emph{the} BFV functional of the HPSM. For completeness, we write out \eqref{SBFVfield} also explicitly, using the actual form of the constraints \eqref{G} and of the structure functions \eqref{f}.
This gives
\begin{align}
S_{BFV}[X,p,b,c] &=
\oint_{S^1} \rd \sigma
\left[ c_i (\partial \xzero^i + \pi^{ij} p_{j})
+ \tfrac{1}{2} 
(\pi^{ij}{}_{,k}
+ \pi^{il} \pi^{jm} H_{klm} )
%A_1^{+k} 
b^k c_i c_j
\right].
\label{BFVBRSTcharge}
\end{align}

\subsection{In terms of superfields}

We now reformulate $S_{BFV}$ in a more elegant form using superfields. For this purpose, we extend $S^1$ to a super-circle 
$S^{1,1} \cong \Pi T S^1 \cong T[1]S^1\cong S^1 \times \R[1]$ parametrized by the coordinates $(\sigma,\theta)$ of degree zero and one, respectively.
%pair the spatial coordinate $\sigma$ with a super-coordinate $\theta$, considering them as coordinates on $S^{1,1} \cong \Pi T S^1 \cong T[1]S^1\cong S^1 \times \R[1]$. 
This permits us to combine the previously introduced fields into the following superfields of degree $0$ and $1$, respectively:
\begin{eqnarray}
&& \tilx^i (\sigma, \theta) := \xzero^i (\sigma)+ \theta \, b^i (\sigma),
%= \xzero^i (\sigma)+ \theta^1 A_1^{+i} (\sigma),
\nonumber \\
&& \tila_i (\sigma, \theta) := - c_i (\sigma) + \theta \, p_{ i} (\sigma).
\label{superfield1d}
\end{eqnarray}
Now the BFV symplectic form \eqref{BFVsymplecticform} and the BFV-BRST charge \eqref{BFVBRSTcharge} become
%more compactly in terms of these super-fields. One obtains
\begin{framed}
\begin{eqnarray}
\omega_{BFV} = \int_{T[1]S^1} \rd \sigma \rd \theta
\left(\delta \tilx^i \wedge \delta \tila_{i}
+ \tfrac{1}{2} H_{ijk}(\tilx) \tilde{\rd} \tilx^i
\delta \tilx^j \wedge \delta \tilx^k \right),
\label{superBFVsymplecticform}
\end{eqnarray}
\begin{eqnarray}
S_{BFV}
%&=& \int_{T[1]S^1}
%\!\!\!\! 
%\mu_1 \, {\cal Q}_{BFV}
%\nonumber \\
&=&
\int_{T[1]S^1} 
\!\!\!\!\!\!\!\! 
\rd \sigma \rd \theta
\left(\tila_i \tilde{\rd} \tilx^i
+ \tfrac{1}{2} \pi^{ij}(\tilx) \tila_i \tila_j
+ \tfrac{1}{2} \pi^{il} \pi^{jm} H_{klm}(\tilx)\,
\tila_i \tila_j \, \tilde{\varepsilon}\tilx^{k} 
\right)
\label{superBFVBRSTcharge}
\end{eqnarray}
\end{framed}
Here certainly $S_{BFV}=S_{BFV}[\tilx,\tila]$. Moreover, $ \tilde{\rd}= \theta \partial$ is the de Rham differential on $S^1$, viewed as a super-derivative on $T[1]S^1$, and 
$\tilde{\varepsilon}$ denotes the second natural derivative operator on the super-circle, the Euler vector field,  $\tilde{\varepsilon}= \theta \tfrac{\partial}{\partial \theta}$.
Thus, we find that the BFV symplectic form \eqref{superBFVsymplecticform} is a super-extension of the symplectic form \eqref{classicalsymplecticform} of the classical Hamiltonian theory while the BFV-BRST charge \eqref{superBFVBRSTcharge} is quite different from a  super-extension of the classical action \eqref{classicalactionofHPSM}. It is also remarkable that the Euler vector field $\tilde{\varepsilon}$ enters the super-description of $S_{BFV}$ explicitly, a phenomenon we did not see elsewhere before. For $H=0$, this last term disappears and we obtain the super-BFV formulation of the PSM, see Appendix \ref{sec:PSMBFV}.

\subsection{Two global descriptions of the BFV phase space}
We defined the classical phase space as  bundle maps from $TS^1$ to $T^*M$ or, equivalently:
\beq  {\cal M}_{Ham} = \mathrm{Hom}(T[1]S^1, T^*[1]M)  \cong  C^\infty(S^1,T^*M) \, . \label{Mclass}
\eeq
Morphisms between graded manifolds are degree-preserving maps. Thus, for example, a function $x^i$ of degree zero on $T^*[1]M$, which is just a (locally defined) coordinate function on $M$, gives rise to a degree zero element in $C^\infty(T[1]S^1)$, 
i.e.\ a (possibly only locally defined) function $X^i$ on $S^1$. Likewise, 
 a momentum coordinate $p_i$ on the cotangent bundle over $M$, which has degree one on $T^*[1]M$, becomes a 1-form on $S^1$. Since we assume $S^1$ to be parametrized by the coordinate $\sigma \sim \sigma + 2\pi$, we can identify this 1-form with a function on $S^1$, the coefficient of $\rd \sigma$, which we denoted by $p_i(\sigma)$ in Section \ref{Hamform}. This explains  \eqref{Mclass}. 

We saw above that the BFV phase space can be parametrized by fields of the form \eqref{superfield1d}. These correspond to the following global description
\beq  {\cal M}_{BFV} = \underline{\mathrm{Hom}}(T[1]S^1, T^*[1]M) \, . \label{MBFV}   
\eeq
Like $x^i$ also the field $ \tilx^i$ has degree zero. However, the underlining implies that now we do not only obtain degree zero elements $X^i$ in $C^\infty(T[1]S^1)$, but we need to consider a formal Taylor expansion in the odd coordinate $\theta$ as in \eqref{superfield1d} and, since $\theta$ has degree one, there now  exists a field $b^i$ of degree minus one  (which is not an element of  the non-negatively graded $C^\infty(T[1]S^1)$). 

In \cite{Ikeda-Strobl} we studied constrained systems on $T^*M$ whose constraint algebra follows the one of a Lie algebroid $E \to M$. We found that the BFV extension of such a \emph{mechanical} system is formulated globally on $T^*E[1]$. This permitted several geometrical interpretations like relating a covariantized momentum to the horizontal lift of a vector field from $M$ to $E$.
According to \eqref{Mclass}, we are considering loops inside $T^*M$ and the constraints reflect the Lie algebroid structure on $E=T^*M$, see \eqref{constraintalgebra} and \eqref{anchorfield}. We thus may ask ourselves, if possibly the BFV phase space \eqref{MBFV} can be identified with loops in this point particle BFV extension $T^*E[1]$, which here equals $T^*(T^*[1]M)$. 

This is indeed the case as the following consideration shows. For all graded manifolds $X,Y,Z$, one has the well-known identification 
\beq \underline{\mathrm{Hom}}(X \times Y,Z)\cong \underline{\mathrm{Hom}}(X,\underline{\mathrm{Hom}}(Y,Z)).
\eeq
Since, after the choice of a coordinate $\sigma$, we can identify $T[1]S^1$ with $S^1 \times \R[1]$, this implies
\beq {\cal M}_{BFV} \cong \underline{\mathrm{Hom}}(S^1,\underline{\mathrm{Hom}}(\R[1],T^*[1]M)).
\eeq
On the other hand (see, e.g., \cite{math0307303}), $\underline{\mathrm{Hom}}(\R[1],Z) \cong T[-1]Z$ for every graded manifold $Z$. Together with the isomorphism $T[k] (T^*[n]M) \cong T^*[k+n](T^*[n]M)$, which holds true for every $k,n \in \mathbb{Z}$, 
we thus find ${\cal M}_{BFV} \cong \underline{\mathrm{Hom}}(S^1, T^*(T^*[1]M))$ or, as claimed above,  
\beq {\cal M}_{BFV} \cong C^\infty(S^1, T^*(E[1])) \label{MBFVE}
\eeq 
for the Lie algebroid $E \cong T^*M$. 

\subsection{Manifestly target space covariant form}\label{sec:manifestlyBFV}
The BFV-functional \eqref{BFVBRSTcharge} contains a partial derivative of the Poisson bivector field, which is a coordinate-dependent expression. One may thus wonder, if this functional---and thus also \eqref{superBFVBRSTcharge} to which it is equivalent---is valid only within local coordinate patches. This is in fact not the case, as we will now show. The remedy comes from the fact that the momentum variable $p$ does not transform like a covector on $M$, but receives ghost-contributions: This can be seen by performing a coordinate change on $M$, requiring $b$ and $c$ to transform like vectors and covectors, respectively, and by lifting this operation to all of ${\cal M}_{BFV}$ as a canonical transformation, i.e.\ so as to leave \eqref{BFVsymplecticform} invariant. 

The identification \eqref{MBFVE} provides a more geometrical understanding of the situation. $(x,c)$ are coordinates on $E[1]$ (we drop the $S^1$-dependence in this discussion for simplicity). The coordinate $p_i$ was originally a momentum function on $T^*M$; as such it is in one-to-one correspondence with the (locally defined) vector field  $\partial_i$ on $M$. To lift it to a vector field on  \eqref{MBFVE}, one needs a connection $\nabla$. The corresponding covariant derivative, rewritten as a covariant momentum variable on $T ^*E[1]$, takes the form \cite{Ikeda-Strobl}:
\beq p_i^\nabla := p_i + \Gamma_{ji}^k  b^j c_k \, , \label{covp}
\eeq 
where certainly $\Gamma_{ji}^k=\Gamma_{ji}^k(X) \equiv X^*\Gamma_{ji}^k$. 

It is now a straightforward calculation to verify that the BFV-functional \eqref{BFVBRSTcharge} can be rewritten identically into
%Let us now go back to the expression \eqref{SBFVfield} for the BFV functional. The constraints $J^i$ in Equation \eqref{G} depend on the momenta $p_i$, which, for a more covariant expression, we want to replace by their covariant extensions \eqref{covp}. We learned in Section \ref{sec:cov}, moreover, that the use of a connection, permits to turn the structure functions of a Lie algebroid into a tensorial object, which we called $T$ here. It is now a to our mind beautiful observation that  \eqref{SBFVfield} can be identically rewritten into the explicitly target space covariant form 
\begin{align}
S_{BFV} [X,p^{\nabla},b,c ]
&=
\int_{S^1} \rd \sigma
\left[ c_i (\partial \xzero^i + \pi^{ij} p_{j}^{\nabla})
- \tfrac{1}{2} T_k^{ij}b^k c_i c_j \right].
\label{covariantBFVBRSTcharge2prime}
\end{align}
We see that in the constraints $J^i$ the momentum is covariantized and, at the same time, the structure functions $f_k^{ij}$ of the functional turned into the $E$-torsion $T$, cf.\ \eqref{Tcomp}.

If one prefers a notation without indices, then \eqref{covariantBFVBRSTcharge2prime} looks as follows:
\begin{align}
S_{BFV} [X,p^{\nabla},b,c ]
&=
\int_{S^1} \rd \sigma
\left[ \langle c , \partial \xzero \rangle + (\pi \circ X)(c, p^{\nabla})
- \tfrac{1}{2} \langle b , (T\circ X)(c,c) \rangle\right],
\label{covariantBFVBRSTcharge2}
\end{align}
The above holds true for an arbitrary connection $\nabla$. Connections of the form \eqref{Gamma}, i.e.\ restricted to have torsion \eqref{torsionpiH}, are singled out by the fact that it is precisely such a choice of connection that makes $T$ independent of $H$, see \eqref{nablapi}---and thus also $S_{BFV}$ when expressed in the above fields, where $H$ then only enters the definition of $p^{\nabla}$.

One may now worry about the BFV Poisson brackets in these new variables. While, as representing covariant derivatives, the bracket between the momenta $p_i^\nabla$ produce the curvature of $\nabla$,
\beq \{p_i^\nabla(\sigma), p_j^\nabla (\sigma')\} = 
- \delta(\sigma-\sigma') \left[H_{ijk}(X(\sigma)) \partial X^k(\sigma) + R^k_{lij}(X(\sigma)) \, b^l(\sigma)c_k(\sigma) \right]\, , \label{old}
\eeq 
these momenta do no more Poisson commute with, for example, $c_i$ and the corresponding brackets do not look very covariant. However, the situation changes, if the momenta are permitted to act on basis vectors. 

Consider $c = c_i \otimes \underline{\rd x^i}$ and $b = b^i \otimes \underline{\partial_i}$. The basis $(\underline{\rd x^i})_{i=1}^n$ of $X^* T^*M$ and its dual $(\underline{\partial_i})_{i=1}^n$ depend implicitly on $X$ and should not Poisson commute with the momentum.\footnote{$\underline{\rd x^i} \in X^* T^*M$ must not be confused with $X^* \rd x^i \equiv \rd X^i \in \Omega^1(S^1)$.}
It is natural to require (see \eqref{conn} and take into account that the Hamiltonian vector field of $p_i$ corresponds to minus $\partial_i$ on $M$)
\bea \{p^{\nabla}_i(\sigma),\underline{\rd x}^j\vert_{\sigma'}\} &:= & \delta(\sigma-\sigma') \, \Gamma^j_{ki}(X(\sigma)) \, \underline{\rd x}^k\vert_{\sigma} \, , \nonumber \\
\{ p^{\nabla}_i(\sigma),\underline{\partial}_j \vert_{\sigma'}\} &:= & - \delta(\sigma-\sigma') \, \Gamma^k_{ji}(X(\sigma)) \,  \underline{\partial}_k\vert_{\sigma} \, .\label{kommtnoch}
\eea
All other fields BFV commute with the basis vectors (since they have vanishing brackets with $X$) as well as do basis vectors among themselves. Then, by means of the Leibniz rule with respect to the tensor product, we have (for whatever choice of arguments) 
\beq \{p^{\nabla}, b\} = 0 \,\, , \qquad \quad \{p^{\nabla}, c\}=0 .
\eeq 
The brackets between the momenta pick up an extra torsion contribution, but still stay covariant; \eqref{old} turns into 
\beq
\{p^\nabla(\sigma), p^\nabla (\sigma')\} = 
 \delta(\sigma-\sigma') \left[ \langle c, \iota_b R \rangle + \langle p^{\nabla}, \Theta \rangle-\iota_{\partial X}H \right]\, , \label{ppnabla}
\eeq 
where,  on the right-hand side, $\Theta$ denotes the torsion of $\nabla$  and $\langle \cdot , \cdot \rangle$ stands for the contraction of the two objects left and right of the comma (which is unique in the two cases here); furthermore, $b$, $c$, $p^{\nabla}$, and $\partial X$ are evaluated at $\sigma$ and $R$, $\Theta$, and $H$  at $X(\sigma)$.

 Also all other brackets are inherently covariant now. For example, the brackets between $b$ and $c$ take the form: 
\begin{eqnarray}
\{c(\sigma), 
%A_{1}^{+j}
b(\sigma^{\prime}) \}&=& 
\delta(\sigma - \sigma^{\prime}) \otimes \mathrm{id}\vert_{X(\sigma)} \, ,
\end{eqnarray}
where $\mathrm{id} = \underline{\rd x^i}  \otimes  \underline{\partial_i}$ denotes the identity endomorphism. 

\newpage 
%%%%%%%%%%%%%%%%%%%%%%%%%%%%%%%%%%%%%%%%%%%%%%%%%%%%%%%%%%%%%%%%%%%%%%%%%%%
%%%%%%%%%%%%%%%%%%%%%%%%%%%%%%%%%%%%%%%%%%%%%%%%%%%%%%%%%%%%%%%%%%%%%%%%%%%
%%%%%%%%%%%%%%%%%%%%%%%%%%%%%%%%%%%%%%%%%%%%%%%%%%%%%%%%%%%%%%%%%%%%%%%%%%%
\section{The BV formulation with a connection}\label{sec:BV1}
In this section we construct the BV extension of the $H$-twisted Poisson sigma model ab initio. This is done also for pedagogical reasons. Similarly to the previous section, we intend to present the resulting BV action in a local patch (on the target manifold) without connection coefficients, in terms of superfields combining them into a joint object, provide an expression that uses the connection coefficients, and reformulate the latter functional in terms of manifestly covariant component fields.

But instead of starting with the gauge symmetries in their ''naive form'', see Eqs.\ \eqref{trafoX}, \eqref{gaugetransformation02a}, and \eqref{trafoAf}, it turns out to be advantageous to already include the connection coefficients from the very beginning. The reason for this is twofold: First, in contrast to the Poisson sigma model, which is the HPSM for $H=0$, we cannot put the connection coefficients in  \eqref{deltaAnabla} to zero altogether; it is only the torsion-free part of $\nabla$ that is at our disposal for a choice, its torsion is fixed, see Eq.\ \eqref{Gamma}. Thus, one can view \eqref{deltaAnabla} even as a slightly more concise form of writing \eqref{trafoAf},
where we only abbreviate the expression  $- \tfrac{1}{2} \pi^{km} H_{mij}$ by  the symbol $\Gamma^k_{ij}$. The local formulas without connection coefficients can, moreover, be obtained from those ones by this mere replacement and we will do so at some point. Second, and more importantly, the use of a connection from the very beginning provides the possibility of a geometrical interpretation of otherwise lengthy expressions. This will become particularly obvious when encountering the tensor $S$ below, which has a clear geometrical meaning relating the connection $\nabla$ to the Lie algebroid structure on $T^*M$---when written locally for a vanishing torsion-free part of the connection coefficients, this becomes an otherwise completely meaningless, lengthy, and non-covariant expression in terms of $H$, $\pi$, and their derivatives.

\subsection{On-shell closed BRST transformations}
\label{sec:BRSTtransformation}
It is standard wisdom in determining the BRST transformations that, in a first step, one replaces the infinitesimal gauge parameters $\epsilon$ by odd and anti-commuting ghost fields $c$. As argued   above, it is advantageous to use the target space covariant gauge transformations  \eqref{deltaAnabla} for this purpose right away. We denote the anti-commuting BRST  operator by the conventional letter $s \equiv s_{BRST}$; we thus obtain (see also Equation \eqref{trafoX}):
\begin{eqnarray}
s X^i &=& - \pi^{ij}c_j,
\label{BRSTofX}
\\
s A_{i} &=&\rd  c_i + f_i^{jk} A_{j} c_k - \Gamma_{il}^k F^l c_k 
.
\label{BRSTofA}
\end{eqnarray}
Here $f_i^{jk}$ are again the structure functions of the Lie algebroid $T^*M$ in a holonomic basis, given by Equation \eqref{f}, $\Gamma_{il}^k$ are the connection coefficients of an auxiliary connection, whose torsion-free part can be chosen arbitrarily, but whose torsion is given by \eqref{torsionpiH}, see \eqref{Gamma}. All these coefficients are understood to be pulled back by $X\colon \Sigma \to M$ certainly. As before, the 1-forms $F^i$ are defined in \eqref{eom1}; they are part of the field equations and vanish on-shell. 

We are left with defining the action of the BRST  operator on the ghosts $c_i$. We put
\begin{equation}
s c_i := - \tfrac{1}{2} f_i^{jk} c_j c_k \, . \label{BRSTofc}
\end{equation}
There are several arguments leading to this choice. Let us provide an intuitive one which will at the same time prove  that the above formulas imply 
\begin{eqnarray} s^2 X^i &=& 0 \,  , \label{ssquareX} \\ s^2 c_i &=& 0 \,  \label{ssquarec}
\end{eqnarray}
without the need of a calculation. 

Let us, for this purpose, first recall Vaintrob's characterization \cite{Vaintrob} of a Lie algebroid in terms of a BRST  like operator $Q$: Let $E$ be a vector bundle over $M$ and  $x^i$ and $\xi^a$ be local coordinates on $E[1]$ of degree zero and one, respectively, corresponding to a choice of coordinates on $M$ and a choice of a basis $e^a$ of sections of $E^*$. A general degree one vector field then is of the form 
\be  Q = \rho_a^i \xi^a \frac{\partial}{\partial x^i} - \tfrac{1}{2} C^a_{bc} \xi^b \xi^c \frac{\partial}{\partial \xi^a} \label{Q}
\ee
where $\rho_a^i$ and $C^a_{bc}$ are functions on $M$. Equipping $E$ with a Lie algebroid structure is \emph{equivalent} to equipping $E[1]$ with a degree one vector field squaring to zero, $Q^2=0$. For the dictionary, one uses $\rho(e_a) = \rho_a^i \partial_i$ and $[e_a,e_b] = C^c_{ab} e_c$, where $e_a$ is a basis of sections of $E$ dual to $e^a$. 

For the Lie algebroid structure on $E=T^*M$ induced by a twisted Poisson structure $(M,\pi,H)$, the vector field $Q$ above takes the form
\be  Q = \pi^{ij} \xi_i \frac{\partial}{\partial x^j} - \tfrac{1}{2} f^{jk}_i \xi_j \xi_k \frac{\partial}{\partial \xi_i} \label{Qtwisted} \, ,
\ee
where $\xi_i$ denote the odd coordinates on $T^*[1]M$ corresponding to the holonomic vector fields $\partial_i$ on $M$ and the structure functions coincide with the symbols $f^{jk}_i$ as used above due to  \eqref{ijE}. Equation \eqref{Qtwisted} is equivalent to 
\begin{eqnarray}Q \, x^i &=& -\pi^{ij} \xi_j  \label{Qx}\\ Q \, \xi_i &=& - \tfrac{1}{2} f^{jk}_i \xi_j \xi_k \label{Qxi}
\end{eqnarray}
and the conditions $Q^2=0$ now follow from the Lie algebroid property. Comparison of \eqref{BRSTofX} with \eqref{Qx}, on the one hand, suggests \eqref{BRSTofc} due to its analogy with 
\eqref{Qxi} and, on the other hand, this choice is now seen to imply the identities \eqref{ssquareX} and \eqref{ssquarec}, since the additional $\sigma$-dependence %(in part explicit, in part induced by the pullback by $X\colon \Sigma \to M$) 
%in the formulas for the odd BRST  operator $s$ %on the field space 
does not play a role in these calculations. 

One now is left to check if $s$ also squares to zero when acting on $A_i$. In fact, a somewhat lengthy but straightforward calculation yields
\begin{eqnarray}
s^2 A_i &=& - \tfrac{1}{2} S_{ij}{}^{kl} F^j c_k c_l ,  
\label{BRST2A}
\end{eqnarray}
where we used the abbreviation\footnote{Indices between parenthesis and square brackets are antisymmetrized and symmetrized over, respectively. Thus, for example, $v_{(ij)}= \tfrac{1}{2} (v_{ij} + v_{ji})$ and $v_{[ij]}= \tfrac{1}{2} (v_{ij} - v_{ji})$.}
\begin{equation} S_{ij}{}^{kl} 
= - f_{(i}^{kl}{},_{j)} 
+ \Gamma_{(ij)}^m f_m^{kl} + 2 \Gamma_{m(j}^{[k} f_{i)}^{l]m} 
%+ \Gamma_{mj}^l f_i^{km}
+ 2 \pi^{m[k}  \Gamma^{l]}_{ij},_m
%+ \pi^{ml} \partial_m \Gamma^k_{ij}
+ 2  \pi^{m[k},_j \Gamma^{l]}_{im}
%+ \partial_j \pi^{ml} \Gamma^k_{im}
- 2 \Gamma^{[k}_{im}\Gamma^{l]}_{nj}  \pi^{mn} 
%- \Gamma^l_{im} \pi^{mn} \Gamma^k_{nj} 
\, . \label{Slong}
\end{equation}

We see that the BRST  operator does not square to zero. It does so only on-shell, by use of the equations of motion \eqref{eom1}. We thus need an extension of the BRST  formalism tailored for such situations and this is precisely provided by the BV formalism.

But before turning to this task in the subsequent subsection, we want to make several remarks about the coefficients appearing in \eqref{Slong}. First, we observe that even in the case when we work in a local patch where the torsion-free part of the connection vanishes, this expression does not become much friendlier, in particular so, when spelled out explicitly in terms of $\pi$ and $H$:
\beqa  \label{K}
S_{ij}{}^{kl}\vert_{\stackrel{\circ}{\Gamma_{ij}^k} =0} &=&
- f_{(i}^{kl}{},_{j)} - \tfrac{1}{2}  \pi^{c[k} \pi^{l]a} \pi^{bd} H_{iab} H_{jcd} \\
&=&
- \pi^{kl},_{ij}
- \tfrac{1}{2} \left[\left( \pi^{km} \pi^{ln} H_{imn} \right)\!,_j
+ \left(
\pi^{km} \pi^{ln} H_{jmn} \right)\!,_i  + \pi^{c[k} \pi^{l]a} \pi^{bd} H_{iab} H_{jcd}\right]\! . \nonumber
\eeqa
Only for $H=0$ this expression becomes simple, then coinciding with the second partial derivatives of the coefficients of the Poisson bivector field. 
%That it does not vanish also in this case, demonstrates the need of BV already for the PSM; only in the case when the bivector $\pi$ is \emph{linear} and the theory reduces to a BF-gauge theory for some Lie algebra, the BRST  procedure is sufficient (at least, if one does not use connection coefficients so as to make the tensor $S$ vanish, see below). 

While partial derivatives of tensor components and thus expressions such as \eqref{K} do not have a coordinate-independent meaning, the expressions \eqref{Slong} assemble into a tensor field,
\be S= \nabla T + 2 \mathrm{Alt} (\iota_\rho  R_\nabla)  . \label{S}
\ee
Here $R_\nabla \in \Omega^2(M,\mathrm{End}E)$ is the curvature of $\nabla$, $\rho = e^a \otimes \rho(e_a) \in \Gamma(E^* \otimes TM)$, $\iota_\rho$ stands for $e^a \otimes \iota_{\rho(e_a)}$, and $\mathrm{Alt}$ denotes an antisymmetrization over $E^* \otimes E^*$. Here $E=T^*M$. 

The expression \eqref{S} was found already in \cite{Mayer-Strobl} in the context of the commutator of gauge transformations of the more general form \eqref{deltaAa} for some Lie algebroid $E$,
\begin{equation} [\delta_\epsilon , \delta_{\epsilon'} ] A^a = \delta_{[\epsilon , {\epsilon'} ]} A^a +  S^a_{ibc} \, (\rd X^i - \rho^i_d A^d) \epsilon^b  {\epsilon'}^c. \label{commutator}
\end{equation}
Later this tensor was considered also in \cite{Abad-Crainic} and called ``basic curvature''. 
Its vanishing describes the compatibility \cite{Blaom} of the Lie algebroid structure on $E$ with a connection $\nabla$ (see \cite{Kotov-Strobl2} for the details). This provides a geometrical interpretation of the coefficient functions appearing in \eqref{BRST2A}.

The specialization of \eqref{commutator} to our situation with $E=T^*M$ and \eqref{ijE} yields \begin{equation} [\delta_\epsilon , \delta_{\epsilon'} ] A_i =   \delta_{[\epsilon , {\epsilon'} ]} A_i +  S_{ij}^{kl} \, F^j \epsilon_k  {\epsilon'}_l  \, ,\label{commutatorhere}
\end{equation}
where $[\epsilon , {\epsilon'} ] _i = f^{jk}_i \epsilon_j\epsilon'_k$ (for the case that $\epsilon$ and $\epsilon'$ do not depend on $X$, which we will assume for simplicity here). 

This is a good moment to again return to the sequence \eqref{seq}: The choice of a splitting of this sequence in the category of sets (or infinite-dimensional manifolds)---which, as we explained in Sec.\ \ref{sec:cov}, is provided by the choice of an admissible  connection $\nabla$---is a splitting of groups \emph{iff} the tensor \eqref{S} vanishes. In other words, for the PSM and for the HPSM there exist (true) splittings of \eqref{seq} if and only if $E=T^*M$ admits a connection of the prescribed torsion \eqref{torsionpiH} which is compatible with the Lie algebroid structure. % (as mentioned above, the compatibility is equivalent to the vanishing of the tensor $S$ or, according to \eqref{commutatorhere}, that the generators $\delta_\epsilon$ close ``off-shell'' already).

The calculation \eqref{commutatorhere} provides further justification for the findings of this subsection: First, we see that the structure functions of the commutators of the generators \eqref{deltaAnabla} of the gauge transformations are given by the (pullback of) the Lie algebroid structure functions $f^{jk}_i$. Since these generators are (essentially) irreducible, this enforces the choice \eqref{BRSTofc} made above according to the usual rules of BRST (see, e.g., \cite{Henneaux-Teitelboim}). Second, the fact that the generators close on-shell only implies that the corresponding BRST operator does not square to zero off-shell. Moreover, the violation \eqref{BRST2A} is governed precisely by the right-hand side of \eqref{commutatorhere}. % (this is not only true for the HPSM, but a general fact). 
The need for the BV-formalism can thus be concluded already from this commutator of gauge transformations. 

We remark as an aside that if we turn off the connection coefficients $\stackrel{\circ}{\Gamma_{ij}^k}$ for a moment and put $H:=0$, we obtain the well-known result that,  for the PSM, the violation of the commutator of generators to close is given by the second partial derivatives of the bivector field $\pi^{ij}$ (see Eq.\ \eqref{K}). Thus, without connection coefficients, we recover the folklore that one needs BV for the PSM precisely except for linear Poisson structures, in which case the PSM is equivalent to a YM-type BF-theory. However, we now see that connection coefficients can be used to make the simpler BRST procedure applicable to the PSM also for some non-linear Poisson structures:  this is possible if there exists a torsion-free connection which is compatible with the Lie algebroid structure on $T^*M$ in the sense of \cite{Blaom}.

In the following we will make use of this class of models, in fact the more general one with twisting $H$, for a first bite on the BV formalism.

%%%%%%%%%%%%%%%%%%%%%%%%%%%%%%%%%%%%%%%%%%%%%%%%%%%%%%%%%%%%%%%%%%%%

 \subsection{The BV extension with connection coefficients}\label{sec:normalBVformalism}
We now turn to  the construction of the BV formulation of the $H$-twisted PSM. We will see that the chosen geometrical setting will lead to formulas which are  surprisingly simple and, to our knowledge, also new even for the ordinary PSM. 

In general, the BV formalism is an extension of the BRST  method which is applicable for BRST operators $s$ satisfying 
\begin{equation} s^2 \approx 0 , \label{approx}
\end{equation}
where the sign $\approx$ means ``upon usage of the field equations''. 
Here the field equations are given by \eqref{eom1} and \eqref{eom2} and the above equation follows from  \eqref{BRST2A} (together with  \eqref{ssquareX} and \eqref{ssquarec}). 

It is illustrative, however, to already  translate the BRST situation with \begin{equation} s^2 = 0 \label{s^2}
\end{equation}
into the  language of BV. The BV extension of the classical action is particularly simple in this case: it consists of adding the Hamiltonian lift of $s$ to an appropriately shifted cotangent bundle of the space of fields introduced in the previous subsection. Showing this in detail also has the advantage that we can introduce much of the BV framework in a simple context and settle for conventions and notation. In a second step we then need  to discuss only the modifications required for the general HPSM, where,  instead of \eqref{s^2}, we have only \eqref{approx}.

 \eqref{s^2} can be achieved for the HPSM as long as we can find a connection 
 $\nabla$ which is compatible with the Lie algebroid structure that is  induced on $E=T^*M$ by  $(M,\pi,H)$. This in turn is tantamount to the vanishing of the ``basic curvature'' tensor \eqref{S} (for its components  
see \eqref{Slong} together with \eqref{f}),
 \beq S=0 . \label{S=0}\eeq
In general, this will certainly be possible only for a rather restricted class of twisted Poisson structures, but, in any case, we will assume this situation to be given now.

To start with, consider the space of classical fields, 
 \beq \cM_{cl} := \{ a \colon T \Sigma \to T^*M \} \cong \{ \left(X \colon \Sigma \to M, \, A \in \Omega^1(\Sigma,X^*TM)\right)\}\, , \label{Mcl} \eeq
 where $a$ is a vector bundle morphism but not (necessarily) also a morphism of Lie algebroids---to which, however,  it reduces precisely on-shell, i.e.\ when restricting to such couples $(X,A)$ which satisfy the field equations \eqref{eom1} and \eqref{eom2} (see \cite{Kotov-Schaller-Strobl} for a proof of this statement). Add to this space the ghost fields $c$ introduced in the previous subsection, which we will declare to carry \emph{ghost number} gh plus one:
  $$\cM_{BRST} := \{ \left(X \colon \Sigma \to M, \, A \in \Omega^1(\Sigma,X^*T^*M), \, c \in C^\infty(\Sigma, X^*T^*[1]M)\right)\}\, .$$
The space of fields of the HPSM in the BV formulation is now simply
\begin{equation} \cM_{BV} := T^*[-1]\cM_{BRST}. \, \label{MBV}
\end{equation}
Evidently $\cM_{BV}$ is  canonically equipped with a degree minus one---and thus odd---symplectic form,
 \begin{eqnarray}
\omega_{BV} &=& \int_{\Sigma} 
\left(\delta X^i \wedge \delta X^+_i + \delta A_i \wedge \delta A^{+i}
+ \delta c_i \wedge \delta c^{+i}\right).
\label{componentBVsymplecticform}
\end{eqnarray}
The conjugate momenta, called antifields in the BV formalism conventionally, take values in the respective dual spaces; in particular, they are differential forms of complementary form degree (and their ghost degree is determined using \eqref{MBV}). $c^+$, for example, is a 2-form taking values in $X^*T^*[-1]M$; without the shift in degree, the fields $c^+$ would have ghost degree minus one, but this is shifted to the negative by one unit, thus leading to: 
\beq \mathrm{fdeg}(c^{+i})=2 \, , \quad \mathrm{gh}(c^{+i})=-2. \label{c+}
\eeq
For the remaining fields the bigrading is:
\begin{align} \mathrm{fdeg}(X^{i})&=0 \, , & \mathrm{gh}(X^{i})&=0 \nonumber 
\\
 \mathrm{fdeg}(A^{+i})&=1 \, , & \mathrm{gh}(A^{+i})&=-1 \nonumber
\\
 \mathrm{fdeg}(X^+_{i})&=2 \, , & \mathrm{gh}(X^+_{i})&=-1\label{degrees}
 \\\mathrm{fdeg}(A_{i})&=1 \, , & \mathrm{gh}(A_{i})&=0\nonumber
 \\\mathrm{fdeg}(c_i)&=0 \, , & \mathrm{gh}(c_{i})&=1 .\nonumber
\end{align}
It is important to mention that in  this section we use the Deligne sign convention for the commutativity of our fields, relating to the bigrading as follows:
 \be \phi \wedge \psi = (-1)^{\mathrm{fdeg}(\phi)\mathrm{fdeg}(\psi)+\mathrm{gh}(\phi)\mathrm{gh}(\psi)
%+\mathrm{anti}(\phi)\mathrm{anti}(\psi)
} \, \psi \wedge \phi . \label{oldproduct}
\ee  

The induced odd Poisson bracket of ghost number minus one is called the BV bracket. 
On the components of the fundamental fields the non-vanishing brackets read as follows: 
\begin{eqnarray}
&& \sbv{\xzero^i(\sigma)}{\xzero_{01 j}^{+}(\sigma^{\prime})}
= \delta^i{}_j\delta^2(\sigma- \sigma^{\prime}), \nonumber\\
&& \sbv{A_{0j}(\sigma)}{A_1^{+i}(\sigma^{\prime})} \nonumber
= \delta^i{}_j\delta^2(\sigma- \sigma^{\prime}),
\\
&& \sbv{A_{1j}(\sigma)}{A_0^{+i}(\sigma^{\prime})}\label{a}
= - \delta^i{}_j\delta^2(\sigma- \sigma^{\prime}),
\\
&& \sbv{c_j(\sigma)}{c_{01}^{+i}(\sigma^{\prime})}
= \delta^i{}_j\delta^2(\sigma- \sigma^{\prime}). \nonumber
%\delta^i{}_j\delta(\sigma- \sigma^{\prime}),
\end{eqnarray}
The BV bracket is a Gerstenhaber bracket. 
%In terms of test differential forms, these equations can be rewritten as follows: 
%\begin{eqnarray} \left( \int_{\Sigma} \langle \alpha \wedge A \rangle,
%\int_{\Sigma} \langle \beta \wedge  A^{+} \rangle\right)_{BV}
%&=& \int_{\Sigma} \langle \alpha \wedge \beta \rangle ,
%\\
% \left( \int_{\Sigma}  \langle \tau \:  c \rangle,
%\int_{\Sigma} \langle \nu \: c^{+}\rangle \right)_{BV}
%&=& \int_{\Sigma}  \langle \tau \: \nu \rangle ,
%\\
% \left( \int_{\Sigma} \gamma  \: \xzero^*\!f , 
%\int_{\Sigma} \langle \mu \: \xzero^{+}  \rangle \right)_{BV}
%&=& \int_{\Sigma} \gamma \, \langle  \rd \! f \: \mu \rangle 
% .
%\end{eqnarray}
%Here 
%$\alpha \in \Omega^1(\Sigma,X^* TM)$, $\beta \in \Omega^1(\Sigma,X^* T^*M)$, $\tau \in \Omega^2(\Sigma,X^* TM)$,  $\nu \in \Omega^0(\Sigma,X^* T^*M)$,   $\gamma \in \Omega^2(\Sigma)$,   $f \in C^\infty(M)$, $\rd \! f \in \Omega^0(\Sigma,X^* T^*M)$, $\mu \in \Omega^0(\Sigma,X^* TM)$,   and the brackets $\langle \ldots \rangle$ indicate contraction of dual spaces.
For fields of fixed form- and ghost-degree, it satisfies the following properties\footnote{We suppress arguments in these formulas, but one may imagine that $\phi$, $\psi$, and $\xi$ depend on $\sigma$, $\sigma^{\prime}$, and $\sigma^{\prime\prime}$, respectively.} \cite{Cattaneo:2000rt,Ikeda:2001fq}
\begin{eqnarray}
\sbv{\phi}{\psi}&=&-(-1)^{\mathrm{fdeg}(\phi)\mathrm{fdeg}(\psi)+(\mathrm{gh}(\phi) + 1)(\mathrm{gh}(\psi) +1)} \sbv{\psi}{\phi}, \nonumber
\\
 \!\!\!\!\!\!\!\!\!\!\!\!\!\!\! \!\!\!\!\!\!\!\!\!\!\!\!\!\!\! \!\!\!\!\!\!\!\!\!\!\!\!\!\!\!\sbv{\phi}{\psi \: \xi}&=&\sbv{\phi}{\psi} \: \xi+ (-1)^{\mathrm{fdeg}(\phi)\mathrm{fdeg}(\psi)+(\mathrm{gh}(\phi) + 1)\: \mathrm{gh}(\psi)} \psi \:\sbv{\phi}{\xi},\label{81}
\\
\sbv{\phi}{\sbv{\psi}{\xi}}
&=&\sbv{\sbv{\phi}{\psi}}{\xi} \nonumber \\ &&+(-1)^{\mathrm{fdeg}(\phi)\mathrm{fdeg}(\psi)+(\mathrm{gh}(\phi) + 1)(\mathrm{gh}(\psi) +1)} \: \sbv{\psi}{\sbv{\phi}{\xi}}.\nonumber
\end{eqnarray}
%\begin{align}
%\sbv{\phi(\sigma)}{\psi(\sigma^{\prime})}=-(-1)^{\mathrm{fdeg}(\phi)\mathrm{fdeg}(\psi)+(\mathrm{gh}(\phi) + 1)(\mathrm{gh}(\psi) +1)} \sbv{\psi(\sigma^{\prime})}{\phi(\sigma)}, 
%\\
% \!\!\!\!\!\!\!\!\!\!\!\!\!\!\! \!\!\!\!\!\!\!\!\!\!\!\!\!\!\! \!\!\!\!\!\!\!\!\!\!\!\!\!\!\!\sbv{\phi(\sigma)}{\psi \xi(\sigma^{\prime})}=\sbv{\phi(\sigma)}{\psi(\sigma^{\prime})} \xi(\sigma^{\prime})
%\nonumber \\  \!\!\!\!\!\!\!\!\!\!\!\!\!\!\!
%+ (-1)^{\mathrm{fdeg}(\phi)\mathrm{fdeg}(\psi)+(\mathrm{gh}(\phi) + 1)\mathrm{gh}(\psi)} \psi(\sigma^{\prime}) \sbv{\phi(\sigma)}{\xi(\sigma^{\prime})},
%\\
%\left(\sbv{\phi(\sigma)}{\sbv{\psi(\sigma^{\prime})}{\xi(\sigma^{\prime\prime})}}\right)_{BV}
%=\sbv{\sbv{\phi(\sigma)}{\psi(\sigma^{\prime})}}{\xi(\sigma^{\prime\prime})} \nonumber \\ \!\!\!\!\!\!\!\!\!\!\!\!\!\!\!\!\!\!\!\!\!\!\!\!\!\!\!\!\!\!\!\!\!\!\!+(-1)^{\mathrm{fdeg}(\phi)\mathrm{fdeg}(\psi)+(\mathrm{gh}(\phi) + 1)(\mathrm{gh}(\psi) +1)} \sbv{\psi(\sigma^{\prime})}{\sbv{\phi(\sigma)}{\xi(\sigma^{\prime\prime})}}.
%\end{align}
In cases where \eqref{s^2} holds true, the BV extension of the classical action now simply looks as follows: 
\beq  S_{BV}^{S=0}  = S_{cl} + \int_{\Sigma} \left(
\xzero^+_i \, s \xzero^i 
+ A^{+i} \,  s A_i - c^{+i} \, s c_i 
\right).  \label{minimal}
\eeq 
Here we denoted the classical action \eqref{classicalactionofHPSM} by $S_{cl}$ for clarity, so as to avoid any confusion with the tensor $S$ defined in equation \eqref{S} above. The action of $s$  on the fundamental fields can be read off from  equations \eqref{BRSTofX}, \eqref{BRSTofA}, and \eqref{BRSTofc}. 

It is evident that this choice for the BV functional satisfies the classical master equation
\beq (S_{BV}, S_{BV})_{BV} = 0 \, . \label{master}
\eeq 
This results from the following three properties: \begin{itemize}
\item The classical action contains no momenta (antifields) and thus commutes with itself.
\item The gauge invariance of the classical action implies its BRST  invariance, $s S_{cl}=0$.
\item The property \eqref{s^2} ensures that $s$ is (graded) self-commuting, $[s,s]=0$. The corresponding parts of the BV action self-commute therefore, since, at least up to a sign, the bracket of the Hamiltonian lifts of vector fields yields the Hamiltonian lift of their (graded) Lie bracket.
\end{itemize}

Let us now turn to the general case, where $s$ squares to zero on-shell only, see \eqref{approx}. Then the last point above fails and
the violation to \eqref{master} is quadratic in the BV momenta or antifields. 
The BV procedure now amounts to the addition of further contributions to the minimal BRST  extension of the classical action:  
\begin{equation}\label{expansion}
\Sbvcov = S_{BV}^{(0)} + S_{BV}^{(1)} + S_{BV}^{(2)} + \cdots
\end{equation}
Here the superscript counts the degree in the antifields, while each term is of total ghost number zero.\footnote{The polynomial \label{antifieldnumber} degree of the antifields, used here to split terms, must not be confused with the antighost number  used in the literature (also called antifield number sometimes): Conventionally, the ghost number is understood as being the total degree of an additional bigrading, composed of the \emph{pure ghost number} pgh and the \emph{antighost number} agh. By definition, $\mathrm{gh} = \mathrm{pgh} - \mathrm{agh}$. While $\mathrm{agh}(X^+)=\mathrm{agh}(A^+)=1$, one agrees on $\mathrm{agh}(c^+)=2$ and the last term in \eqref{minimal}, now inside $S_{BV}^{(1)}$---see \eqref{SBV01}---has antighost number two. But while the BV bracket is of degree minus one in the degree of BV momenta (the polynomial degree of the antifields), it is not homogeneous with respect to the antighost number. Thus, the counting in the main text above seems more useful for our purposes. } 
In this notation, 
\begin{eqnarray} S_{BV}^{(0)} &=& S_{cl} \nonumber\\
S_{BV}^{(1)} &=& \int_{\Sigma} (-1)^{\mathrm{gh}(\Phi)} \Phi^+ s \Phi ,
 \label{SBV01} \end{eqnarray} 
where $\Phi$ denotes all fundamental fields in $\cM_{BRST}$, 
i.e.\ the classical fields together with the ghost fields.  %Thus, 
%for the choice of a connection compatible with the Lie algebroid structure---if it exists!---simply $\Sbvcov= S_{BV}^{(0)} + S_{BV}^{(1)}$, see equation \eqref{minimal}. 

In general, the expansion \eqref{expansion} does not terminate, see \cite{Henneaux-Teitelboim}.  Here it does already at level two,
\beq \Sbvcov = S_{BV}^{(0)} + S_{BV}^{(1)} + S_{BV}^{(2)} ,
\label{S_BV} \eeq 
and it does so with an astonishingly simple addition when expressed within the geometrical approach using connections: 
\beq S_{BV}^{(2)}
=
\int_{\Sigma} \tfrac{1}{4} S_{nk}{}^{ij}(X) A^{+n} A^{+k} c_i c_j  \, . \label{SBV2}\eeq

We now will prove that this choice of the BV extension of \eqref{classicalactionofHPSM} indeed satisfies the classical master equation \eqref{master}. $(S_{BV}^{(0)}, S_{BV}^{(0)})_{BV} = 0$ and  $(S_{BV}^{(0)}, S_{BV}^{(1)})_{BV} =0$ follow from the above two items as before. $(S_{BV}^{(2)}, S_{BV}^{(2)})_{BV}=0$ is also evident. 
$(S_{BV}^{(1)}, S_{BV}^{(1)})_{BV}$ and $(S_{BV}^{(0)}, S_{BV}^{(2)})_{BV}$ are both proportional to $\int_{\Sigma} S_{ij}{}^{kl}(X) \frac{\delta S_{BV}^{(0)}}{\delta A_n} A^{+j} c_k c_l$ and cancel (this is, in fact, what suggests to consider \eqref{SBV2}):
\begin{eqnarray}
(S_{BV}^{(1)}, S_{BV}^{(1)})_{BV} + 2(S_{BV}^{(0)}, S_{BV}^{(2)})_{BV} =0.
\end{eqnarray}
We are thus left with verifying that $(S_{BV}^{(1)}, S_{BV}^{(2)})_{BV}$ vanishes. A direct, somewhat involved calculation yields:
\begin{equation}
(S_{BV}^{(1)}, S_{BV}^{(2)})_{BV} = 
\tfrac{1}{3!} \int \left[
\pi^{lm} \nabla_m S_{nk}{}^{ij}
+ T_m^{jl} S_{nk}{}^{mi} + T_n^{mi} S_{mk}{}^{jl} - T_k^{lm} S_{nm}{}^{ij} 
\right] A^{+n} A^{+k} c_i c_j c_l . 
\end{equation}
Here now the Bianchi identity of the basic curvature $S$ comes at help, which for $E=T^*M$ takes the form (see Appendix \ref{sec:BianchiofS} for a derivation):
\begin{eqnarray}
&& \pi^{m[l} \nabla_m S_{nk}{}^{ij]} + T_m^{[jl} S_{nk}{}^{i]m} + T_n^{m[i} S_{mk}{}^{jl]} +T_k^{m[l} S_{nm}{}^{ij]} 
= 0.
\label{BianchiofS}
\end{eqnarray}
Evidently, the validity of this identity is even equivalent to 
$(S_{BV}^{(1)}, S_{BV}^{(2)})_{BV} = 0$. This concludes our proof of \eqref{master}.

Collecting the three contributions to \eqref{S_BV}---as given in \eqref{SBV01} and \eqref{SBV2}---we see that the BV extension of the HPSM with connection coefficients can be put into the form:\footnote{Here we add the superscript $\nabla$ since it depends on a connection and is a global result. The traditional expressions, resulting from this functional by the choice of connection coefficients with vanishing symmetric part within a local patch on $M$, will be simply denoted by   $S_{BV}$. Somewhat miraculously, these expressions will turn out to be globally well-defined as well, but with a different lift of the coordinate transformations to the field space.}
\begin{framed}
\begin{eqnarray} 
S_{BV}^\nabla[X,A,c,X^+,A^+,c^+] &=&  \int_{\Sigma} \left(A_i \wedge \bbd \xzero^i + \tfrac{1}{2} \pi^{ij} A_i \wedge A_{j} \right)
+ \int_{N} H
\nonumber \\ && 
\!\!\!\!\!\!\!\!\!\!\!\!\!\!\!\!\!\!\!\!\!\!\!\!\!\!\!\!\!\!\!\!\!\!\!\!\!\!\!\!\!\!\!\!\!\!\!\!\!\!+ \int_{\Sigma} 
\left[
- \pi^{ij} \xzero^+_{i} c_j
+ A^{+i} \wedge \left(\rd c_i +
f_i^{jk} A_{j} c_k - \Gamma_{ij}^k F^j c_k
\right)
+ \tfrac{1}{2} f_k^{ij} c^{+k} c_i c_j
\right]
\nonumber \\ &&
+  \int_{\Sigma} 
\tfrac{1}{4} S_{nk}{}^{ij} A^{+n} \wedge A^{+k} c_i c_j. \label{Mainresult2}
\end{eqnarray}
\end{framed}
In the formula above the first line corresponds to the classical action, the second line is the Hamiltonian lift of the (only on-shell-nilpotent) BRST  differential constructed in the previous subsection, and the last line contains the lift of the basic curvature tensor \eqref{S} to the BV field space. 

If one intends to write $S_{BV}$ explicitly in terms of only $\pi$, $H$, and $\nabla \sim \Gamma$, one can do so upon usage of \eqref{Slong} and \eqref{f} together with $F^j = \rd X^j + \pi^{jk} A_k$.
The current expression is not only more favorable for the geometrical meaning of the second order contribution, however: Except for the classical action in the first line of \eqref{Mainresult2}, all structural functions entering the BV extension are determined by the Lie algebroid structure and the connection on the underlying bundle only. Thus this result may be pathing the way also for the construction of globally well-defined BV extensions of more general gauge theories based on (possibly also higher) Lie algebroids symmetries.

\subsection{Manifestly target space covariant form}\label{sec:Manifest}

One expects that the extension \eqref{Mainresult2} is covariant with respect to changes of coordinates on the target manifold $M$: After all, we took care of a proper formulation of the gauge symmetry generators in Sec.\ \ref{sec:cov} which underly the BRST construction of Sec.\ \ref{sec:BRSTtransformation} and the addition of the quadratic term is governed completely by a tensorial object, the basic curvature tensor $S$ of $(E,\nabla)$. In the following, we will make this covariance explicit here. 

Let us view the ghosts $c$, the classical fields $A$, and their   conjugate antifields as covariant objects to start with:
\begin{align} A&\in  \Omega^1(\Sigma,X^* T^*M) \, , & A^+&\in  \Omega^1(\Sigma,X^* T^*[-1]M) \, ,\nonumber
\\c&\in  \Omega^0(\Sigma,X^* T^*[1]M) \, , & c^+&\in  \Omega^2(\Sigma,X^* T^*[-2]M) \, .\label{cc+}
\end{align}
This implies that if we change coordinates $x^i$ to $\widetilde{x}^i$ and denote the corresponding Jacobian matrix by 
\beq M^i_j (x):= \frac{\partial  \widetilde{x}^i(x)}{\partial x^j} \, ,\label{Jacobian}
\eeq 
then, for example, the components of the ghost antifields are related to those in the previous holonomic basis according to 
\beq  \widetilde{c}^{+i} = M^i_j  \,  c^{+j} \, , \label{tensorial}
\eeq 
where we wrote $M^i_j$ for $X^*M^i_j\equiv M^i_j(X)$ to simplify the notation. These transformations are lifted to a canonical transformation of the BV symplectic form 
\eqref{componentBVsymplecticform},
\begin{eqnarray}
\omega_{BV} &=& \int_{\Sigma} 
\left(\delta \widetilde{X}^i \wedge \delta \widetilde{X}^+_i + \delta \widetilde{A}_i \wedge \delta \widetilde{A}^{+i}
+ \delta \,\widetilde{c}_i \wedge \delta \, \widetilde{c}^{+i}\right),
\label{componentBVsymplecticformtilde}
\end{eqnarray}
 if the fields $X^+$ transform as follows:
\beq  \widetilde{X}^{+}_{i} = (M^{-1})^j_i X^+_j 
- (M^{-1}){}_s^l  M^s_k{}_{,i} \, (A^{+k} A_l + c^{+k} c_l). \label{X+transform} 
\eeq
In fact, these combined transformations also leave the canonical 1-form $\sum \Phi^+ \delta \Phi$ of \eqref{MBV} invariant.
Inspection of \eqref{Mainresult2} shows that its first and third line are inherently covariant. One may now in principle check by a direct calculation that also the second line transforms into itself under the above canonical transformation. 

However, there is a simpler way for checking covariance. By the same scheme as for BFV, see \eqref{covp}, we are led to consider  the following covariantized BV-momentum: 
\beq \label{neuesX+}
\xzero^{+\!\nabla}_i := \xzero^{+}_i + \Gamma_{ji}^k (A^{+j} \wedge A_k + c^{+ j} c_k).\eeq 
This  is a tensorial field; more precisely, this means
\beq 
\xzero^{+\!\nabla}\in \Omega^2(\Sigma,X^*T^*[-1]M) \, . \label{tensor}
\eeq 
Now one verifies that \eqref{Mainresult2} can be rewritten identically as follows: 
\beq 
S_{cl}+ \int_{\Sigma} 
\left[
- \pi^{ij} \xzero^{+ \nabla}_{i} c_j + A^{+i} \wedge \left(\mathrm{D} c_i - T_i^{jk} A_{j} c_k 
\right)
- \tfrac{1}{2} T_k^{ij} c^{+k} c_i c_j
\right] + \int_{\Sigma} 
\tfrac{1}{4} S_{nk}{}^{ij} A^{+n} \wedge A^{+k} c_i c_j.\label{universal}
\eeq 
Here $Dc$ denotes the covariant exterior derivative on $c$, $
\mathrm{D} c_i = \rd c_i - \Gamma_{il}^k \rd \xzero^l c_k$. 
We believe, that this form of the BV action will be similar to the BV extensions of other Lie algebroid based gauge theories, where then $S_{cl}$ denotes the corresponding classical action and the last term receives appropriate adaptations (the (H)PSM is a Chern-Simons type Lie algebroid gauge theory, possible for a two-dimensional $\Sigma$ and $E=T^*M$ only \cite{StroblAYM}).

To sum up, we rewrite  the action \eqref{Mainresult2} or \eqref{universal} in manifestly frame-independent terms, clarifying or recalling some of the notation right thereafter:

\begin{framed}
\begin{eqnarray} 
S_{BV}^\nabla[X,A,c,\xzero^{+\!\nabla},A^+,c^+] &=&  \int_{\Sigma} \left[\langle A ,%\stackrel{\wedge}{,} 
\bbd \xzero \rangle + \tfrac{1}{2} (\pi \circ X)(A, A) \right]
+ \int_{N} X^*H
\nonumber \\ && \!\!\!\!\!\!\!\!\!\!\!\!\!\!\!\!\!\!\!\!\!\!\!\!\!\!\!\!\!\!\!\!\!\!\!\!\!\!\!\!\!\!\!\!\!\!\!\!\!\!\!\!\!\!\!\!\!
+ \int_{\Sigma} 
\left[
 \langle A^{+} , \mathrm{D} c 
-  (T\circ X) (A, c) \rangle-  (\pi\circ X) (\xzero^{+\!\nabla},  c)  - \tfrac{1}{2}\langle c^+, (T\circ X) (c, c) \rangle
\right]
\nonumber \\ &&
+  \int_{\Sigma}  \tfrac{1}{4} 
\langle A^+, (S\circ X)(A^+, c,c) \rangle
. \label{Mainresult2prime}
\end{eqnarray}
\end{framed}
Here the couple $(\pi,H)$ satisfies the defining equation \eqref{PiPi} of a twisted Poisson structure,  $T$ denotes the $E$-torsion (Eqs.\ \eqref{defEtorsion}, \eqref{Tcomp}, and \eqref{nablapi}) of the corresponding Lie algebroid $E=T^*M$ equipped with a connection $\nabla$ whose torsion is restricted by \eqref{torsionpiH}  and  $S$ is its basic curvature (Eqs.\ \eqref{S} and \eqref{Slong}). The wedge product between the differential 1-forms on $\Sigma$ is understood; for example,  \beq (\pi \circ X)(A, A) \equiv (X^*\pi^{ij}) \, A_i \wedge A_j .\eeq
%stands, more explicitly, for $(\pi \circ X)(A, A)$,  as well. 
As usual for a Wess-Zumino term, in the last term of the first line, the 3-form $H$ is pulled back by  $X \colon N \to M$ where $\Sigma= \partial N$; in all other terms---as well as in the Euler-Lagrange equations of the whole functional---one has  $X \colon \Sigma \to M$. 

Using the covariant field $\xzero^{+\!\nabla}$ has the price of loosing the Darboux form of the BV brackets. In complete analogy with the BFV situation in Sec.\ \ref{sec:manifestlyBFV}, one may, however, also show that the BV brackets can be cast into a manifestly covariant form when taking into account a non-trivial action of $(\xzero^{+\!\nabla}, \cdot )_{BV}$ on the basis vectors of the sections.  We leave the details to the reader (or refer to Sec.\ \ref{sec:circle} for equivalent formulas in a somewhat more adapted notation). Like this one obtains a BV formulation of the HPSM which is  manifestly covariant.

\subsection{Independence of the auxiliary connection}\label{sec:independence}
At the end of this section we want to address the dependence of the functional $S_{BV}^\nabla$ on the choice of the auxiliary connection $\nabla$. More precisely, 
the connection $\nabla$ underlying \eqref{Mainresult2} and \eqref{Mainresult2prime} always has the following form 
\beq \nabla = \:\stackrel{0}{\nabla} - \tfrac{1}{2} \langle \pi,H\rangle \label{ournabla}
\eeq
for some torsion-free connection $ \stackrel{0}{\nabla}$ that we can choose at will.
We want to show here that different choices of this auxiliary connection $ \stackrel{0}{\nabla}$ lead to equivalent BV formulations for the HPSM. In other words, since the space of connections is  affine, we want to show that the theory remains invariant with respect to the shifts \beq  \stackrel{0}{\nabla} \:\mapsto\: \stackrel{0}{\nabla} +\,\, t \, 
, \label{newnablabla}
\eeq
for arbitrary $t \in \Gamma(TM \otimes S^2T^*M)$. Since in the manifestly covariant BV formulation of Sec.\ \ref{sec:Manifest}, the argument field $\xzero^{+\!\nabla}$ of \eqref{Mainresult2prime} depends on $\nabla$, and, what is worse, also the fundamental BV brackets do, it is preferable here to consider \eqref{Mainresult2} instead: its fields do not depend on the choice of connection and $\omega_{BV}$ is in Darboux form. 

Let $\nabla' := \nabla + t$. The searched-for statement will follow by showing that,  for every  $t$, there is a  redefinition of fields $(X,A,c,X^+,A^+,c^+) \mapsto (X',A',c',X^+{}',A^+{}',c^+{}')$ which preserves the BV symplectic form, $\omega_{BV}=\omega_{BV}'$, and 
for which, given the relation between the fields, one has
\beq S_{BV}^\nabla[X,A,c,X^+,A^+,c^+] = S_{BV}^{\nabla'}[X',A',c',X^+{}',A^+{}',c^+{}']. \label{redefinition}
\eeq 

The most general bi-degree preserving map with $X^{\prime} = X$ is of the form\begin{align} 
A^{+i} &= f_1{}^i_j(X) A^{+\prime j}, \nonumber \\
c^{+i} &= f_2{}^i_j(X) c^{+\prime j} + \frac{1}{2} t_1{}^i_{jk}(X) A^{+\prime j} A^{+\prime k}, \nonumber \\
c_{i} &= f_3{}^j_i(X) c^{\prime}_j, \label{fieldchange} \\
A_i &= f_4{}_i^j(X) A^{\prime}_j + t_2{}_{ki}^j(X) A^{+\prime k} c^{\prime}_j, \nonumber\\
X^+_i & := f_5{}_i^j(X) X^{+\prime}_j + t_3{}_{ij}^k(X) A^{+\prime j} A^{\prime}_k + t_4{}_{ij}^k(X) c^{+\prime j} c^{\prime}_k
+ \frac{1}{2} t_5{}_{ikl}^j(X) A^{+\prime k} A^{+\prime l} c^{\prime}_j
\nonumber
\end{align}
for some set of (local) coefficient functions $(f_a{}^i_j, t_a{}_{ij}^k)_{a=1,\ldots, 5; \, i,j,k = 1 \ldots n}$. The functions $f_a{}^i_j$ are  of no relevance here and  we put $f_a{}^i_j:= \delta^i_j$ for all $a$. The remaining transformations are BV symplectic iff  
\begin{align} &t_1{}^i_{jk}=t_2{}^i_{jk}= t_2{}^i_{kj} \, ,\nonumber \\
& t_3{}_{ij}^k=t_4{}_{ij}^k=0 \, , \\
&t_5{}_{ikl}^j = - t_2{}^j_{kl,i}\, .\nonumber
\end{align}
Identifying the symmetric tensor $t_2$ with  $t$ in \eqref{newnablabla}, the transformations \eqref{fieldchange} become
\begin{align} 
A^{+i} &= A^{+\prime i}, \nonumber \\
c^{+i} &= c^{+\prime i} - \tfrac{1}{2} t{}^i_{jk}(X) A^{+\prime j} A^{+\prime k}, \label{fieldchange2} \\
c_{i} &= c^{\prime}_i, \nonumber \\
A_i &= A^{\prime}_i - t{}_{ki}^j(X) A^{+\prime k} c^{\prime}_j, \nonumber\\
X^+_i & := X^{+\prime}_i 
%- t{}_{ij}^k(X) A^{+\prime j} A^{\prime}_k + t_4{}_{ij}^k(X) c^{+\prime j} c^{\prime}_k
+ \tfrac{1}{2} t{}_{kl,i}^j(X) A^{+\prime k} A^{+\prime l} c^{\prime}_j.
\nonumber
\end{align}
One verifies by a straightforward calculation that indeed \eqref{fieldchange2} 
leads to \eqref{redefinition}. This concludes the proof.

\newpage 

%%%%%%%%%%%%%%%%%%%%%%%%%%%%%%%%%%%%%%%%%%%%%%%%%%%%%%%%%%%%%%%%%%%%%%%%%%%

\section{The BV formulation with superfields}\label{sec:BV2}
In this section we will derive the superfield formulation of the BV action for the HPSM. We will start with the traditional setting for the BV construction: namely the BV functional that one obtains when considering the gauge symmetries without the use of an auxiliary connection. This functional is obtained merely by putting to zero the connection coefficients in the functional $S_{BV}^\nabla$ within some $U \subset M$ and for some choice of coordinates $x^i$ in $U$.  As discussed in Sec.\ \ref{sec:cov}, \emph{a priori} such a functional is valid only within the local patch $U$ and for the choice of coordinates made. We then will show that this functional can be recast into one expressed merely in terms of superfields---in fact, those superfields that are also used for the PSM in its AKSZ formulation. It is a non-trivial fact that all the terms containing $H$ can be recombined into a smaller number of them in this way. The resulting BV formulation leads to another global extension of the functional, without the use of the connection and which,  correspondingly, we will denote simply by $S_{BV}$. The different gluing needed on $M$ to patch local expressions for the BV functional together can be regarded as a different (canonical) lift of the group of diffeomorphisms of $M$ to the space of BV fields. 

In the context of the superfield formulation, it will be useful to change conventions  from Deligne to Bernstein-Leites for the commutativity relations of the bigraded fields. We will comment on this step in detail.

At the end, finally, we will show that both formulations, the one obtained in this section and the one of the previous section, are BV equivalent: there is a BV canonical transformation, depending on the choice of a connection $\nabla$ on $M$, which maps $S_{BV}$ into $S_{BV}^\nabla$. We do not see any a priori reason why this should be the case: in general, one only knows that different BV extensions of the same classical functional should be quasi-isomorphic. Here we obtain an isomorphism \eqref{iso} on the nose. 

As mentioned above, to obtain the BV functional underlying the naive gauge symmetry generators the gauge transformations \eqref{trafoX} and \eqref{gaugetransformation02a}, we merely need to put the torsion free part of the connection coefficients  $\stackrel{\circ}{\Gamma_{ij}^k} = 0$ in all of the formulas using coordinates on $M$ in the previous section.

\subsection{Traditional BV formulation}\label{sec:flatBVformalism}
In this subsection we present the result of the BV action in a local chart where the torsion free part of the connection coefficients has been put to zero,  $\stackrel{\circ}{\Gamma_{ij}^k} = 0$. This corresponds to a more traditional setting, where one starts with the gauge transformations \eqref{trafoX} and \eqref{gaugetransformation02a}; they are absolutely fine as generators of non-trivial gauge transformations and differ from the covariant counterpart using a connection only by trivial gauge transformations---see, in particular, the discussion around \eqref{Lambda}. However, as we pointed out in section \ref{sec:cov}, they have the deficiency of not behaving well with respect to the gluing of patches on the target manifold. 
Correspondingly, at least a priori, such formulas are valid only if one restricts to one coordinate system in a patch $U$ on $M$. However, as it turns out, this can be cured by additional ghost contributions to the transformation formulas of fields which are present only in the BV extension.

In any case, we obtain the formulas of relevance here by merely setting the connection coefficients within the previous expressions equal to their torsion part, $\Gamma_{ij}^k \mapsto - \tfrac{1}{2} \pi^{kl} H_{ijl}$. Then the BRST transformations become
\begin{eqnarray}
s X^i &=& - \pi^{ij}c_j,
\nonumber
\\
s A_{i} &=&\rd c_i + f_i^{jk} A_{j} c_k + \tfrac{1}{2} \pi^{kl} H_{ijl} F^j c_k,
\label{BRSTofXAcflat}
\\
s c_i &=& - \tfrac{1}{2} f_i^{jk} c_j c_k \, . \nonumber
\end{eqnarray}
As before, the operator $s$ squares to zero except when acting on the fields $A_i$,
\begin{eqnarray}
s^2 A_i &=& \left(\tfrac{1}{2} f_{(i}^{kl}{},_{j)}+ \tfrac{1}{4}  \pi^{ck} \pi^{la} \pi^{bd} H_{iab} H_{jcd}\right) F^j c_k c_l .  
\label{BRST2Aflat}
\end{eqnarray}
Here we made use of \eqref{K} inside \eqref{BRST2A}. The expression becomes even longer---and less transparent---when spelling out the Lie algebroid structure functions $f_{k}^{ij}$  in terms of $\pi$ and $H$, cf.\ the second line of \eqref{BRST2A}. In this form, any geometric interpretation of the coefficient functions in \eqref{BRST2Aflat} is lost. But, as before, the right-hand side of this equation vanishes on-shell, showing that one needs the BV formalism. 

All what we said about the space of fields of the BV theory remains unaltered. We are thus left only with specifying the BV action. It takes the form, \begin{eqnarray}
S_{BV} &=&  \int_{\Sigma} \left(A_i \wedge \rd \xzero^i 
+ \tfrac{1}{2} \pi^{ij} A_i \wedge A_{j} \right)
+ \int_{N} H
\nonumber \\ && 
+ \int_{\Sigma} 
\left[
- \pi^{ij} \xzero^+_{i} c_j
+ A^{+i}\wedge  \left(
\rd c_i + f_i^{jk} A_{j} c_k + \tfrac{1}{2} \pi^{kl} H_{ijl} F^j c_k
\right)
+ \tfrac{1}{2} f_k^{ij} c^{+k} c_i c_j
\right]
\nonumber \\ &&
- \int_{\Sigma} 
\tfrac{1}{4} \left(f_{n}^{ij}{},_{k} + \tfrac{1}{2}  \pi^{ci} \pi^{ja} \pi^{bd} H_{nab} H_{kcd} \right)
A^{+n} \wedge A^{+k} c_i c_j. \label{flatBVactionfirst}
\end{eqnarray}
for $S_{BV}=S_{BV}[X,A,c,X^+,A^+,c^+]$, i.e.\ when written as a functional of the same variables as those used in $S_{BV}^\nabla$ in \eqref{Mainresult2}. 
If we had constructed the BV extension ab initio starting with the \eqref{trafoX} and \eqref{trafoAf} and not deducing it from \eqref{Mainresult2}, the first two lines would have followed essentially in the same way as before: One takes the classical action, adds the generating formulas for the gauge symmetry, with ghosts $c$ replacing the gauge parameters $\epsilon$, and extend this by the Lie algebroid structure functions on the ghosts. This yields the first two lines of the above expansion, corresponding still to \eqref{SBV01}, also for this somewhat more traditional presentation of the gauge symmetries. One may still have been led to consider adding a term like the one in the last line and then be surprised that, after some calculation using the structural identities of a Lie algebroid, that this is already sufficient. The point is that with the geometric interpretation of $S$ as a kind of curvature, it is near at hand that it satisfies also some kind of Bianchi identity, and it is precisely the equation that needs to hold for the process to terminate. 

%The situation becomes even less transparent, if one would not only ignore the need for connection coefficients in the generators of non-trivial gauge symmetries, but also would not notice the Lie algebroid structure governing the HPSM. In this case  the extension process is started with the symmetry generators of the form \eqref{trafoX} and \eqref{gaugetransformation02a}.
 Expressed in terms of the original quantities $\pi$ and $H$---and the original fields $A$, $X$, amended by the ghosts $c$ together with all their antifields---
the final BV extension  reads as follows:
\begin{eqnarray} 
S_{BV} &=&  \int_{\Sigma} \left(A_i \wedge \rd \xzero^i 
+ \tfrac{1}{2} \pi^{ij}A_i\wedge  A_{j} \right)
+ \int_{N} H
\nonumber \\ && 
+ \int_{\Sigma} 
\left[
- \pi^{ij} \xzero^+_{i} c_j
+ A^{+i}\wedge  \left(
\rd c_i + \pi^{jk},_i A_{j} c_k + \tfrac{1}{2} \pi^{jk} H_{ijl} ( \rd \xzero^l 
- \pi^{lm} A_{m}) c_k
\right)
\right]
\nonumber \\ && 
 + \int_{\Sigma} \tfrac{1}{2} (\pi^{ij},_k + \pi^{il} \pi^{jm} H_{klm}) c^{+k} c_i c_j
\nonumber \\ &&
- \int_{\Sigma} 
\tfrac{1}{4} \left(\pi^{ij},_{nk} + (\pi^{il} \pi^{jm} H_{klm}),_n + \tfrac{1}{2}  \pi^{ci} \pi^{ja} \pi^{bd} H_{nab} H_{kcd} \right)
A^{+n} \wedge A^{+k} c_i c_j
.
\label{flatBVaction}
\end{eqnarray}
When seeing $S_{BV}$ in this form,  it is maybe not so astonishing that,  in the case of a non-vanishing 3-form $H$, it was not found earlier. %Already only the \emph{verification} of the validity of the  master equation for the above expression---merely by means of the fundamental equation \eqref{PiPi} and its derivatives---is a highly non-trivial endeavor. 
Note also that, according to the traditional expansion along the antifield number---see footnote \ref{antifieldnumber}---the first and second line of \eqref{flatBVaction} have antifield number zero and one, respectively, while the  third and the fourth line both have antifield number two. 

%In Sec.\ \ref{sec:BFV}, we considered several versions of the BFV extension, one in components without connection coefficients, one using a connection to represent the result in inherently covariant terms using covariantized fields, and a final one in terms of superfields (in terms of the original fields, but combined into superfields). It turned out that this was all the same functional (up to a change of ``coordinates in field space''), just rewritten in different ways. In particular, the connection $\nabla$ used to obtain the form \eqref{covariantBFVBRSTcharge2} was absolutely arbitrary. The only place where the special role of $\nabla$ to have torsion \eqref{torsionpiH} 
%was that then and only then the tensor $T$ becomes independent of $H$.

%
%The situation in the present section is substantially different: First, the preferred class of connections with the given torsion, see Eq.\ \eqref{Gamma}, became clearly visible in the context of the parametrization of the generators of nontrivial gauge symmetries, see \eqref{deltaAnabla}.
As a side remark, we find it interesting to compare the local expressions for $S^{\nabla}_{BV}$ and $S_{BV}$, both  understood as functionals of the same variables, i.e.\ of $(X,A,c)$ and their antifields $(X^+,A^+,c^+)$):\begin{eqnarray}
(\Delta S)_{BV} :=S^{\nabla}_{BV} - S_{BV} 
&=& 
\int_{\Sigma} 
%\!\!\!\!
%d^2 \sigma d^2 \theta \,
\left[
A^{+i} \stackrel{\circ}{\Gamma^k_{ij}} F^j c_k
+ \tfrac{1}{4} (S_{nk}{}^{ij} - S_{nk}{}^{ij}|_{\stackrel{\circ}{\Gamma}=0}
) A^{+n} A^{+k} c_i c_j
\right]
\nonumber \\ 
&& \!\!\!\!\!\!\!\!\!\!\!\!\!\!\!\!\!\!\!\!\!\!\!\!\!\!\!\!\!\!\!\!\!\!\!\!\!\!\!\!\!\!\!\!\!\!\!\!\!=\int_{\Sigma} 
%\!\!\!\!
%d^2 \sigma d^2 \theta \,
\left[
A^{+i} \stackrel{\circ}{\Gamma^k_{ij}} F^j c_k
+ \tfrac{1}{4} (\stackrel{\circ}{S_{nk}{}^{ij}} 
+ \pi^{ij}{}_{,nk} + \stackrel{\circ}{\Gamma^m_{nk}} \pi^{ia} \pi^{jb} H_{mab}
) A^{+n} A^{+k} c_i c_j
\right], \label{noncov}
\end{eqnarray}
where 
$\stackrel{\circ}{S_{nk}{}^{ij}} = S_{nk}{}^{ij}|_{H=0}$. There are two observations one can make here: First, this expression is evidently not manifestly covariant with respect to a change  of coordinates on the target manifold $M$ (under the transformations provided above): This is simply the case since connection coefficients do not transform tensorially and, for example, 
there are no contributions from the second term on the right-hand side of \eqref{noncov} which can cancel the unwanted contributions from the first term.

Second, $S_{BV}$ and $S_{BV}^\nabla$  result as an extension of the same classical action but for a different choice  of generators of the gauge symmetries. As such these generators differ by terms vanishing on-shell. Still, the difference \eqref{noncov} is not simply BV-exact. In fact, $(\Delta S)_{BV}$ only satisfies the standard Maurer-Cartan equation, with both terms non-vanishing in general:
\beq 0 = \left(S_{BV},(\Delta S)_{BV}\right)_{BV} + \tfrac{1}{2}\left((\Delta S)_{BV},(\Delta S)_{BV}\right)_{BV} .
\eeq 

And still, as we will see below, there is a precise way in which $S_{BV}$ will become covariant with respect to changes of coordinates on $M$ and thus globally well-defined. Moreover, it will  be even  possible to \emph{identify} $S_{BV}^\nabla$ and $S_{BV}$ by some non-trivial BV-isomorphism. But this is not something to be expected, at least not due to some general properties we are aware of. According to current knowledge, both these facts appear to be coincidences that occur for the topological $H$-deformation of the PSM. It will be interesting, however, to see which parts of them hold true also for more general settings in the BV construction.

%%%%%%%%%%%%%%%%%%%%%%%%%%%%%%%%%%%%%%%%%%%%%%%%%%%%%%%%%%%%%%%%%%%%%%%%%%%%%%%
\subsection{Reformulation in terms of superfields}\label{superfieldBVformalism}
The space of classical fields \eqref{Mcl} can be seen as the space of degree-preserving maps from $T[1] \Sigma$ to $T^*[1]M$, i.e.
\beq {\cal M}_{cl} = \mathrm{Hom}(T[1] \Sigma,T^*[1]M) \, . \label{Hom}
\eeq 
It is far less evident, but a consequence of the AKSZ formalism of the PSM, that all the fields in the BV phase space \eqref{MBV} can be combined simply into all (not necessarily degree-preserving) maps:\footnote{Note the analogy of \eqref{Hom} and \eqref{Homunderline} with its Hamiltonian counterpart, \eqref{Mclass} and \eqref{MBFV}.}
\beq {\cal M}_{BV} \cong \underline{\mathrm{Hom}}(T[1] \Sigma,T^*[1]M) .
\label{Homunderline}\eeq 
This holds still true for the HPSM since the presence of the 3-form $H$ does not modify the BV phase space.

Let us use coordinates $(\sigma^{\mu}, \theta^{\mu})$ on $T[1]\Sigma$ of degree $(0,1)$. In this picture,  differential forms on $\Sigma$ become functions on $T[1]\Sigma$. The fields of \eqref{Homunderline} are two-dimensional superfields, encompassing all the previously introduced ones as follows:\footnote{We will clarify the relative signs in these formulas after explanations about the conventions. In this context, the new notation with underlining the fields will become clearer as well; it is in part due to different conventions about their commutativity properties.} %Equation \eqref{superfield2d}.
\begin{eqnarray}
\bx^i (\sigma, \theta)
&\equiv & \uX^i (\sigma) - \uAplus^i (\sigma, \theta)
+ \ucplus^i (\sigma, \theta)
\nonumber \\ &:= & \xzero^i (\sigma) - \theta^{\mu} A_{\mu}^{+i} (\sigma)
+ \tfrac{1}{2} \theta^{\mu} \theta^{\nu} c_{\mu\nu}^{+i} (\sigma),
%= \xzero^i + A^{+i} + c^{+i},
\label{2Dsuperfield1} \\
\ba_i (\sigma, \theta) 
&\equiv &
- \uc_i (\sigma) + \uA_i (\sigma, \theta)
+ \uXplus\!\!\!_i (\sigma, \theta)
\nonumber \\ &:=&
- c_i (\sigma) + \theta^{\mu} A_{\mu i} (\sigma)
+ \tfrac{1}{2} \theta^{\mu} \theta^{\nu} \xzero_{\mu\nu i}^{+} (\sigma).
\label{2Dsuperfield2}
\end{eqnarray}
In this context it is useful to introduce a total degree as the sum of the two previously used degrees: 
\beq \label{deg} 
\mathrm{deg}(\phi):= \mathrm{fdeg}(\phi)+\mathrm{gh}(\phi). \eeq Then $\mathrm{deg}(\bx^i)= 0$ and $\mathrm{deg}(\ba_i)= 1$, in coincidence with the fact that the sum of the individual degrees of \eqref{c+} and the first two lines of \eqref{degrees} give zero and of the last three lines in \eqref{degrees} give one. In accordance with  \eqref{Hom}, the classical fields correspond precisely to the  functions on $T[1]\Sigma$ of (fdeg) degree zero and of (fdeg) degree one inside the superfields \eqref{2Dsuperfield1} and \eqref{2Dsuperfield2}, respectively.\footnote{Note that fdeg now counts the polynomial degree of $\theta$ for functions on $T[1]\Sigma$, not to be confused with a form degree of graded differential forms on this supermanifold. In particular, on $T[1]\Sigma$ one now has $\mathrm{deg}(\theta^\mu) = \mathrm{fdeg}(\theta^\mu) = 1$ but $\mathrm{deg}(\rd \sigma^\mu) = \mathrm{fdeg}(\rd \sigma^\mu) = 0$ where $\rd \sigma^\mu \in \Omega^1(T[1]\Sigma)$. It is thus also important that we change the notation consistently from the previous sections to the present one: what previously was a differential form  on $\Sigma$ like $\alpha = \rd \sigma^\mu \, \alpha_\mu$ will now be underlined and written as the function $\underline{\alpha} = \theta^\mu \, \alpha_\mu(\sigma)$, to avoid any confusion of the above sort.}

In fact, in the present context it is much more convenient to change the sign conventions from Deligne to Bernstein-Leites. For the product of two fields $\uphi$ and $\upsi$, viewed as functions on $T[1]\Sigma$ of some particular ghost degrees (and of some fixed total degrees according to \eqref{deg}), we pose:
\begin{eqnarray} \label{newcomm}
\uphi \, \upsi = (-1)^{\mathrm{deg}(\uphi)\mathrm{deg}(\upsi)} \upsi \, \uphi .
\end{eqnarray}
In view of \eqref{oldproduct}, this has the potential to lead to contradictions when transcribing formulas from the previous section into this one by mere transcription. So, some care is needed for the passage. 

Let $\phi$ and $\psi$ be fields within Sections \ref{sec:normalBVformalism}, \ref{sec:Manifest}, and \ref{sec:flatBVformalism}, viewed as differential forms on $\Sigma$ and subject to the Deligne sign rule  \eqref{oldproduct}. Denote by $\uphi$ and $\upsi$ the corresponding 
superfields on $T[1]\Sigma$. Let us \emph{define} a new product for these superfields by the following formula:
\begin{eqnarray}
\phi \wedge \psi %\rightarrow 
\mapsto
(-1)^{\mathrm{gh}(\phi)\mathrm{fdeg}(\psi)} \, \uphi \, \upsi .
\label{signchangingrule}
\end{eqnarray}
First of all, it is easy to verify that this product indeed satisfies \eqref{newcomm}. This conversion rule is consistent with associativity moreover: For example,
for a product of three fields, $\phi \wedge \psi \wedge \xi \equiv (\phi \wedge \psi) \wedge \xi = \phi \wedge (\psi \wedge \xi)$, we have 
\begin{eqnarray}
\phi \wedge \psi \wedge \xi %\rightarrow 
\mapsto
(-1)^{\mathrm{gh}(\phi)\mathrm{fdeg}(\psi)
+ \mathrm{gh}(\phi)\mathrm{fdeg}(\xi) + \mathrm{gh}(\psi)\mathrm{fdeg}(\xi)} \, \uphi \, \upsi \, \uxi,
\end{eqnarray}
independently of in which order one applies \eqref{signchangingrule} to the two wedge products. Now it is easy to verify that the commutation relation \eqref{oldproduct} for the old product implies \eqref{newcomm} for the new one.

In a similar fashion, we now define a new bracket:
\begin{eqnarray}
~(\phi, \psi)_{BV} &%\rightarrow
\mapsto
& (-1)^{(\mathrm{gh}(\phi) + 1)\mathrm{fdeg}(\psi)} (\uphi, \upsi). \label{newbracket}
\end{eqnarray}
The identities  \eqref{81} for the old BV bracket then imply the following properties for the new one:\footnote{As before, one may imagine that $\uphi$, $\upsi$, and $\uxi$ depend on $(\sigma, \theta)$, $(\sigma^{\prime}, \theta^{\prime})$, and  $(\sigma^{\prime\prime}, \theta^{\prime\prime})$, respectively.}
\begin{eqnarray}
&& \left(\uphi, \upsi\right)
= - (-1)^{(\mathrm{deg}(\uphi) + 1)(\mathrm{deg}(\upsi) + 1)}
\left(\upsi, \uphi \right),  \nonumber
\\
&& \left( \uphi, \upsi \, \uxi \right)
= (\uphi,\upsi) \, \uxi + (-1)^{(\mathrm{deg}(\uphi) + 1)\mathrm{deg}(\upsi)} \, \upsi \, (\uphi, \uxi),\label{equation}
\\
&& 
\left(\uphi, \left(\upsi, 
 \uxi \right) \right)
=\left( \left( \uphi, \upsi \right), \uxi \right) 
 +(-1)^{(\mathrm{deg}(\uphi) + 1)(\mathrm{deg}(\upsi) +1)} \left(\upsi, \left( \uphi, \uxi \right) \right). \nonumber
\end{eqnarray}

Denote by $\sd = \theta^{\mu} \partial_{\mu}$ the degree one vector field on $T[1]\Sigma$ corresponding to the de Rham differential $\rd$ on $\Sigma$. Applying the sign changing rule \eqref{signchangingrule}, 
the total BV action \eqref{flatBVaction} can be rewritten as:
\begin{eqnarray}
S_{BV} 
&=& 
\int_{T[1]\Sigma} 
\!\!\!\!
d^2 \sigma d^2 \theta \,
\left(\uA_i \, \sd \uX^i + \tfrac{1}{2} \pi^{ij}  \uA_i \, \uA_j \right)
\nonumber \\ &&
+ \int_{T[1]N} 
\!\!\!\!
d^3 \sigma d^3 \theta \,
\tfrac{1}{3!} H_{ijk}\,  \sd X^i \, \sd X^j \, \sd X^k
%\nonumber \\ && 
- \int_{T[1]\Sigma} 
\!\!\!\!
d^2 \sigma d^2 \theta \,
\pi^{ij} \uXplus\!\!\!_i \, \uc_j
\nonumber \\ && 
+\int_{T[1]\Sigma} 
\!\!\!\!
d^2 \sigma d^2 \theta \,
\uAplus^i \left[\sd \uc_i + \pi^{jk},_i \uA_j \, \uc_k 
+ \tfrac{1}{2} \pi^{kl} H_{ijl} (\sd \uX^j - \pi^{jl} \uA_l) 
%F^j 
\, \uc_k
\right]
\nonumber \\ &&
+ \int_{T[1]\Sigma} 
\!\!\!\!
d^2 \sigma d^2 \theta \,
\tfrac{1}{2} (\pi^{ij},_k + \pi^{il} \pi^{jm} H_{klm}) \, \ucplus^k \uc_i \, \uc_j
\nonumber \\ &&
+ \int_{T[1]\Sigma} 
\!\!\!\!
d^2 \sigma d^2 \theta \,
\tfrac{1}{4} \left[\pi^{ij},_{nk} + (\pi^{il} \pi^{jm} H_{klm}),_n \right] \uAplus^n \uAplus^k \uc_i \, \uc_j
\nonumber \\ &&
+ \int_{T[1]\Sigma} 
\!\!\!\!
d^2 \sigma d^2 \theta \,
\tfrac{1}{8}   \pi^{ci} \pi^{ja} \pi^{bd} H_{nab} H_{kcd} \, \uAplus^n \uAplus^k \uc_i \, \uc_j
\, .
\label{supercoordBVaction}
\end{eqnarray}
Here $d^2 \sigma d^2 \theta$ denotes the Berezinian measure on $T[1]\Sigma$. In particular, for every odd coordinate $\theta$: $\int d \theta \,  \theta =1$. Note also that,  for every 2-form $\alpha = \tfrac{1}{2} \rd \sigma^\mu \wedge \rd \sigma^\nu \alpha_{\mu\nu}$ on $\Sigma$, one has
\beq 
\int_\Sigma \alpha = \int_{T[1]\Sigma} 
\!\!\!\!
d^2 \sigma d^2 \theta \, \tfrac{1}{2} \theta^\mu \theta^\nu \alpha_{\mu\nu} \, .\label{superintegration}
\eeq 

In addition to $\sd$, there is a second canonical vector field on $T[1]\Sigma$, 
the Euler vector field ${\bm \varepsilon}$:
\begin{eqnarray}
\bm{\varepsilon} = \theta^{\mu} \tfrac{\partial}{\partial \theta^{\mu}},
\end{eqnarray}
the eigenvalues of which count the polynomial degree in $\theta^{\mu}$.

This now puts us into the position to reexpress the BV data in terms of the superfields  $\bx$ and $\ba$. We first present the result for $S_{BV}[\bx,\ba]=S_{BV}[X,A,c,X^+,A^+,c^+]$:
\begin{framed}
\begin{eqnarray}
S_{BV} 
&=& 
\int_{T[1]\Sigma} 
\!\!\!\!
d^2 \sigma d^2 \theta \,
\left[\ba_i \, \sd \bx^i 
+ \tfrac{1}{2} \pi^{ij}(\bx)\, \ba_i \ba_{j} \right]
+ \int_{T[1]N} \!\!\!\! d^3 \sigma d^3 \theta \, H(\bx)
\nonumber \\ && 
+ \int_{T[1]\Sigma} \!\!\!\! d^2 \sigma d^2 \theta \, \left[\tfrac{1}{4} 
(\pi^{il} \pi^{jm} H_{lmk})(\bx) \, \ba_i \ba_j \bm{\varepsilon} \bx^k 
%\rc{+} 
+ \tfrac{1}{2} (\pi^{il} H_{jkl}) (\bx)\, \ba_i (\sd \bx^j) \bm{\varepsilon} \bx^k 
\right]
\nonumber \\ &&
+ \int_{T[1]\Sigma} \!\!\!\! d^2 \sigma d^2 \theta \, \left[ \tfrac{1}{8} (\pi^{im} \pi^{jn} \pi^{pq}  H_{mql}H_{npk})(\bx)\,
\ba_i \ba_j (\bm{\varepsilon} \bx^k) \bm{\varepsilon} \bx^l
\right].
\label{superfieldBVaction}
\end{eqnarray}
\end{framed}
Here the derivations $\sd$ and $ \bm{\varepsilon}$ only act on the field right next to them; in two terms we emphasized this by putting brackets around the respective expression. A term such as $(\pi^{il} \pi^{jm} H_{lmk})(\bx)$ stands for $\pi^{il}(\bx)\pi^{jm} (\bx)H_{lmk}(\bx)$, where each of the corresponding functions is formally Taylor expanded by means of \eqref{2Dsuperfield1}.

For convenience of the reader we now provide some of the details for seeing explicitly  
 that \eqref{superfieldBVaction} follows from \eqref{supercoordBVaction}. For the first two terms, the rewriting is the same as for the PSM. In particular, the first term, 
\begin{align}
& \ba_i \sd \bx^i 
= \uA_i \sd X^i + \uc_i \sd \uAplus^i
\nonumber \\ 
& \qquad\qquad = \uA_i \sd X^i + \uAplus^i \sd \uc_i - \sd (\uAplus^i \uc_i), \label{firstterm}
\end{align}
gives rise to two of the terms in \eqref{supercoordBVaction} after noting that the last term above drops out since $\Sigma$ has no boundary. The second term in  \eqref{superfieldBVaction} is developed as follows:
\begin{align}
& \tfrac{1}{2} \pi^{ij}(\bx) \ba_i \ba_j
= - \pi^{ij}(X) \uXplus_i \uc_j + \tfrac{1}{2} \pi^{ij}(X) \uA_i \uA_j 
\nonumber \\ &
\qquad + \pi^{ij},_k \uAplus^k \uA_i \uc_j
+ \tfrac{1}{2} \pi^{ij},_k(X) \ucplus^k \uc_i \uc_j 
%\nonumber \\ & \qquad 
- \tfrac{1}{4} \pi^{ij},_{nk}(X) \uAplus^n \uAplus^k \uc_i \uc_j \, .
\end{align}
The remaining parts are the new ones. For the Wess-Zumino term there is not much to do, since every $\sd$ already produces one $\theta$:
\begin{align}
& \tfrac{1}{3!} H_{ijk}(\bx) \sd \bx^i \sd \bx^j \sd \bx^k
= \tfrac{1}{3!} H_{ijk}(X) \sd X^i \sd X^j \sd X^k .
\end{align}
Since also $\bm{\varepsilon}$ yields terms of at least order one in $\theta$, the following two terms give only one contribution to the expansion each:
\begin{align}
&(\pi^{il} H_{jkl}) (\bx)\, \ba_i (\sd \bx^j) \bm{\varepsilon} \bx^k  
= \pi^{il} (X)H_{jkl} (X) \uc_i \,  (\sd X^j) \uAplus^k ,
 \\ &(\pi^{im} \pi^{jn} \pi^{pq}  H_{mql}H_{npk})(\bx)\,
\ba_i \ba_j (\bm{\varepsilon} \bx^k) \bm{\varepsilon} \bx^l = \pi^{im} \pi^{jn} \pi^{pq}  H_{mql}H_{npk}\,
\uc_i \uc_j \uAplus^k \uAplus^l. \nonumber
\end{align}
There is one $H$-dependent term of the superfield formulation that unites three terms of \eqref{supercoordBVaction} or \eqref{flatBVaction}. It is the remaining term of \eqref{superfieldBVaction}:  
\begin{align}
& \pi^{il} \pi^{jm} H_{lmk}(\bx) \ba_i \ba_j \bm{\varepsilon} \bx^k
= 2 \pi^{il} \pi^{jm} H_{lmk} \uAplus^k \uA_i \uc_j
\nonumber \\ 
&
\qquad + 2 \pi^{il} \pi^{jm} H_{lmk} \ucplus^k \uc_i \uc_j
+ ( \pi^{il} \pi^{jm} H_{lmk}),_n \uAplus^n \uAplus^k \uc_i \uc_j. \label{lastterm}
\end{align}
The result \eqref{superfieldBVaction} now follows by term-wise comparison of Eqs.\ \eqref{firstterm}-\eqref{lastterm} with \eqref{supercoordBVaction}, in part after renaming and permuting indices. 

\vspace{5mm}

We are left with showing that the BV symplectic form \eqref{componentBVsymplecticform} combines into the natural symplectic form of \eqref{Homunderline}:
\begin{eqnarray} \label{omega}
\omega &=& \int_{T[1]\Sigma} d^2 \sigma d^2 \theta \,
\delta \bx^i \delta \ba_i.
\label{superBVsymplecticformxy}
\end{eqnarray}
Here $\delta$ denotes the de Rham differential on field space again. Note that now, in addition to the form degree fdeg on $\Sigma$ and the ghost degree gh of an object, the commutativity properties will be also determined by the form degree Fdeg on field space; instead of \eqref{deg}, we now have   \beq \label{tdeg} 
\mathrm{tdeg}(\phi):= \mathrm{Fdeg}(\phi) + \mathrm{fdeg}(\phi)+\mathrm{gh}(\phi) \eeq as the total degree determining the commutativity properties of quantities like $\delta \bx^i$ and $\delta \ba_i$ in the Bernstein-Leites conventions. %yyyy
This in turn requires a  generalizition of the sign conversion rule \eqref{signchangingrule}.

If $\phi$ and $\psi$  now denote homogeneous elements on this bigger space, including differential forms on the mapping space. Then, in the Deligne conventions, the wedge product has the following property:
\begin{eqnarray} \label{hihi}
\phi \wedge \psi = (-1)^{\mathrm{fdeg}(\phi)\mathrm{fdeg}(\psi)+\mathrm{gh}(\phi)\mathrm{gh}(\psi) + \mathrm{Fdeg}(\phi)\mathrm{Fdeg}(\psi)
} \, \psi \wedge \phi .
\end{eqnarray}
 We now put 
\begin{eqnarray}
\phi \wedge \psi 
&\mapsto&
(-1)^{\mathrm{gh}(\phi)\mathrm{fdeg}(\psi) + \mathrm{gh}(\phi)\mathrm{Fdeg}(\psi) + \mathrm{Fdeg}(\phi)\mathrm{fdeg}(\psi)} \, \uphi \, \upsi.
\label{signchangingrule12}
\end{eqnarray}
This evidently generalizes \eqref{signchangingrule}, to which it reduces if Fdeg vanishes. A direct calculation, using \eqref{hihi}, now shows that indeed 
\begin{eqnarray}
\uphi  \, \upsi = (-1)^{(\mathrm{gh}(\uphi) + \mathrm{fdeg}(\uphi) + \mathrm{Fdeg}(\uphi))((\mathrm{gh}(\upsi) + \mathrm{fdeg}(\upsi) + \mathrm{Fdeg}(\upsi)} \upsi \, \uphi 
= (-1)^{\mathrm{tdeg}(\uphi) \, \mathrm{tdeg}(\upsi)} \upsi \,  \uphi ,
\end{eqnarray}
in agreement with the  Bernstein-Leites conventions. As before, one may prove, moreover, that the sign changing rule \eqref{signchangingrule12} is consistent with associativity.

By substituting the expansions \eqref{2Dsuperfield1} and \eqref{2Dsuperfield2} into \eqref{superBVsymplecticformxy}, we obtain
\begin{eqnarray} 
\omega &=& \int_{T[1]\Sigma} d^2 \sigma d^2 \theta \,
\delta \bx^i \delta \ba_i,
\nonumber \\
&=& \int_{T[1]\Sigma} d^2 \sigma d^2 \theta \,
(\delta X^i \delta \uXplus_i - \delta \uAplus^{i} \delta \uA_i
- \delta \ucplus^{i} \delta \uc_i ).
\nonumber\\
&=& \int_{T[1]\Sigma} d^2 \sigma d^2 \theta \,
(\delta X^i \delta \uXplus_i - \delta \uA_i \delta \uAplus^{i}
- \delta \uc_i \delta \ucplus^{i}).
\end{eqnarray}
Reversing the map \eqref{signchangingrule12},  we now obtain
\begin{eqnarray} 
\omega  
&\mapsto & \int_{\Sigma} 
(\delta X^i \wedge \delta X^+_i + \delta A_i \wedge \delta A^{+i}
+ \delta c_i \wedge \delta c^{+i}),
\end{eqnarray}
in agreement with \eqref{componentBVsymplecticform}. 
Note that using Bernstein-Leites sign conventions also for differential forms on field space, the BV symplectic form $\omega$ has symmetry properties as they are consistent with the one for the bracket introduced above, see, e.g., \eqref{equation}.

We remark in parenthesis that the requirement that the fields of the previous subsections recombine into \eqref{superBVsymplecticformxy} together with the desire that the superfields $\bx$ and $\ba$ contain the classical fields $X$ and $A$ without any additional prefactors, completely fixes the signs in the expansion \eqref{2Dsuperfield1} and \eqref{2Dsuperfield2}.

The BV brackets induced from the BV symplectic form \eqref{superBVsymplecticformxy} are
\begin{eqnarray}
 (\bx^i(\sigma, \theta), \ba_j(\sigma^{\prime}, \theta^{\prime})) 
&= &\delta^i{}_j \delta^2(\sigma- \sigma^{\prime})
\delta^2(\theta - \theta^{\prime}),
\nonumber
\\
(\bx^i(\sigma, \theta), \bx^j(\sigma^{\prime}, \theta^{\prime})) 
&=& 0 ,\label{superfieldXABVbracket}\\
 (\ba_i(\sigma, \theta), \ba_j(\sigma^{\prime}, \theta^{\prime})) &=&0.\nonumber
\end{eqnarray} 
Since for an odd variable $\theta$ one has $\delta(\theta)=\theta$, here 
$\delta^2(\theta - \theta^{\prime})
= (\theta^0 - \theta^{\prime 0})(\theta^1 - \theta^{\prime 1})
$. We recover the BV brackets \eqref{a} of the component fields by expanding both sides of the first equation of \eqref{superfieldXABVbracket} in polynomials of $\theta^{\mu}$ and $\theta^{\prime \nu}$.
For example, the second BV bracket in \eqref{a}  between $A_{0j}$ and $A_1^{+i}$ follows in this way by collecting the $\theta^1 \theta^{\prime 0}$-terms on both sides: on the left-hand side this gives
$- \theta^1 \theta^{\prime 0} (A_1^{+i}(\sigma), A_{0j}(\sigma^{\prime}))$, while on the right-hand side of \eqref{superfieldXABVbracket} one finds
$\theta^1 \theta^{\prime 0} \delta^i_j \delta^2(\sigma- \sigma^{\prime})$; using the first equation in \eqref{equation} and the inverse to \eqref{newbracket}, this indeed yields the second equation of \eqref{a}.

While the validity of the master equation for the functional \eqref{superfieldBVaction},
\beq  (S_{BV},S_{BV})=0 ,
\eeq 
follows by construction, it can be also readily verified by means of the more compact bracket relations \eqref{superfieldXABVbracket} for the superfields.

%%%%%%%%%%%%%%%%%%%%%%%%%%%%%%%%%%%%%%%%%%%%%%%%%%%%%%%%%%%%%%%%%%%%%%%%%%%
\subsection{Covariance with superfields} \label{sec:Covsu}
The functional $S_{BV}$, written in terms of superfields as in \eqref{superfieldBVaction}, looks manifestly covariant with respect to a change of coordinates on $M$, provided only we can establish that $\sd \bx^i$ and  $\bm{\varepsilon} \bx^i$ transform like the components of a vector field on $M$. This, however, follows from the fact that both $\sd$ and $\bm{\varepsilon}$ are derivations. 

In the following we will provide further details about this. Let $U\cong \R^n$ and $\widetilde U \cong \R^n$ be two local charts on $M$ with coordinates $x^i$ and $\widetilde{x}^i$, respectively. Denote be $f \colon U\cap \widetilde U \to U\cap \widetilde U$ the transition function, such that $\widetilde{x}^i = f^i(x)$, which is sometimes also simply written as $\widetilde{x}^i = \widetilde{x}^i(x)$. Denote by $(M^i_j)$ the Jacobian matrix of this transformation (see, equivalently, Eq.\ \eqref{Jacobian})
\beq M^i_j(x) := \frac{\partial f^i}{\partial x^j} \, .
\eeq 

We implement this change of coordinates on the superfields $(\bx,\ba)$ as follows:
\begin{eqnarray}
\widetilde{\bx}^i &=& \widetilde{x}^i(\bx) \equiv f^i(\bx)\, ,
\qquad
\widetilde{\ba}_i = \ba_j \,  \frac{\partial x^j}{\partial \widetilde{x}^i}(\widetilde{x}^i(\bx)) 
=\ba_j \, (M^{-1})_i^j(\bx) \, , 
\label{targetdiffeo}
\end{eqnarray}
where the functions on the right-hand side are again formally Taylor expanded with respect to the decomposition \eqref{2Dsuperfield1}. Explicitly this becomes 
\begin{align}
\widetilde{\bx}^i &= f^i(X) - M^i_j(X) \uAplus^j + M^i_j(X) \ucplus^j
+ \tfrac{1}{2} M^i_{j,k}(X) \uAplus^j \uAplus^k,
%\label{superdiffeo1}
\nonumber
\\
 \widetilde{\ba}_i %&=& 
%\ba_j \frac{\partial \bx^j}{\partial \widetilde{\bx}^i}
%\ba_j  \, (M^{-1})_i^j(\bx) ???
%\nonumber \\ 
&= \left(-\uc_j + \uA_j + \uXplus\!\!\!_j \right)
%\nonumber \\ & \quad \times 
\left[(M^{-1})^j_i + (M^{-1})^j_{i,k} \, \left(- \uAplus^k + \ucplus^k \right)
+ \tfrac{1}{2} (M^{-1})^j_{i,kl}  \, \uAplus^k \uAplus^l \right], 
\label{superdiffeo12}
\end{align}
where also in the second equation all matrices and their derivatives depend on $X$. 
Sorting terms according to their order in $\theta^\mu$, these equations imply the following transformation properties for the component fields
\cite{Losev} 
\begin{eqnarray}
\widetilde{\xzero}^{i} &= & f^{i}(\xzero), \nonumber%\label{n1}
\\
\widetilde{\uAplus^i} &=& M^i_j \, \uAplus^j,\nonumber%\label{n2}
\\
\widetilde{\ucplus^i} &=& M^i_j \, \ucplus^j + \tfrac{1}{2} M^i_{j,k} \, \uAplus^j \uAplus^k,\nonumber
\\
\widetilde{\uc}_{i} &=& (M^{-1}){}_i^j \, \uc_j,\label{n3}%\label{n4}
\\
\widetilde{\uA_i}\nonumber% \label{n5}
&=& (M^{-1}){}_i^j  \, \uA_j + (M^{-1})^j_{i,k} \, %M^k_l \, 
\uAplus^k \uc_j ,
%M^{-1}{}_i^m M^{-1}{}_l^j \partial_k M^l_m  \, A^{+k} c_j,
\\
\widetilde{\uXplus\!\!\!_i}\nonumber
&=& (M^{-1}){}_i^j \, \uXplus\!\!\!_j - (M^{-1})^j_{i,k} %M^k_l 
\, (\uAplus^k \uA_j + \ucplus^k \uc_j) 
 %\widetilde{\partial}_k M^{-1}{}_i^j M^k_l (A^{+l} A_j - c^{+l} c_j) 
\nonumber \\ && 
-  \tfrac{1}{2} (M^{-1})^j_{i,kl} %M^k_mM^l_n 
\, \uAplus^k \uAplus^l \uc_j .\nonumber
%- \frac{1}{2} \widetilde{\partial}_i \widetilde{\partial}_k M^{-1}{}_l^j 
%M^k_m M^l_n A^{+m} A^{+n} c_j,
\end{eqnarray}
%where $M^i_j = M^i_j(X)$ and we use the formula,
%\begin{eqnarray}
%M^i_l \widetilde{\partial}_j (M^{-1})^l_k M^j_m M^k_n = \partial_m M^i_n.
%\end{eqnarray}
Since matrices multiplied by their inverses give the identity also if expanded in terms of a formal Taylor series, it follows from \eqref{targetdiffeo} that one has, e.g., $\widetilde{\pi}^{ij}(\widetilde{\bx})\, \widetilde{\ba}_i \widetilde{\ba}_{j} = {\pi}^{ij}({\bx})\, {\ba}_i {\ba}_{j}$, where  ${\pi}^{ij}$ and $\widetilde{\pi}^{ij}$ denote the components of $\pi$ in the coordinate system $(x^i)$ and $(\widetilde{x}^i)$, respectively. We already argued that $\bm{\varepsilon} \bx^i$ transforms as one expects, but we verify this also explicitly by the above formulas: 
\begin{eqnarray}
\widetilde{\bm{\varepsilon} \bx^i} &=& - \widetilde{\uAplus^i} + 2 \widetilde{\ucplus^i}
\nonumber \\ &=& 
- M^i_j \uAplus^{j} + 2 M^i_j \ucplus^{j} + \partial_j M^i_k \uAplus^{j} \uAplus^{k}
\nonumber \\ &=& M^i_j(\bx) (\bm{\varepsilon} \bx^j). 
\end{eqnarray}
So $S_{BV}$ is covariant with respect to local diffeomorphisms of $M$. This also applies to the Liouville 1-form  $$ \vartheta := \int_{T[1]\Sigma} d^2 \sigma d^2 \theta \,
 \ba_i \, \delta \bx^i  \, , $$ 
and to the BV symplectic form \eqref{omega}, $\omega = - \delta \vartheta$. 

Thus, both $(S_{BV},\vartheta)$ and $(S_{BV}^\nabla,\vartheta)$ have a global meaning. They are defined independently of a choice of local coordinates on $M$ (and also on $\Sigma$, certainly). 

In comparison to the previous section, the implementation of the diffeomorphisms of $M$ on the space of fields is different. This becomes particularly obvious if one compares the transformations \eqref{n3} with those found in Sec.\ \ref{sec:Manifest}.  All fields except for $X$, $\uc$, and $\uAplus$ have more complicated transformation rules now, see Eqs.\ \eqref{cc+} and \eqref{X+transform}. On the other hand, from the perspective of the combined superfields, the present transformations are again simple and natural, see \eqref{targetdiffeo}.

%%%%%%%%%%%%%%%%%%%%%%%%%%%%%%%%%%%%%%%%%%%%%
%%%%%%%%%%%%%%%%%%%%%%%%%%
\subsection{Closing the circle} \label{sec:circle}
Regarding $S_{BV}$ as a functional of the fields $(X,A,c)$ and their antifields, we may---in view of their in part unorthodox transformation properties \eqref{n3} and the non-covariant component expansion given in \eqref{flatBVaction}---attempt to adapt the approach of Secs.\ \ref{sec:manifestlyBFV} and \ref{sec:Manifest} and introduce tensorial fields. Note, that although the fields in this section seem to be the same as in the previous one, Sec.\ \ref{sec:BV1}, they transform in a different way under changes of coordinates on $M$. This implies that identifying them (directly) does not have a coordinate independent meaning. 

More explicitly, according to  \eqref{n3}, we see that $X$, $\uc$, and $\uAplus$ transform in the obvious way, but, e.g., the classical field $A \sim \uA$ receives ghost contributions when changing coordinates on $M$, i.e.\ it  is no more a purely classical field after such a change. On the other hand, in Sec.\ \ref{sec:manifestlyBFV} we experienced a similar feature in the context of the (classical) momentum $p\sim A_1$ conjugate to the field $X$. 
We may thus attempt a similar procedure  (see, in particular, \eqref{covp}) to arrive at an inherently 
covariant action functional in the sense of a tensorial behavior of the fields, at the expense of again introducing a connection. 

Let $\stackrel{0}{\nabla}$ denote an arbitrary torsion-free connection on $M$. It is not too difficult to verify that, upon usage of the third and fifth equation of \eqref{n3}, the new fields 
\begin{align} 
\ucplus^{{\nabla}i} & := \ucplus^{i} + \tfrac{1}{2} \!\stackrel{\!\!\circ}{\Gamma_{jk}^i} \uAplus^{j} \uAplus^{k}, \nonumber \\
\uA^{{\nabla}}_i & := \uA_i \, - \stackrel{\!\!\circ}{\Gamma_{ki}^j} \uAplus^k \uc_j, \label{transformationofBVfields1}
\end{align}
transform tensorially with respect to coordinate changes on $M$, i.e.\ that one has, e.g., $\widetilde{\ucplus^{\nabla i}} = M^i_j \, \ucplus^{\nabla j}$. In other words, $\ucplus^{{\nabla}}$ and $\uA^{{\nabla}}$ are sections in pullback bundles.

The transformation property of $\uXplus\!\!\!_{i}\,$ is more involved, see the last equation in  \eqref{n3}. The following field provides a covariant redefinition of it,
\begin{align} 
\widehat{\uXplus^{{\!\nabla}}_{\!\!\!\!i}} & := \uXplus\!\!\!_{i} + \tfrac{1}{2}\big( \partial_{l} \!\stackrel{\!\!\circ}{\Gamma_{ki}^j} - \stackrel{\!\!\circ}{\Gamma_{ik}^s}\stackrel{\!\!\circ}{\Gamma_{sl}^j} 
%+ \stackrel{\!\!\circ}{\Gamma_{is}^j}\stackrel{\!\!\circ}{\Gamma_{kl}^s} 
\big)\,\uAplus^{k} \uAplus^{l} \uc_j 
+ \stackrel{\!\!\circ}{\Gamma_{ij}^k} \left( \uAplus^{j} \uA_k^{} + \ucplus^{j} \uc_k \right),
%\\
%\\
%\uXplus^{{\!\nabla}}_{\!\!\!\!i} & := \rc{\widehat{\uXplus^{{\!\nabla}}_{\!\!\!\!i}} - \tfrac{1}{2}\pi^{km}H_{mij} \left( \uAplus^{j} \uA_k^{{\nabla}} - \ucplus^{{\nabla}j} \uc_k \right), }
\label{transformationofBVfields2}
\end{align}
so that also  $\widehat{\uXplus^{{\!\nabla}}} := \widehat{\uXplus^{{\!\nabla}}_{\!\!\!\!i}} \otimes \udx^i$ is also a globally well-defined section. It turns out to be useful to (covariantly) redefine  this field in the following way:
\begin{align} 
\uXplus^{{\!\nabla}}_{\!\!\!\!i} & := \widehat{\uXplus^{{\!\nabla}}_{\!\!\!\!i}} + \tfrac{1}{2}\!\stackrel{\circ}{R_{kil}^j} \uAplus^{k} \uAplus^{l} \uc_j + \tfrac{1}{2}\pi^{km}H_{mij} \left( \uAplus^{j} \uA_k^{{\nabla}} + \ucplus^{{\nabla}j} \uc_k \right),
\label{transformationofBVfields3}
\end{align}
where $\stackrel{\circ}{R}$ denotes the curvature of $\stackrel{0}{\nabla}$. The reason for this  will become clear below.

Using these variables, by construction, it must be possible to rewrite \eqref{flatBVactionfirst} in terms of purely tensorial coefficient functions. Denoting again by $\nabla$ the connection induced by the above torsion-free one and equation \eqref{Gamma}, the crucial observation is that with the above formulas one finds\footnote{By this notation we mean that when replacing $\uA$, $\uXplus$, and $\ucplus$ by the expressions inverse to Eqs.\ \eqref{transformationofBVfields1} and \eqref{transformationofBVfields3} in 
$S_{BV}\!\left[X, \uA,  \uc , \uXplus,\uAplus , \ucplus \right]$, we obtain the expression given in this formula and denoted by $S_{BV}\!\left[X, \uA^{{\nabla}},  \uc \,, \uXplus^{\!{\nabla}}\!,\uAplus , \ucplus^{{\nabla}} \right]$; in other words, it describes the same abstract functional $S_{BV}$, but written in different fields (coordinates), which are specified behind it in the squared brackets.}
\begin{align}
& S_{BV}\left[X, \uAplus^{\nabla}, \ucplus^{{\nabla}}, \uc^{\nabla}, \uA^{{\nabla}}, \uXplus^{{\nabla}} \right] 
\nonumber \\ &=
\int_{T[1]\Sigma} 
\!\!\!\!
d^2 \sigma d^2 \theta \,
\left( \uA^{{\nabla}}_i \sd \xzero^i 
+ \tfrac{1}{2} \pi^{ij} \uA^{{\nabla}}_i \uA^{{\nabla}}_j
\right)
\nonumber \\ &
+ \int_{T[1]N} 
\!\!\!\!
d^3 \sigma d^3 \theta \,
\tfrac{1}{3!} H_{ijk} \sd X^i \sd X^j \sd X^k
%\nonumber \\ && 
- \int_{T[1]\Sigma} 
\!\!\!\!
d^2 \sigma d^2 \theta \,
\pi^{ij} \uXplus^{\bar{\nabla}}_i \uc_j
\nonumber \\ &
+ \int_{T[1]\Sigma} \!\!\!\! d^2 \sigma d^2 \theta \,
\left(\uAplus^k \left(\mathrm{D} \uc_i - T_k^{ij} \uA^{{\nabla}}_i \uc_j \right)
\right)
%\nonumber \\ &
- 
\int_{T[1]\Sigma} 
\!\!\!\!
d^2 \sigma d^2 \theta \,
\tfrac{1}{2} T_k^{ij} \ucplus^k \uc_i \uc_j
\nonumber \\ &
- \int_{T[1]\Sigma} 
\!\!\!\!
d^2 \sigma d^2 \theta \,
\tfrac{1}{4} S{}_{nk}^{ij} \uAplus^{n} \uAplus^{k} \uc_i \uc_j
\, ,
\label{SBVsurprise}
\end{align}
Replacing superfields by the corresponding differential form fields, this functional is now the \emph{same} as the manifestly covariant BV functional found in the previous section, displayed in Eqs.\ \eqref{universal} and \eqref{Mainresult2prime}.

But also the BV brackets agree with the covariantized ones in the last section. To make this more explicit, let $\uXplus^{\nabla}$ act non-trivially on the basis vectors of the covariantized fields, in complete analogy with \eqref{kommtnoch}: 
\begin{eqnarray}
\ssbv{\upartial_i|_{\sigma}}{\uXplus^{\!\nabla}_{\!\!\!j}(\sigma^{\prime}, \theta^{\prime})}
&=& \Gamma_{ji}^k \,\, \upartial_k|_{\sigma} \,\delta^2(\sigma- \sigma^{\prime})\delta^2(\theta^{\prime}), \nonumber
\\
\ssbv{\udx^i|_{\sigma}}{\uXplus^{\!\nabla}_{\!\!\!j}(\sigma^{\prime}, \theta^{\prime})}
&=& - \Gamma_{kj}^i \, \, \udx^k|_{\sigma} \,\delta^2(\sigma- \sigma^{\prime})\delta^2(\theta^{\prime}), \label{neu}
\end{eqnarray}
where $\Gamma_{ij}^k\equiv \Gamma_{ij}^k(X(\sigma))$. 
With the obvious notation $\uAplus = \uAplus^i \otimes \upartial_i$, $\uA^{\nabla} = \uA^{\nabla}_i \otimes \udx^i$ etc, this convention implies that $\uXplus^{\!\nabla}$ BV-commutes with all the  covariantized fields, i.e.\
\begin{eqnarray}
\ssbv{\uAplus(\sigma, \theta)}{\uXplus^{\!\nabla}(\sigma^{\prime}, \theta^{\prime})} &= &
0,\nonumber \\
\ssbv{\uA^{\nabla}(\sigma, \theta)}{\uXplus^{\!\nabla}(\sigma^{\prime}, \theta^{\prime})} &=& 0,
\end{eqnarray} 
etc,  except for with itself and the fields $X^i$ certainly. The latter bracket can be expressed, for example, in the following way: For every function $f \in C^\infty(M)$ one has
\beq \ssbv{f(X(\sigma))}{\uXplus^{\!\nabla}(\sigma^{\prime}, \theta^{\prime})}
= \delta^2(\sigma- \sigma^{\prime})\delta^2(\theta^{\prime})\, \underline{\rd f} ,
\eeq
where $\underline{\rd f} =  f_{,i}(X(\sigma)) \, \udx^i|_{\sigma}$.  The only other non-vanishing brackets between fundamental fields now take the form 
\begin{eqnarray}
&& \ssbv{\uA^{\nabla}(\sigma, \theta)}{\uAplus(\sigma^{\prime}, \theta^{\prime})}
= \delta^2(\sigma- \sigma^{\prime})\delta^2(\theta- \theta^{\prime})\, \mathrm{id},
\nonumber
\\
&& \ssbv{\uc(\sigma, \theta)}{\ucplus^{\nabla}(\sigma^{\prime}, \theta^{\prime})}
= \delta^2(\sigma- \sigma^{\prime})\delta^2(\theta- \theta^{\prime})\, \mathrm{id},
\end{eqnarray}
where $\mathrm{id}= \udx^i\otimes \upartial_i  \cong  \upartial_i \otimes \udx^i$, as well as 
%\beq \ssbv{\uXplus_{i}^{\nabla}(\sigma, \theta)}{\uXplus_{j}^{\nabla}(\sigma^{\prime}, \theta^{\prime})} = R_{kij}^l (- \uAplus^{k} \uA^{\nabla}_l - \ucplus^{\nabla k} \uc_l) \delta^2(\sigma- \sigma^{\prime})\delta^2(\theta- \theta^{\prime}). \eeq 
%\beq \rc{\ssbv{\uXplus^{\nabla}(\sigma, \theta)}{\uXplus^{\nabla}(\sigma^{\prime}, \theta^{\prime})} = [R_{kij}^l (- \uAplus^{k} \uA^{\nabla}_l - \ucplus^{\nabla k} \uc_l) + \Theta_{ij}^k \uXplus_k^{\nabla}] \udx^i \otimes \udx^j \delta^2(\sigma- \sigma^{\prime})\delta^2(\theta- \theta^{\prime}).} \eeq 
\beq 
\ssbv{\uXplus^{\nabla}(\sigma, \theta)}{\uXplus^{\nabla}(\sigma^{\prime}, \theta^{\prime})} = \delta^2(\sigma- \sigma^{\prime})\delta^2(\theta- \theta^{\prime})[- \langle \uA^{\nabla}, \iota_{\uAplus} R \rangle
- \langle \uc, \iota_{\ucplus^{\nabla}} R \rangle
+ \langle \uXplus^{\nabla}, \Theta \rangle ].
\eeq 
This last equation is analogous to Eq.\ \eqref{ppnabla} for the covariantized momenta in the BFV formulation of the theory.

Comparison of the BV-functional \eqref{SBVsurprise} and the manifestly covariant BV brackets above with the corresponding formulas established in Sec.\ \ref{sec:Manifest} proves the BV-isomorphism announced in \eqref{iso}. 

Here some clarification about the notation is in order. To avoid any confusion, in \eqref{iso} we denote  the BV space of fields obtained in Sec.\ \ref{sec:BV1} by $\cM_{BV}$, its symplectic form by $\omega_{BV}$ and the BV functional by $S_{BV}^\nabla$. It is important to emphasize here that by this we mean the abstract objects, independent of any choice of concrete fields parametrizing them. For example, we can present $\omega_{BV}$ in Darboux form as in \eqref{componentBVsymplecticform}, in the variables used in Sec.\ \ref{sec:normalBVformalism}, or by means of the covariantized fields as in Eqs.\ \eqref{neuesX+}, \eqref{tensor}  introduced in Sec.\ \ref{sec:Manifest}. The same holds true for the functional $S_{BV}^\nabla$, which in terms of the fields used in Sec.\ \ref{sec:normalBVformalism} is displayed in  \eqref{Mainresult2} while for the fields of  Sec.\ \ref{sec:Manifest} it takes the form \eqref{Mainresult2prime}. In addition, for a comparison and isomorphism,  the algebraic structure in Sec.\ \ref{sec:BV1} needs to be modified to the new product and BV bracket (inverse to the BV symplectic form, which is adapted correspondingly) according to \eqref{signchangingrule} and \eqref{newbracket}, respectively.

Likewise, we \emph{define} the BV phase space of the present section by the right-hand side of Eq.\ \eqref{Homunderline}, i.e.\
\beq \cM := \underline{\mathrm{Hom}}(T[1] \Sigma,T^*[1]M) \, , 
\eeq 
which is equipped canonically with a (weakly) symplectic form $\omega$, see Eq.\ \eqref{omega}, inherited from the cotangent bundle structure of the target space and the measure on the source space. In these canonical Darboux coordinates, the BV functional, denoted by $S_{BV}$, takes the form \eqref{superfieldBVaction}. There is a canonical lift of $\mathrm{Diff}(M)$ to the (shifted) cotangent bundle $T^*[1]M$  and ultimately also to $\cM$---as summarized most compactly %in the superfield language 
in \eqref{targetdiffeo}. 

On the other side again, we constructed the $\mathrm{Diff}(M)$-action on $\cM_{BV}$ by hand, implementing it on the fields \eqref{cc+} in terms of linear representations. The group action needed to be non-linear only on the BV momentum $X^+$, conjugate to the $X$-field, so as to leave  $\omega_{BV}$ invariant. Covariantization by means of \eqref{neuesX+}, \eqref{tensor}, replaces this by a linear representation as well---at the expense of giving up Darboux coordinates, but still remaining a BV-symplectic group action certainly. 

We used this covariantized form of the BV formulation of the HPSM as our guiding principle for obtaining the isomorphism \eqref{iso}, which we require to be equivariant with respect to $\mathrm{Diff}(M)$. Indeed, the change of ``coordinates'' (or fields) on $\cM$,
\begin{eqnarray}
X &\mapsto& X , \nonumber \\
\uA&\mapsto& \uA^{{\nabla}} , \nonumber \\
 \uc&\mapsto&  \uc , \nonumber \\
  \uXplus&\mapsto&\uXplus^{\!{\nabla}}  ,  \\
  \uAplus 
  &\mapsto& \uAplus  , \nonumber \\
  \ucplus &\mapsto&  \ucplus^{{\nabla}} , \nonumber 
\end{eqnarray}
obtained, in each local chart $U\subset M$, by combining the formulas \eqref{transformationofBVfields1}, \eqref{transformationofBVfields2}, and \eqref{transformationofBVfields3}, leads to a coordinate representation of the BV data which is \emph{identical} to the one in Sec.\ \ref{sec:Manifest} (after the changes in conventions). Similarly, on both sides the $\mathrm{Diff}(M)$-action on the $X$-fields is lifted in an identical way to a linear representation on the remaining BV-fields. 

To sum up, with this choice of adapted coordinates on both sides, i.e.\ for $\cM_{BV}$ on one side and $\cM$ on the other side, the isomorphism \eqref{iso}, composed with the chart maps, becomes, essentially, nothing but the identity map. In the original Darboux coordinates, on the other hand, the identification  of  $\cM$ with $\cM_{BV}$ is far from obvious, however, as expressed best maybe by the difference of the respective $\mathrm{Diff}(M)$-action (compare, e.g., Eqs.\ \eqref{tensorial} and \eqref{X+transform} with the third and sixth Eq.\ in \eqref{n3}).

%This globally well-defined map intertwines two different BV-symplectic lifts of $\mathrm{Diff}(M)$ to $\cM_{BV}$, which, on one side, is given locally by equations of the form \eqref{targetdiffeo}, \eqref{superdiffeo12}, or \eqref{n3} and on the other side by a purely tensorial representation of the diffeomorphism group---such as, e.g., \eqref{tensorial}---as follows from Eqs.\ \eqref{cc+} and \eqref{tensor}. \vspace{1cm}

%%%%%%%%%%%%%%%%%%%%%%%%%%%%%%%%%%%%%%%%%%%%%%%%%%%%%%%%%%%%%%%%%%%%%%%%%%%
%%%%%%%%%%%%%%%%%%%%%%%%%%%%%%%%%%%%%%%%%%%%%%%%%%%%%%%%%%%%%%%%%%%%%%%%%%%
%%%%%%%%%%%%%%%%%%%%%%%%%%%%%%%%%%%%%%%%%%%%%%%%%%%%%%%%%%%%%%%%%%%%%%%%%%%%%%%%%%%%%%%%%%%%%%%%%%%%%%%%%%%
%%%%%%%%%%%%%%%%%%%%%%%%%%%%%%%%%%%%%%%

%%%%%%%%%%%%%%%%%%%%%%%%%%%%%%%%%%%%%%%%%%%%%%%%%%%%%%%%%%%%%%%%%%%%%%%%%%%
%%%%%%%%%%%%%%%%%%%%%%%%%%%%%%%%%%%%%%%%%%%%%%%%%%%%%%%%%%%%%%%%%%%%%%%%%%%
%%%%%%%%%%%%%%%%%%%%%%%%%%%%%%%%%%%%%%%%%%%%%%%%%%%%%%%%%%%%%%%%%%%%%%%%%%%%%%%%%%%%%%%%%%%%%%%%%%%%%%%%%%%
%%%%%%%%%%%%%%%%%%%%%%%%%%%%%%%%%%%%%%%%%%%%%%%%%%%%%%%%%%%%%%%%%%%%%
%%%%%%%%%%%%%%%%%%%%%%%%%%%%%%%%%%%%%%%%%%%%%%%%%%%%%%%%%%%%%%%%%%%%%
%%%%%%%%%%%%%%%%%%%%%%%%%%%%%%%  APP.  %%%%%%%%%%%%%%%%%%%%%%%%%%%%%%
%%%%%%%%%%%%%%%%%%%%%%%%%%%%%%%%%%%%%%%%%%%%%%%%%%%%%%%%%%%%%%%%%%%%
\section*{Acknowledgments}
N.I. is grateful  to
the Max Plank institute for mathematics, the university of Geneva,
and the Prague Charles university, where part of this work was carried out during 2017 and 2018, for their hospitality. T.S. is grateful to Satoshi Watamura for his invitation to Sendai in 2016, where this collaboration all began. 
We are grateful to the Erwin Schr\"odinger International Institute for Mathematics and Physics for support within the program ``Higher Structures and Field Theory''.

This work was supported by the LABEX MILYON 
(ANR-10-LABX-0070) of Universit\'e de Lyon, within the program ``Investissements d'Avenir'' (ANR-11-IDEX-0007) operated by the French National Research Agency (ANR).

%%%%%%%%%%%%%%%%%%%%%%%%%%%%%%%%%%%%%%%%%%%%%%%%%%%%%%%%%%%%%%%%%%%%%
%%%%%%%%%%%%%%%%%%%%%%%%%%%%%%%  APP.  %%%%%%%%%%%%%%%%%%%%%%%%%%%%%%
%%%%%%%%%%%%%%%%%%%%%%%%%%%%%%%%%%%%%%%%%%%%%%%%%%%%%%%%%%%%%%%%%%%%
%\newpage
\appendix
%%%%%%%%%%%%%%%%%%%%%%%%%%%%%%%%%%%%%%%%%%%%%%%%%%%%%%%%%%%%%%%%
%%%%%%%%%%%%%%%%%%%%%%%%%%%%%%%%%%%%%%%%%%%%%%%%%%%%%%%%%%%%%%%%%%%%
%%%%%%%%%%%%%%%%%%%%%%%%%%%%%%%%%%%%%%%%%%%%%%%%%%%%%%%%%%%%%%%%%%%%
%%%%%%%%%%%%%%%%%%%%%%%%%%%%%%%%%%%%%%%%%%%%%%%%%%%%%%%%%%%%%%%%%%%%

%%%%%%%%%%%%%%%%%%%%%%%%%%%%%%%%%%%%%%%%%%%%%%%%%%%%%%%%%%%%%%%%%%%%
\section{BV \& BFV for the Poisson sigma model}
\label{sec:PSMBVBFV}
\noindent
The classical action of the Poisson sigma model \cite{Ikeda,Schaller-Strobl} results from putting  to zero the 3-form $H$ in the twisted generalization \eqref{classicalactionofHPSM} studied in the present paper:
\begin{eqnarray}
S = \int_{\Sigma} 
\left(A_i \wedge \rd \xzero^i + \tfrac{1}{2} \pi^{ij}(\xzero) A_i \wedge A_j 
\right).
\label{classicalactionofPSM}
\end{eqnarray}
It agrees with (the classical part of) the AKSZ model \cite{AKSZ} in two dimensions \cite{CFAKSZ}. In this appendix we recall the BV formulation \cite{Cattaneo-Felder} of the action \eqref{classicalactionofPSM}---which is the starting point in the AKSZ approach---as well as the BFV formulation of the PSM.

%%%%%%%%%%%%%%%%%%%%%%%%%%%%%%%%%%%%%%%%%%%%%%%%%%%%%%%%%%%%%%%%%%%%%%
\subsection{Recollection of BV for the PSM}\label{sec:PSMBV}
\noindent
The BV symplectic form is
\begin{eqnarray}
\omega_{BV} &=& \int_{T[1]\Sigma} d^2r\sigma d^2\theta\,
\delta \bx^i \delta \ba_i,
\label{BVsymplecticformPSM}
\end{eqnarray}
where $\bx^i$ and $\ba_i$ can be viewed as unconstrained superfields on $T[1]\Sigma$ of total degree zero and one, respectively, see \eqref{2Dsuperfield1} and \eqref{2Dsuperfield2} in the main text. 

The BV action takes the form of the classical action \eqref{classicalactionofPSM}, written in terms of the superfields:
\begin{equation}
S_{BV}
= \int_{T[1]\Sigma}
%\!\!\!\! 
d^2\sigma d^2\theta\,
 \left(
\ba_i \sd \bx^i +  \tfrac{1}{2} \pi^{ij}(\bx) \ba_i \ba_j
\right).
\end{equation}
Expanded in terms of the original fields, the ghost fields, and their antifields, this becomes:\footnote{Using the superfield expansions \eqref{2Dsuperfield1} and \eqref{2Dsuperfield2}, one ends up with an apparently different sign for the last term. However, this then is still a function integrated over $T[1]\Sigma$; transcribing this as an integral over a differential form on $\Sigma$, this term receives a change in its sign, as explained in detail in Sec.\ \ref{superfieldBVformalism} (see in particular Eq.\ \eqref{signchangingrule}).}
%\begin{eqnarray}
%S_{BV} 
%&=& 
%\int_{T[1]\Sigma} 
%\!\!\!\!
%d^2 \sigma d^2 \theta \,
%\left[(\theta^{\mu} A_{\mu i}) \sd \xzero^i 
%+ \tfrac{1}{2} \pi^{ij}(\xzero) (\theta^{\mu} A_{\mu i})(\theta^{\nu} A_{\nu j})\right.
%\nonumber \\ &&
%\left.
%- \pi^{ij} \left(\tfrac{1}{2} \theta^{\mu} \theta^{\nu} \xzero^+_{\mu \nu i}\right) c_j
%- \theta^{\mu} A_{\mu}^{+i} \left(\sd c_i + \pi^{jk},_i \theta^{\mu} A_{\mu j} c_k \right)
%\right.
%\nonumber \\ &&
%\left.
%+ \tfrac{1}{2} \pi^{ij},_k \left(\tfrac{1}{2} \theta^{\mu} \theta^{\nu} c^{+k}_{\mu \nu} \right) c_i c_j
%+ \tfrac{1}{4} \pi^{ij},_{nk} (\theta^{\mu} A_{\mu}^{+n}) (\theta^{\nu} A_{\nu}^{+k}) c_i c_j
%\right].
%\label{PSMsupercoordBVaction}
%\end{eqnarray}
%In differential forms,
\begin{eqnarray}
S_{BV}&=& \!\!\! \int_{\Sigma} \biggl(A_i \wedge \rd \xzero^i
+ \tfrac{1}{2} \pi^{ij}(\xzero) A_i \wedge A_j 
- 
%\tfrac{1}{2} \pi^{ij}(\xzero)
%A_i A_j - 2 
\pi^{ij}(\xzero) \xzero^{+}_i c_j
%\quad
+ A^{+ i} \wedge \left( \rd c_{i} + \pi^{jk},_i(\xzero) A_j c_k \right)
%\right.
\nonumber \\
&& 
%\left.
+ \tfrac{1}{2} \pi^{ij},_k(\xzero)  c^{+k} c_i c_j
- \tfrac{1}{4} \pi^{ij},_{kl}(\xzero) A^{+k} \wedge A^{+l} c_i c_j
\biggr).
%\nonumber
\label{BVPSM}
\end{eqnarray}
In the AKSZ formalism, it is evident---by its very construction---that this action satisfies the classical master equation $(S_{BV},S_{BV})=0$: Indeed, the AKSZ sigma model can be seen as the Hamiltonian lift 
 of the sum of two commuting differentials to the mapping space $\underline{\mathrm{Hom}}(T[1] \Sigma,T^*[1]M)$, one being the de Rham differential on the source space $T[1]\Sigma$ and the other one the Lie algebroid differential on target space $T^*[1]M$. Note that this construction cannot be applied as such to the $H$-twisted Lie algebroid structure on $T^*[1]M$ since, for $H\neq 0$, the corresponding nilpotent vector field $Q$ is not compatible with the symplectic form on $T^*[1]M$; this, however, is needed for the Hamiltonian lift in the AKSZ formalism, see \cite{AKSZ,CFAKSZ,Ikeda:2012pv,RoytenbergAKSZ}.

%%%%%%%%%%%%%%%%%%%%%%%%%%%%%%%%%%%%%%%%%%%%%%%%%%%%%%%%%%%%%%%%%%%%
\subsection{BFV for the PSM} \label{sec:PSMBFV}
\noindent
The BFV formulation \cite{Schaller-Strobl} of the PSM  permits a superfield description similar to the one above, too. The BFV symplectic form can be written as 
\begin{eqnarray}
\omega_{BFV} &=& \int_{T[1]S^1} d \sigma d \theta \,
\delta \tilx^i \wedge \delta \tila_i.
\label{BFVsymplecticformPSM}
\end{eqnarray} In terms of the expansion \eqref{superfield1d}, this gives rise to the following Poisson brackets for the fields and ghosts:
\begin{eqnarray}
&& \{\xzero^i(\sigma)}{p_{j}(\sigma^{\prime})\}
= \delta^i{}_j\delta(\sigma- \sigma^{\prime}),
%= - \delta^i{}_j\delta(\sigma- \sigma^{\prime}),
\\
&& \{c_j(\sigma)}{
%A_1^{+i}
b^i(\sigma^{\prime})\}
= \delta^i{}_j\delta(\sigma- \sigma^{\prime}).
%= - \delta^i{}_j\delta(\sigma- \sigma^{\prime}),
\end{eqnarray}
Likewise, also the Hamiltonian BFV or BRST charge  permits a compact description in terms of these superfields:
\begin{equation}
S_{BFV}
= \int_{T[1]S^1}
%\!\!\!\! 
d \sigma d \theta \, \left(
\tila_i \bbd \tilx^i +  \tfrac{1}{2} \pi^{ij}(\tilx) \tila_i \tila_j
\right) \, ,
\end{equation}
which, when decomposed into its components takes the form:
\begin{eqnarray} S_{BFV} &=&
\int_{S^1} d \sigma \left( c_i \partial \xzero^i
+ \pi^{ij}(\xzero) c_i p_{j} + \tfrac{1}{2} \pi^{ij},_k(\xzero) b^k c_i c_j
\right).
\label{BFVPSM}
\end{eqnarray}
It satisfies $\{S_{BFV}, S_{BFV}\} =0$.
The BFV symplectic form and the BFV-BRST charge are just a one-dimensional super-extension of the classical symplectic form (on the cotangent space of  loop space) and the classical action, respectively.

%%%%%%%%%%%%%%%%%%%%%%%%%%%%%%%%%%%%%%%%%%%%%%%%%%%%%%%%%%%%%%%%%%%%%%%%%%%
%%%%%%%%%%%%%%%%%%%%%%%%%%%%%%%%%%%%%%%%%%%%%%%%%%%%%%%%%%%%%%%%%%%%%%%%%%%
%%%%%%%%%%%%%%%%%%%%%%%%%%%%%%%%%%%%%%%%%%%%%%%%%%%%%%%%%%%%%%%%%%%%%%%%%%%
\section{Naive generalization of AKSZ does not work for the HPSM}\label{sec:naive}
We already explained at the end of Sec.\ \ref{sec:PSMBV}, why---at least in its direct, unmodified form---the AKSZ procedure cannot be applied to the HPSM for non-vanishing $H$; in fact, the PSM is the most general AKSZ sigma model for a two-dimensional choice of $\Sigma$, see, e.g., \cite{RoytenbergAKSZ}. One may still hope that there may be \emph{some} elegant superfield formalism that yields the BV theory of the HPSM without much explicit work. We want to show here that, at least in the most direct way, this is not the case. 

Rewriting the classical symplectic form 
\eqref{classicalsymplecticform} of the HPSM and its classical action \eqref{classicalactionofHPSM} in terms of super-fields on $T[1]\Sigma$ as before, one would arrive at a BV symplectic form
\begin{eqnarray}
\omega_{BV} &=& \int_{T[1]\Sigma} d^2 \sigma d^2 \theta\,
\left(
\delta \bx^i \delta \ba_i
+  \tfrac{1}{2} H_{ijk}(\bx)  \sd \bx^k \delta \bx^i \delta \bx^j
\right)\label{omegaBVnaive}
\end{eqnarray}
and a BV action
\begin{eqnarray}
S_{BV}
&=& \int_{T[1]\Sigma}
%\!\!\!\! 
d^2 \sigma d^2 \theta\ \left(
\ba_i  \sd \bx^i +  \tfrac{1}{2} \pi^{ij}(\bx) \ba_i \ba_j
\right)
\nonumber \\ &&
+ \int_{T[1]N}
%\!\!\!\! 
d^3 \sigma d^3 \theta \
\tfrac{1}{3!} H_{ijk}(\bx)  \sd \bx^i  \sd \bx^j  \sd \bx^k.
\label{naiveBVPSM}
\end{eqnarray}
However, this naive candidate of the BV action does not satisfy the classical master equation.
In fact, with the induced BV brackets
\begin{eqnarray}
\left(\bx^i(\sigma, \theta),\ba_j(\sigma^{\prime}, \theta^{\prime})\right) 
&=& \delta^i{}_j \delta^2(\sigma- \sigma^{\prime})
\delta^2(\theta - \theta^{\prime}),
\\
\left(\ba_i(\sigma, \theta),\ba_j(\sigma^{\prime}, \theta^{\prime})\right)
&=& -  H_{ijk}(\bx)  \sd \bx^k 
\delta^2(\sigma- \sigma^{\prime}) \delta^2(\theta - \theta^{\prime})
\end{eqnarray} one obtains
\begin{eqnarray}
\left(S_{BV},S_{BV}\right) &=& \int_{T[1]\Sigma}
%\!\!\!\! 
d^2 \sigma d^2 \theta\ 
\left(
\pi^{kl} H_{ijl}  \sd \bx^i  \sd \bx^j \ba_k 
- \pi^{jm} \pi^{kn} H_{imn}  \sd \bx^i \ba_j \ba_k 
\right.
\nonumber \\ &&
\left.
+ \tfrac{1}{3} \pi^{il} \pi^{jm} \pi^{kn} H_{lmn} \ba_i \ba_j \ba_k 
\right) \label{intermediary}
\\
&=& - \int_{T[1]\Sigma}
%\!\!\!\! 
d^2 \sigma d^2 \theta\ 
%\left(
\tfrac{1}{3} H_{ijk} ( \sd \bx^i + \pi^{il} \ba_l)( \sd \bx^j + \pi^{jm} \ba_m)
( \sd \bx^k + \pi^{kn} \ba_n)
%\right)
.
\nonumber
\end{eqnarray}
This is zero only for $H = 0$. Note that the second expression shows that the master equation holds true on-shell (in some sense), 
i.e.\ upon usage of the obvious superfield extension of the field equation \eqref{eom1}. 
A similar statement also follows if one drops the $H$-contribution to \eqref{omegaBVnaive}, taking $\omega_{BV}$ to have Darboux form, the choice made in \cite{Park}. The on-shell validity of the master equation  is, however, not sufficient for the BV formulation. In fact, this is precisely the reason imposing the use of the more elaborate BV formalism over its simpler BRST  version (see also the discussion around \eqref{BRST2A}).

Let us try a more general ansatz. Suppose that one is not sure of how to distribute the 3-form $H$ between the BV symplectic form and the BV action (see also Appendix \ref{appC} for a related question). We thus introduce two parameters $\alpha,\beta \in \R$ and attempt to use 
\begin{eqnarray}
\omega_{BV} &=& \int_{T[1]\Sigma} d^2 \sigma d^2 \theta\,
\left(
\delta \bx^i \delta \ba_i
+ \tfrac{\alpha}{2} H_{ijk}(\bx)  \sd \bx^k \delta \bx^i \delta \bx^j
\right) \label{omegaa}
\end{eqnarray}
as the BV symplectic form and 
\begin{align}
S_{BV}
&= \int_{T[1]\Sigma}
%\!\!\!\! 
d^2 \sigma d^2 \theta\ \left(\ba_i  \sd \bx^i +  \tfrac{1}{2} \pi^{ij}(\bx) \ba_i \ba_j
+  \tfrac{\beta}{2} B_{ij}(\bx)  \sd \bx^i  \sd \bx^j
\right) \label{Sa}
\end{align}
as the BV action. Here we assumed, for simplicity, that $H = \rd B$. The previous ansatz is reproduced for $\alpha=\beta=1$, the considerations of \cite{Park} correspond to $\alpha=0$ and $\beta=1$, 
but one might still hope that for some other choice of parameters there is some decisive cancellation. However, a direct calculation yields 
\begin{eqnarray}
\left(S_{BV},S_{BV}\right) &=& \int_{T[1]\Sigma}
%\!\!\!\! 
d^2 \sigma d^2 \theta\ 
\left(\pi^{ij},_l \pi^{lk} \ba_i \ba_j \ba_k 
\right.
\nonumber \\ &&
\left.
- \alpha \pi^{jm} \pi^{kn} H_{imn}  \sd \bx^i \ba_j \ba_k 
+ (2 \alpha - \beta) \pi^{kl} H_{ijl}  \sd \bx^i  \sd \bx^j \ba_k 
\right).
\label{masterequationmulambda}
\end{eqnarray}
This calculation holds true for \emph{every} choice of the bivector field $\pi$ and the 3-form $H=\rd B$, not necessarily restricted by means of the twisted Poisson condition \eqref{PiPi}---which was used to obtain \eqref{intermediary}, to which it then reduces for the initial choice of constants.

We see that, for $H \neq 0$, \eqref{masterequationmulambda} vanishes if and only if $\alpha = 0$, $\beta = 0$, and $\pi$ Poisson. But in this case $H$ drops out from both \eqref{omegaa} and \eqref{Sa} and we only reproduce the BV formulation of the untwisted Poisson sigma model (see Appendix \ref{sec:PSMBV}). 

In terms of superfields as above and for the case that one writes the BV symplectic form in Darboux coordinates as in \eqref{BVsymplecticformPSM}, the correct BV extension of the HPSM is found to be rather of the involved form \eqref{superfieldBVaction} in the main text. A qualitatively new feature of this functional, which is at least not easy to guess, is the appearance of the Euler vector field $\bm{\varepsilon}$ inside $S_{BV}$---a fact that holds true also  for the simpler BFV functional $S_{BFV}$ of the HPSM, see \eqref{superBFVBRSTcharge}.

%%%%%%%%%%%%%%%%%%%%%%%%%%%%%%%%%%%%%%%%%%%%%%%%%%%%%%%%%%%%%%%%%%%%
%%%%%%%%%%%%%%%%%%%%%%%%%%%%%%%%%%%%%%%%%%%%%%%%%%%%%%%%%%%%%%%%%%%%
%%%%%%%%%%%%%%%%%%%%%%%%%%%%%%%%%%%%%%%%%%%%%%%%%%%%%%%%%%%%%%%%%%%%
\section{BV symplectic form with $H$}\label{appC}

The graded symplectic form $\omega_{BFV}$  of the HPSM 
necessarily containts the 3-form $H$, see \eqref{BFVsymplecticform} or \eqref{superBFVsymplecticform}. This is not the case for the BV symplectic form $\omega_{BV}$, see \eqref{componentBVsymplecticform} or \eqref{superBVsymplecticformxy} (in fact, for degree reasons,  $\omega_{BV}$ is always exact \cite{RoytenbergAKSZ}). One may be curious, however, how the BV functional of the HPSM would look like if one adds an $H$-contribution to the BV symplectic form similar to the one in $\omega_{BFV}$. %To answer this question briefly is the purpose of the appendix. 

Indeed, in terms of the new field\footnote{This transformation is analogous to Eq.\ (5.5) in \cite{Ikeda-Strobl3} and the proof of the statements below follow by the same steps as those provided there.} 
\begin{eqnarray}
\ba_i &=& \cba_{i} 
%\rc{+} 
+ H_{ijk}(\bx) (\varepsilon \bx^j) \bbd \bx^k,
\label{covariantHsuper}
\end{eqnarray}
the BV symplectic form \eqref{superBVsymplecticformxy} 
changes to
\begin{eqnarray}
\omega_{BV}
%&=& \int_{T[1]\Sigma} d^2 \sigma d^2 \theta \, \delta \bx^{i} \delta \ba_i
%\nonumber \\ 
&=& \int_{T[1]\Sigma} d^2 \sigma d^2 \theta \, \left(\delta \bx^{i} \delta \cba_i + \tfrac{1}{2} H_{ijk} \bbd \bx^i \delta \bx^j \delta \bx^k \right).
\label{covHBVsymplectic}
\end{eqnarray}
and the  BV action  \eqref{superfieldBVaction}, 
becomes
\begin{eqnarray}
S_{BV} 
&=& 
\int_{T[1]\Sigma} 
\!\!\!\!
d^2 \sigma d^2 \theta \,
\left[\cba_i \, \sd \bx^i 
+ \tfrac{1}{2} \pi^{ij}(\bx)\, \cba_i \cba_{j} \right]
+ \int_{T[1]N} \!\!\!\! d^3 \sigma d^3 \theta \, H(\bx)
\nonumber \\ && 
+ \int_{T[1]\Sigma} \!\!\!\! d^2 \sigma d^2 \theta \, \left[\tfrac{1}{4} 
(\pi^{il} \pi^{jm} H_{lmk})(\bx) \, \cba_i \cba_j \bm{\varepsilon} \bx^k 
%\rc{-} 
- \tfrac{1}{2} (\pi^{il} H_{jkl}) (\bx)\, \cba_i (\sd \bx^j) \bm{\varepsilon} \bx^k 
\right]
\nonumber \\ &&
+ \int_{T[1]\Sigma} \!\!\!\! d^2 \sigma d^2 \theta \, \left[ \tfrac{1}{8} (\pi^{im} \pi^{jn} \pi^{pq}  H_{mql}H_{npk})(\bx)\,
\cba_i \cba_j (\bm{\varepsilon} \bx^k) \bm{\varepsilon} \bx^l
\right].
\label{AsuperfieldBVaction}
\end{eqnarray}
This answers the question posed in this appendix.

\section{Bianchi identity for the basic curvature $S$}\label{sec:BianchiofS}
In this Appendix we want to provide a non-sophisticated derivation of the Bianchi identity \eqref{BianchiofS} for the tensor $S$, defined in \eqref{S}.\footnote{The tensor $S$ for a general Lie algebroid $E$ appeared first in  \cite{Mayer-Strobl}. Its geometrical meaning is to measure the violation of the compatibility \cite{Blaom} of a connection $\nabla$ on $E$ with the Lie algebroid structure \cite{Kotov-Strobl2}---but see also \cite{Abad-Crainic}, where this tensor was termed ``basic curvature''.}
We do this in terms of a component calculation. In a local coordinate system, the curvature $R$ and the torsion $\Theta$ of the affine $TM$-connection $\Gamma_{ij}^k$ as well as the E-torsion $T$ and the basic curvature $S$ take the following form:
\beqa 
R_{lij}^k &=& 
\partial_i \Gamma_{lj}^k - \partial_j \Gamma_{li}^k
+ \Gamma_{lj}^m \Gamma_{mi}^k - \Gamma_{li}^m \Gamma_{mj}^k,
\\
\Theta^k_{ij} &=& - \Gamma_{ij}^k + \Gamma_{ji}^k = \pi^{kl} H_{ijl},
\\
T_k^{ij} &=& - f_k^{ij} - \pi^{il} \Gamma_{kl}^j - \pi^{lj} {\Gamma_{kl}^i},
\\
S_{ij}{}^{kl} 
&=& \nabla_j 
T_i^{kl}
%{}_{;j} 
- \pi^{mk} R^l_{ijm} + \pi^{ml} R^k_{ijm}
\nonumber \\
&&\!\!\!\!\!\!\!\!\!\!\!\!\!\!\!\!\!\!
\!\!\!\!\!\!\equiv  - f_{(i}^{kl}{},_{j)} 
+ \Gamma_{(ij)}^m f_m^{kl} + 2 \Gamma_{m(j}^{[k} f_{i)}^{l]m} 
%+ \Gamma_{mj}^l f_i^{km}
+ 2 \pi^{m[k}  \Gamma^{l]}_{ij},_m
%+ \pi^{ml} \partial_m \Gamma^k_{ij}
+ 2  \pi^{m[k},_j \Gamma^{l]}_{im}
%+ \partial_j \pi^{ml} \Gamma^k_{im}
- 2 \Gamma^{[k}_{im}\Gamma^{l]}_{nj}  \pi^{mn} 
%- \Gamma^l_{im} \pi^{mn} \Gamma^k_{nj} 
\, . \label{Slocal}
\eeqa
The Bianchi identities for $R_{ijk}^l$ and $\Theta_{ij}^k$ are 
\begin{eqnarray}
&& R_{lij}^k = - R_{lji}^k,
\\
&& R_{[ijk]}^l = \nabla_{[i} \Theta^l_{jk]} - \Theta_{[ij}^m \Theta_{k]m}^l,
\\
&& 
\nabla_{[i} R_{l|jk]}^m
%{}_{;k]} 
+ \Theta_{[ij}^n R_{k]ln}^m =0.
\end{eqnarray}
A twisted Poisson structure on $M$ gives rise to a Lie algebroid structure on $T^*M$, which, by means of a connection with the prescribed torsion as in \eqref{Gamma}, can be expressed as follows: \begin{eqnarray}
&& [\pi^{\sharp}(\alpha), \pi^{\sharp}(\beta)]^{\nabla} 
= \pi^{\sharp}([\alpha, \beta]_{\pi}^{\nabla}),
\label{covariantHTPS1}
\\
&& [[\alpha, \beta]_{\pi}^{\nabla}, \gamma]_{\pi}^{\nabla} +  \mbox{cyclic}(\alpha\beta\gamma) = 0,
\label{covariantHTPS2}
\end{eqnarray}
where $\alpha, \beta, \gamma \in \Omega^1(M)$. Here 
$[X, Y]^{\nabla}$ is the covariantized Lie bracket for vector fields $X, Y \in \mathfrak{X}(M)$, which is defined by replacing the derivative $\partial_i$ by the covariant derivative $\nabla_i$. 
Similarly, $[\alpha, \beta]_{\pi}^{\nabla}$ is the covariantized Koszul bracket   \cite{HellerIkedaWatamura}
\begin{eqnarray}
&& [\alpha, \beta]_{\pi}^{\nabla} 
= L^{\nabla}_{\pi^{\sharp} (\alpha)}\beta - L^{\nabla}_{\pi^{\sharp} (\beta)} \alpha - \nabla(\pi(\alpha, \beta)),
\end{eqnarray}
where  $L_X = \rm{D}\circ \iota_X + \iota_X \circ \rm{D}$ is the covariantized Lie derivative (as before, $\rm{D}$ denotes the exterior covariant derivative induced by the connection $\nabla$).
In local coordinates, \eqref{covariantHTPS1} and \eqref{covariantHTPS2} take the form
\begin{eqnarray}
&& \pi^{l[i} \nabla_l \pi^{jk]}
%{}_{;l}
%+ (ijk \ \mbox{cyclic}) 
=0,
\label{pinablapi}
\\
&& \pi^{m[i} \nabla_m T_l^{jk]}
%{}_{;m} 
- T_{l}^{m[i} T_m^{jk]}
- \pi^{[i|m} \pi^{|j|n} R_{lmn}^{k]} 
%+ (ijk \ \mbox{cyclic}) 
= 0.
\label{pinablaT}
%\\
%&& \pi^{mi} S_{lm}{}^{jk} + \pi^{mi} \nabla_m T_l^{jk}
%- 2 T_{l}^{mi} T_m^{jk} + (ijk \ \mbox{cyclic}) = 0,
\end{eqnarray}
In addition, one may verify the following formulas:
\begin{eqnarray}
&& [\nabla_m, \nabla_k]T_n^{ij}
= \Theta_{km}^l \nabla_l T_n^{ij} 
- R_{nmk}^l T_l^{ij} + R_{lmk}^i T_n^{lj} + R_{lmk}^j T_m^{il}
\label{nablanablaT}
\end{eqnarray}
and (cf.\ Eq.\ \eqref{nablapi})
\begin{eqnarray}
&& - \nabla_k \pi^{ij}
%{}_{;k}
 + \pi^{il} \Theta_{kl}^j = T_k^{ij}.
\label{piThetaT}
\end{eqnarray}
Using \eqref{pinablapi}, \eqref{pinablaT}, \eqref{nablanablaT}, and \eqref{piThetaT}, one now verifies
\begin{eqnarray}
&& \pi^{m[i} \nabla_m S_{nk}{}^{jl]} + T_m^{[ij} S_{nk}{}^{l]m} - T_n^{m[i} S_{mk}{}^{jl]} - T_k^{m[i} S_{nm}{}^{jl]} 
%+ \mathrm{Cycl}(ijl) 
= 0,
\label{BianchiofSapp}
\end{eqnarray}
 which is the searched-for Bianchi identity of the basic curvature $S_{nk}{}^{ij}$.

%%%%%%%%%%%%%%%%%%%%%%%%%%%%%%%%%%
%\newpage

\end{document}